\documentclass[aps,prd,eqsecnum,11pt,showpacs,preprintnumbers,nofootinbib]{revtex4}
\usepackage{amssymb,amsmath,bm,color}
\usepackage{graphicx}
\usepackage{graphics}
\usepackage{float}
\pdfoutput=1
\def\nbox#1#2{\vcenter{\hrule \hbox{\vrule height#2in
\kern#1in \vrule} \hrule}}
\def\sq{\,\raise.5pt\hbox{$\nbox{.09}{.09}$}}
\def\sqb{\,\raise.5pt\hbox{$\overline{\nbox{.09}{.09}}$}}

\newcommand{\bea}{\begin{eqnarray}}
\newcommand{\eea}{\end{eqnarray}}
\newcommand{\be}{\begin{equation}}
\newcommand{\ee}{\end{equation}}
\newcommand{\bes}{\begin{subequations}}
\newcommand{\ees}{\end{subequations}}

\newcommand{\psl}{p \! \! \!  /}
\newcommand{\qsl}{q \! \! \!  /}
\newcommand{\lsl}{l \! \! \!  /}

\def\lag{\langle}
\def\rag{\rangle}

\def\nn{\nonumber \\}
\newcommand{\pq}{p\cdot q}

\newcommand{\rt}[1]{{}}

\begin{document}
{\allowdisplaybreaks

\preprint{\rm LA-UR 08-6329}

\title{The Trace Anomaly and Massless Scalar Degrees of Freedom in Gravity}

\author{Maurizio Giannotti and Emil Mottola}
\affiliation{Theoretical Division, T-2\\
Los Alamos National Laboratory \\
Los Alamos, NM 87545 USA}
\email{maurizio@lanl.gov}
\email{emil@lanl.gov}

\begin{abstract}
\vskip .3cm
The trace anomaly of quantum fields in electromagnetic or gravitational
backgrounds implies the existence of massless scalar poles in physical
amplitudes involving the stress-energy tensor. Considering first the axial 
anomaly and using QED as an example, we compute the full one-loop 
triangle amplitude of the fermionic stress tensor with two current vertices, 
$\langle T^{\mu\nu} J^{\alpha} J^{\beta}\rangle$, and exhibit the scalar pole 
in this amplitude associated with the trace anomaly, in the limit of zero electron 
mass $m\rightarrow 0$. To emphasize the infrared aspect of the anomaly, we 
use a dispersive approach and show that this amplitude and the existence of 
the massless scalar pole is determined completely by its ultraviolet finite terms, 
together with the requirements of Poincar{\'e} invariance of the vacuum, Bose 
symmetry under interchange of $J^{\alpha}$ and $J^{\beta}$, and vector current
and stress tensor conservation. We derive a sum rule for the appropriate positive 
spectral function corresponding to the discontinuity of the triangle amplitude, 
showing that it becomes proportional to $\delta (k^2)$ and therefore contains a 
massless scalar intermediate state in the conformal limit of zero electron mass. 
The effective action corresponding to the trace of the triangle amplitude can be 
expressed in local form by the introduction of two scalar auxiliary fields which 
satisfy massless wave equations. These massless scalar degrees of freedom 
couple to classical sources, contribute to gravitational scattering processes, 
and can have long range gravitational effects. 
\end{abstract}

\date{\today}

\maketitle
\pagebreak

\section{Introduction}

Quantum effects are most commonly associated with short distance physics. The 
basic reason for this is that the fluctuations of most fields have a finite correlation 
length, and hence their correlations fall off rapidly at large distances. For relativistic 
fields which are massive, the correlation length is $m^{-1}$, and the fall off is exponential, 
$e^{-mr}$. This is an elementary example of decoupling. In the limit of very large mass,
$m\rightarrow \infty$, the quantum effects of a heavy field become negligible at any 
finite distance scale.

In the opposite limit of massless fields $m \rightarrow 0$, the correlation 
length becomes infinite. Decoupling no longer holds, and it becomes possible 
for quantum correlations to extend over very great distances on even macroscopic 
scales. Such infrared effects are more pronounced the lower the spacetime 
dimensionality. In $d=2$ the two-point propagator function of a free massless 
scalar field, rather than falling off, grows logarithmically in the separation of the 
points. Massless conformal field theories in $d=2$ have been studied extensively 
by a variety of methods, and it is clear that their fluctuations are non-negligible 
and have important physical consequences in the infrared \cite{CFT}. 

In massless field theories in two dimensions, the conformal group algebra and its 
central extension play important roles. The central term, a Schwinger term in the 
commutation algebra of stress-tensors in flat space, may also be recognized as the 
trace anomaly of the stress-energy tensor defined in curved space \cite{CFT,BirDav}. 
The anomaly in the trace of the stress-energy tensor, corresponds to a well-defined 
additional term in the effective action which has long range effects \cite{Poly}. As one 
illustrative example of these infrared effects, one can show that the curved space 
anomaly modifies the critical scaling exponents of the two-dimensional Ising model 
at its second order phase transition point \cite{KPZ}. This modification of the critical 
exponents is associated with the fluctuations of the spacetime metric at large distance 
scales. The gravitational metric fluctuations may be described by an additional massless 
scalar field, the conformal or Liouville mode, whose dynamics is generated and required 
by the conformal anomaly. 

In two dimensions the central term related to the anomaly in curved space can 
be seen already in the two-point correlation of stress tensors, $\langle T^{\mu\nu}(x) 
T^{\alpha\beta}(y)\rangle$. The meaning of this term is most clearly deduced from 
the momentum space representation, where the corresponding amplitude exhibits 
a massless pole in the conformal limit \cite{BerKohl}. This massless pole 
corresponds to a $\delta (k^2)$ in the corresponding imaginary part, describing 
a propagating massless scalar degree of freedom in the two-particle
intermediate state of the cut diagram. The same kinematics applies to the pole 
and massless scalar state in the correlator of electromagnetic currents 
$\langle J^{\mu}(x) J^{\nu}(y)\rangle$ in the Schwinger model of two-dimensional 
massless electrodynamics \cite{Schw,LSB}. The infrared effects of the anomaly may be understood 
as the result of the fluctuations of this additional massless degree of freedom.

In dimensions greater than two, infrared effects due to anomalies are both more subtle, 
and somewhat less well studied. Since the correlation functions of canonical free theories fall off 
as power laws at large distances for $d>2$, at first sight there would seem to be little 
possibility of enhanced infrared effects in higher dimensions. QCD is  a notable 
counterexample, where the growth of the effective coupling at larger distances leads 
to large quantum fluctuations and infrared confinement. The renormalization group 
flow $\beta(g^2)$ of the coupling arises from the same breaking of scale invariance 
by quantum fluctuations which give rise to the conformal anomaly\,\cite{ChanEll}. 

In the most familiar case of the axial anomaly, a massless pseudoscalar pole does 
appear in the triangle amplitude $\lag J_5^{\mu} J^{\alpha} J^{\beta}\rag$\,\cite{DolZak,Huang}, 
in the chiral limit of vanishing fermion mass, a feature we review in the next section.
This example of the axial anomaly in massless quantum electrodynamics (QED) shows that infrared 
relevant fluctuations due to anomalies can occur in $d=4$, and that triangle amplitudes 
are the simplest ones to reveal these effects. In QCD the lightest pseudoscalar state 
is the pion, whose mass vanishes in the chiral limit of zero quark mass. By identifying 
this state with the massless pole appearing in the perturbative anomaly in the chiral limit, 
the low energy rate of neutral pion decay, $\pi^0 \rightarrow 2 \gamma$ is determined by the short 
distance colored quark degrees of freedom in the one-loop $\lag J_5^{\mu} J^{\alpha} J^{\beta}\rag$ 
amplitude\,\cite{Adler,BarFriGM}. The agreement of the measured rate with the coefficient obtained 
with $N_C =3$ quarks is a striking confirmation of both QCD and the infrared effects 
of the anomaly. This well-known example of anomaly matching\,\cite{tHooft} shows that 
anomalies can provide a mechanism for short distance quantum degrees of freedom 
to have long distance or low energy consequences.

Although the special role of the triangle diagram in $d=4$ has been emphasized in\,\cite{ColGro} 
in the context of the chiral anomaly some time ago, to date there has been no clear indication 
of a massless pole or infrared degrees of freedom in flat space amplitudes involving the 
energy-momentum tensor. It is known that in $d=4$ the trace anomaly in curved space 
involves geometric invariants that are quadratic in the Riemnann curvature tensor\,\cite{BirDav,Duff}.
This has the immediate consequence that the simplest amplitude in four dimensional flat spacetime 
that can show any direct evidence of the full curved space anomaly is the three-point function of stress 
tensors, $\lag T^{\mu\nu}(x) T^{\alpha\beta}(y) T^{\gamma\delta}(z)\rag$, indicating again the
importance of triangle amplitudes in $d=4$. 

In this paper we address the possibility for low energy quantum effects in gravity, analogous to those 
in gauge theories, due to the corresponding trace anomaly, and in particular for additional massless 
scalar degrees of freedom with long range effects which can modify the predictions of classical General 
Relativity on macroscopic and even cosmological scales. We present a complete calculation of the one-loop 
triangle amplitude $\langle T^{\mu\nu}J^{\alpha}J^{\beta}\rangle$ in QED, for all values of the 
kinematical invariants. This amplitude contains the same basic kinematics as both the more 
familiar chiral triangle $\lag J_5^{\mu}J^{\alpha}J^{\beta}\rag$, and the more complicated 
amplitude $\lag TTT\rag$ involving three stress tensors. The $\lag TJJ\rag$ amplitude is 
sensitive to the trace anomaly of the one-loop stress tensor expectation value in a background 
electromagnetic potential $A_{\mu}$ (rather than a gravitational background curvature). 
By calculating this amplitude for arbitrary electron mass $m$, both the decoupling limit 
$m\rightarrow \infty$, and the conformal limit $m\rightarrow 0$, where the massless pole 
in the amplitude appears can be studied. It is the latter limit that reveals the consequences
for low energy gravity.

Following methods that have been used previously for the chiral anomaly\,\cite{Rosenb,Horejb},
we show that the $\lag TJJ\rag$ amplitude can be determined completely from general principles 
of Poincar{\'e} invariance of the vacuum, Bose symmetry under interchange of $J^{\alpha}$ and 
$J^{\beta}$, and the Ward identities of vector current and stress-energy conservation, once its finite 
tensor components are given. These finite components can be determined unambiguously from the 
imaginary part of the cut triangle amplitude, with the real part obtained by dispersion relations 
which require no subtractions (other than charge renormalization in one particular component). 
This dispersive approach based upon the finite parts of the amplitude emphasizes the infrared 
aspect of the anomaly, making it clear that the anomaly is finite, well defined and uniquely
determined, independent of UV regularization scheme, provided only that the amplitude is 
defined in a way consistent with the non-anomalous low energy symmetries of the theory.

In the conformal limit of massless QED, the two-particle intermediate state of the 
cut triangle diagram has a delta function contribution at $k^2 = 0$. Because this state 
couples to the stress tensor, it contributes to gravitational scattering amplitudes
at arbitrarily low energies. We demonstrate that the trace part of the $\lag TJJ\rag$
amplitude containing this massless intermediate scalar state and its gravitational couplings 
may be described by the introduction of local massless scalar degrees of freedom, which 
render the trace part of the one-loop effective action local. The auxiliary field description 
introduced recently in the context of curved space\,\cite{MotVau} reproduces the trace part 
of the amplitude exactly in flat space, and the massless pole in the trace part of the flat space 
$\lag TJJ\rag$ amplitude is precisely the propagator of these scalar fields.

In QED the scalar state may also be understood as a two-particle correlation of $e^+e^-$ which 
in the massless limit move collinearly at the speed of light in a total spin-$0$ configuration.
When the electron mass is non-zero, the singularity at $k^2 = 0$ is replaced by a resonance 
with a width of order $m^2$. However, the corresponding spectral function obeys a
sum rule, which shows that although broadened, and eventually decoupled for larger
$m$, the scalar state survives deformations away from the conformal limit.  In this sense
it behaves analogously to the pion in QCD.

The paper is organized as follows. In the next section we review the axial anomaly in QED 
in four dimensions, using the spectral representation and dispersion relations to emphasize 
its infrared character, exhibiting the massless $0^-$ intermediate state, and the finite spectral 
sum rule in this case. In section 3 we give the auxiliary field description of the chiral amplitude, 
showing that the massless pseudoscalar state can be described by a local effective field theory. 
In section 4 we turn to the main task of evaluating the $\lag TJJ\rag$ amplitude in QED. 
Imposing the Ward identities, we show that the full amplitude is determined by its finite terms 
and imaginary parts for any $m$ and its three kinematic invariants, independently of any specific 
UV regularization method. In section 5 we evaluate its trace, isolate the anomaly and discuss its 
relation to the $\beta$ function and scaling violation. In section 6 we give the spectral representation 
of the $\lag TJJ\rag$ amplitude, derive the corresponding finite sum rule, and show that a 
$\delta(k^2)$ appears in the appropriate spectral function in the conformal limit of massless electrons. 
In section 7 the foregoing results are compared with the auxiliary field representation of the anomaly 
given in\,\cite{MotVau}, and shown to coincide exactly in the trace sector. In section 8 we show how 
the anomalous amplitude contributes to gravitational scattering of photons by a source, prove that 
the anomaly pole induces a massless scalar interaction and propagating intermediate state in this 
scattering process, and provide the effective action description of the scattering by scalar exchange. 
Finally, section 9 contains a concise summary of our results. Technical details of extracting the finite
parts of the $\lag TJJ\rag$ amplitude are given in Appendix A, while the proofs of some identities
needed in the text are given in Appendix B.

\section{The Axial Anomaly in QED}
\label{sec:chiral}}

In order to exhibit the relationship between anomalies and massless degrees of freedom, we review 
first the familiar case of the axial anomaly in QED in this section\,\cite{Adler,BellJack,Jackcurr}. Although 
the triangle anomaly has been known for quite some time, the general behavior of the amplitude off 
the photon mass shell, its spectral representation, the appearance of a massless pseudoscalar 
pole, and its infrared aspects generally have received only limited attention\,\cite{DolZak,Horej}. 
It is this generally less emphasized infrared character of the axial anomaly upon which we focus here.

The vector and axial currents in QED are defined by\footnote{We use the conventions 
that $\{\gamma^{\mu}, \gamma^{\nu}\} = -2\,g^{\mu\nu} = 2\,{\rm diag}\ (+---)$, 
so that $\gamma^0 = (\gamma^0)^{\dagger}$, and $\gamma^5 \equiv i\gamma^0 \gamma^1 
\gamma^2 \gamma^3 = (\gamma^5)^{\dagger}$ are hermitian, and 
tr$(\gamma^5\gamma^{\mu}\gamma^{\nu}\gamma^{\rho}\gamma^{\sigma})
= -4i \epsilon^{\mu\nu\rho\sigma}$, where $\epsilon^{\mu\nu\rho\sigma}
= - \epsilon_{\mu\nu\rho\sigma}$ is the fully anti-symmetric Levi-Civita tensor, 
with $\epsilon_{0123} =+1$.} 
\bes\bea
J^{\mu}(x) = \bar\psi (x) \gamma^{\mu} \psi (x)\,.\\
J_5^{\mu}(x) = \bar\psi (x) \gamma^{\mu} \gamma^5 \psi (x)\,.
\eea\label{currents}\ees
The Dirac eq.,
\be
-i \gamma^{\mu} ( \partial_{\mu} - ieA_{\mu})\psi + m \psi = 0\,.
\label{Dirac}
\ee
implies that the vector current is conserved,
\be
\partial_{\mu} J^{\mu} = 0\,,
\label{vecons}
\ee
while the axial current apparently obeys 
\be
\partial_{\mu} J_5^{\mu} = 2 i m\, \bar\psi\gamma^5 \psi \qquad {\rm (classically)}.
\label{axclass}
\ee
In the limit of vanishing fermion mass $m\rightarrow 0$, the classical Lagrangian has a
$U_{ch}(1)$ global symmetry under $\psi \rightarrow e^{i\alpha\gamma^5}\psi$, in 
addition to $U(1)$ local gauge invariance, and $J_5^{\mu}$ is the Noether current
corresponding to this chiral symmetry. As is well known, both symmetries cannot be 
maintained simultaneously at the quantum level. Let us denote by $\lag J_5^{\mu}(z) \rag_{_A}$ 
the expectation value of the chiral current in the presence of a background electromagnetic 
potential $A_{\mu}$. Enforcing $U(1)$ gauge invariance (\ref{vecons}) on the full quantum theory 
leads necessarily to a finite axial current anomaly,
\be
\partial_{\mu}\lag J_5^{\mu}\rag_{_A}\Big\vert_{m=0} = \frac{e^2}{16\pi^2} \
\epsilon^{\mu\nu\rho\sigma}F_{\mu\nu}F_{\rho\sigma} = \frac{e^2}{2\pi^2}\,{\bf E \cdot B}\,,
\label{axanom}
\ee
in a background electromagnetic field. 

\begin{figure}[htp]
\includegraphics[width=40cm, viewport=80 550 1000 660,clip]{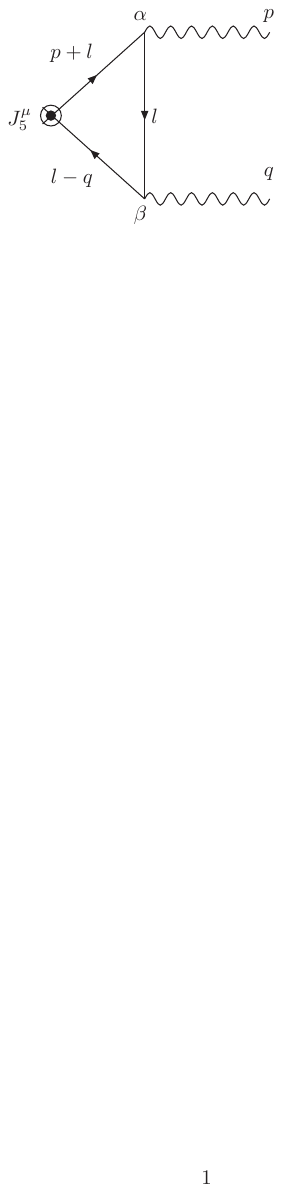}
\caption{The Axial Anomaly Triangle Diagram}
\label{Fig:tri}
\end{figure}

The second variation of $\lag J_5^{\mu}(z) \rag_{_A}$ with $A_{\mu}$ then set to zero,
\be
\Gamma^{\mu \alpha\beta}(z:x,y) \equiv  -i\frac{\delta^2 \lag J_5^{\mu}(z) \rag_{_A}}
{\delta A_{\alpha}(x) \delta A_{\beta}(y)}\Bigg\vert_{A=0}
= -i(ie)^2  \lag {\cal T} J_5^{\mu}(z) J^{\alpha}(x) J^{\beta}(y)\rag\big\vert_{A=0}
\label{axvarpos}
\ee
is thus the primary quantity of interest. By translational invariance of the Minkowski vacuum at 
$A=0$, this amplitude depends only upon the coordinate differences $x-z$ and $y-z$.
Hence with no loss of generality we may fix $z=0$. Taking the Fourier
transform of (\ref{axvarpos}) and removing the factor of total momentum conservation,
$(2\pi)^4 \delta^4(k - p - q)$, we obtain
\bea
&&\Gamma^{\mu \alpha\beta}(p,q) \equiv - i \int d^4x\int d^4y\, e^{ip\cdot x+ i q\cdot y}\  
\frac{\delta^2 \lag J_5^{\mu}(0) \rag_{_A}}{\delta A_{\alpha}(x) \delta A_{\beta}(y)} 
\nonumber\Bigg\vert_{A=0}\\
&&= ie^2 \int d^4 x \int d^4 y\, e^{i p\cdot x + i q \cdot y}\ \lag {\cal T}
J_5^{\mu}(0) J^{\alpha}(x) J^{\beta}(y)\rag\big\vert_{A=0}\,.
\label{GJJJ}
\eea
At the lowest one-loop order it is given by the triangle diagram of Fig. \ref{Fig:tri}, plus
the Bose symmetrized diagram with the photon legs interchanged. The chiral current 
expectation value in position space can be reconstructed from this momentum space amplitude by
\be
\lag J_5^{\mu}(z) \rag_{_A} = \frac{i}{2} \int \frac{d^4 p}{(2\pi)^4} \int \frac{d^4 q}{(2\pi)^4}
\int d^4x \int d^4 y \, e^{-i p\cdot (x-z)}\, e^{- i q \cdot (y-z)}\, \Gamma^{\mu \alpha\beta}(p,q)
\,A_{\alpha}(x) A_{\beta}(y) + \dots
\ee
up to second order in the gauge field background $A_{\mu}$. 

Elementary power counting indicates that the triangle diagram of Fig.  \ref{Fig:tri} is superficially linearly divergent. 
The formal reason why (\ref{vecons}) and (\ref{axclass}) cannot both be maintained at the quantum level is that 
verifying them requires the ability to shift the loop momentum integration variable $l$ in the triangle amplitude. 
Because the diagram is linearly divergent, such shifts are inherently ambiguous, and can generate finite 
extra terms. It turns out that there is no choice for removing the ambiguity which satisfies both the vector and 
chiral Ward identities simultaneously, and one is forced to choose between them. Thus although the ambiguity 
results in a well-defined finite term, the axial anomaly has most often been presented as inherently a problem 
of regularization of an apparently ultraviolet linearly divergent loop integral\,\cite{Adler,BellJack,Jackcurr}.

There is an alternative derivation of the axial anomaly that emphasizes instead its infrared character. 
The idea of this approach is to use the tensor structure of the triangle amplitude to extract its well-defined 
ultraviolet {\it finite} parts, which are homogeneous of degree three in the external momenta $p$ and $q$. Then the remaining parts of the full amplitude may be determined by the joint requirements of Lorentz covariance, Bose 
symmetry under interchange of the two photon legs, and electromagnetic current conservation,
\be
p_\alpha \Gamma^{\mu\alpha\beta}(p,q)=0 = q_\beta\Gamma^{\mu\alpha\beta}(p,q)\,,
\label{chicons}
\ee
at the two vector vertices. By this method the full one-loop triangle contribution to $\Gamma^{\mu\alpha\beta}(p,q)$,
becomes completely determined in terms of well-defined ultraviolet finite integrals which require no further regularization
\cite{Rosenb, Horejb}. The divergence of the axial current may then be computed unambiguously, and one obtains
(\ref{axanom}) in the limit of vanishing fermion mass\,\cite{Adler}. There is of course no contradiction between these 
two points of view, since it is the same Ward identities which are imposed in either method, and in the conformal limit 
of vanishing fermion mass the infrared and ultraviolet behavior of the triangle amplitude are one and the same. 

\begin{table}[htp]
$$
\begin{array}{|@{\hspace{.8cm}}c @{\hspace{.8cm}}|@{\hspace{.8cm}} c @{\hspace{.8cm}}| 
@{\hspace{.8cm}} c @{\hspace{.8cm}} | @{\hspace{.8cm}} c @{\hspace{.8cm}}|} \hline
\epsilon^{\mu\alpha\beta\lambda} p_{\lambda} & 
\epsilon^{\mu\alpha\rho\sigma}p^{\beta}p_{\rho}q_{\sigma} &
\epsilon^{\mu\beta\rho\sigma}p^{\alpha}p_{\rho}q_{\sigma} & 
\epsilon^{\alpha\beta\rho\sigma}p^{\mu}p_{\rho}q_{\sigma} \\
\hline
\epsilon^{\mu\alpha\beta\lambda} q_{\lambda} &
\epsilon^{\mu\alpha\rho\sigma}q^{\beta}p_{\rho}q_{\sigma} &
\epsilon^{\mu\beta\rho\sigma}q^{\alpha}p_{\rho}q_{\sigma} &
\epsilon^{\alpha\beta\rho\sigma}q^{\mu}p_{\rho}q_{\sigma} \\
\hline
\end{array}
$$
\caption{The $8$ three index psuedo-tensor monomials into which $\Gamma^{\mu\alpha\beta}(p,q)$ 
can be expanded \label{eightmono}}
\end{table}

Since we will apply this method to the trace anomaly amplitude in the next section, let us first review the calculation 
in the axial current case. One first uses the Poincar{\'e} invariance of the vacuum to assert that $\Gamma^{\mu\alpha\beta}(p,q)$ 
can be expanded in the set of all three-index pseudotensors constructible from the $p$ and $q$, with the correct 
Lorentz transformation properties. There are exactly eight such pseudotensors, which are listed in Table \ref{eightmono}, 
the first two of which are linear in $p$ or $q$, while the remaining six are homogeneous of degree three in the external 
momenta. Since the amplitude $\Gamma^{\mu\alpha\beta}(p,q)$ has mass dimension one, any regularization 
ambiguity can appear only in the coefficients of the tensors which are linear in momenta, {\it i.e.} 
$\varepsilon^{\mu\alpha\beta\lambda} p_{\lambda}$ and $\varepsilon^{\mu\alpha\beta\lambda} q_{\lambda}$. 
The coefficients of these tensors have mass dimension zero and are therefore potentially logarithmically
divergent. On the other hand, the remaining six tensors in Table \ref{eightmono}, homogeneous of degree three 
in $p$ and $q$ can appear in $\Gamma^{\mu\alpha\beta}(p,q)$ multiplied only by scalar loop integrals with 
{\it negative} mass dimension, $-2$, which are completely convergent in the ultraviolet. If these scalar coefficient 
functions can be extracted unambiguously, then vector current conservation can be used to determine the coefficients 
of the remaining two tensors of dimension one. Indeed the general amplitude satisfying (\ref{chicons}) must be a linear 
combination of only the six linear combinations defined below and listed in Table \ref{chitens}. Since the tensors 
$\varepsilon^{\mu\alpha\beta\lambda} p_{\lambda}$ and $\varepsilon^{\mu\alpha\beta\lambda} q_{\lambda}$ appear 
only in those linear combinations which satisfy (\ref{chicons}), their coefficients are determined unambiguously 
by the finite coefficients multiplying the tensors of degree three.

\begin{table}[hbp]
$$
\begin{array}{|@{\hspace{.8cm}}c @{\hspace{.8cm}}|@{\hspace{.8cm}}c @{\hspace{.8cm}}|} 
\hline
i & \tau_i^{\mu\alpha\beta}(p,q) \\ \hline \hline
1 & -\pq \,\epsilon^{\mu\alpha\beta\lambda} p_\lambda  -
p^\beta \,\upsilon^{\mu\alpha}(p,q) \\ \hline
2 & p^2 \epsilon^{\mu\alpha\beta\lambda} q_\lambda \, +
 p^{\alpha} \upsilon^{\mu\beta}(p,q) \\ \hline
3 & p^\mu\,\upsilon^{\alpha\beta}(p,q)\\ \hline \hline
4 & \pq \,\epsilon^{\mu\alpha\beta\lambda} q_\lambda  +
q^{\alpha} \,\upsilon^{\mu\beta}(p,q) \\ \hline
5 & -q^2 \varepsilon^{\mu\alpha\beta\lambda} p_\lambda \,
-q^{\beta} \upsilon^{\mu\alpha}(p,q)\\ \hline
6 & q^\mu\,\upsilon^{\alpha\beta}(p,q) \\ \hline
\end{array}
$$
\caption{The $6$ third rank pseudotensors obeying (\ref{taucons}) \label{chitens}}
\end{table}

\begin{figure}[htp]
\includegraphics[width=40cm, viewport=100 570 900 690,clip]{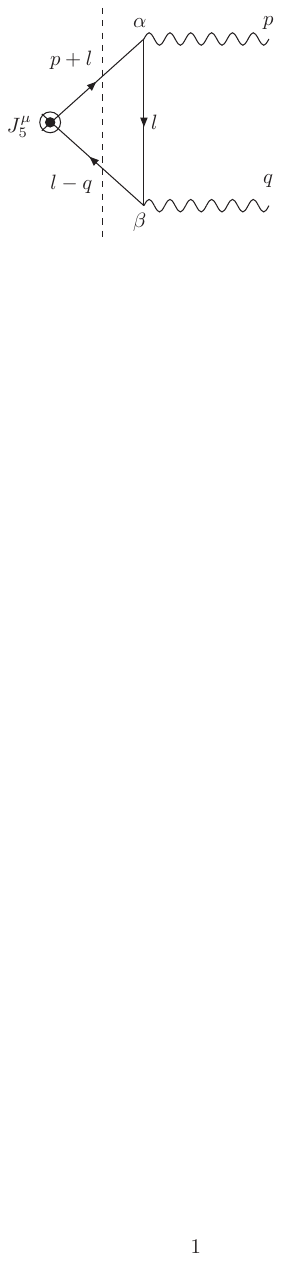}
\caption{Discontinuity or Imaginary Part of the Triangle Diagram obtained by cutting two lines}
\label{Fig:ImTri}
\end{figure}

To make the procedure of extraction of finite terms of the amplitude completely unambiguous, 
one may first calculate the imaginary part of the cut triangle amplitude in Fig. \ref{Fig:ImTri} 
at timelike $k^2$, which is finite, and then construct the real part by a dispersion relation. For 
the mass dimension $-2$ terms, the dispersion relations constructing the real parts of the amplitude 
from its imaginary parts are finite and require no subtractions\,\cite{DolZak,Horej}.

To construct the tensors satisfying (\ref{chicons}), let us define first the two index pseudotensor,
\be
\upsilon^{\alpha\beta}(p,q) \equiv \epsilon^{\alpha\beta \rho\sigma} p_{\rho}q_{\sigma}\,,
\label{Wdef}\ee
which satisfies
\bes\bea
\upsilon^{\alpha\beta}(p,q) &=& \upsilon^{\beta\alpha}(q,p)\,,\\
p_{\alpha} \upsilon^{\alpha\beta}(p,q) = &0& = q_{\beta}\upsilon^{\alpha\beta}(p,q) \,.
\eea\ees
By taking general linear combinations of the eight pseudotensors in Table \ref{eightmono}, we find 
then that there are exactly six third rank pseudotensors, $\tau_i^{\mu\alpha\beta}(p,q)$,
$i= 1, \dots, 6$ which can be constructed from them to satisfy the conditions (\ref{chicons}),
\be
p_\alpha \tau_i^{\mu\alpha\beta}(p,q)=0=\tau_i^{\mu\alpha\beta}(p,q)\,q_\beta = 0 \,,
\qquad i= 1, \dots, 6,
\label{taucons}
\ee
given in Table \ref{chitens}. Hence  we may express the amplitude (\ref{GJJJ}) satisfying  (\ref{chicons}) 
in the form,
\be
\Gamma^{\mu\alpha\beta}(p,q)= \sum_{i=1}^6 f_i\, \tau_i^{\mu\alpha\beta}(p,q)\,,
\label{tencomp}
\ee
where $f_i = f_i(k^2; p^2,q^2)$ are dimension $-2$ scalar functions of the three invariants,
$p^2, q^2$, and $k^2$. 

We note also that the full amplitude (\ref{GJJJ}) must be Bose symmetric, 
\be
\Gamma^{\mu\alpha\beta}(p,q)=\Gamma^{\mu \beta \alpha}(q,p)  \,.
\label{chibose}
\ee
Since $\tau_{i+3}^{\mu\alpha\beta} (p, q) = \tau_{i}^{\mu\beta\alpha} (q,p)$ for $i=1, 2, 3$,
it follows that the six scalar coefficient functions $f_i$ 
also fall into three Bose conjugate pairs, {\it i.e.}
\bes\bea
&& f_1(k^2; p^2,q^2) = f_4(k^2; q^2, p^2)\,,\\
&& f_2(k^2; p^2,q^2) = f_5(k^2; q^2, p^2)\,,\\
&& f_3(k^2; p^2,q^2) = f_6(k^2; q^2, p^2)\,,
\eea\label{Bosepairs}\ees
related by interchange of $p^2$ and $q^2$.

Actually, owing to the algebraic identity obeyed by the $\epsilon$ symbol,
\be
g^{\alpha\beta} \epsilon^{\mu\nu\rho\sigma} + g^{\alpha\mu} \epsilon^{\nu\rho\sigma\beta}
+ g^{\alpha\nu} \epsilon^{\rho\sigma\beta\mu} + g^{\alpha\rho} \epsilon^{\sigma\beta\mu\nu}
+ g^{\alpha\sigma} \epsilon^{\beta\mu\nu\rho} = 0\,,
\label{epsiden}
\ee
in four dimensions, the six tensors $\tau_i$ are not linearly independent, and form an 
overcomplete basis. The identity (\ref{epsiden}) leads to the relations,
\bes\bea
&&\tau_3^{\mu\alpha\beta}(p,q) = \tau_1^{\mu\alpha\beta}(p,q) + \tau_2^{\mu\alpha\beta}(p,q)\,,\\
&&\tau_6^{\mu\alpha\beta}(p,q) = \tau_4^{\mu\alpha\beta}(p,q) + \tau_5^{\mu\alpha\beta}(p,q)\,.
\eea\label{lindep}\ees
Thus the tensors $\tau_3$ and $\tau_6$ could be eliminated completely by means of
(\ref{lindep}), and the full amplitude expressed entirely in terms of the linearly independent and 
complete basis set of only the four tensors, $\tau_1, \tau_2, \tau_4$, and $\tau_5$. Indeed this 
has been the general practice in the literature on the axial anomaly\,\cite{Rosenb,Adler,BellJack}

Eliminating these or any other two tensors is not necessary for our purposes, and we 
choose instead to work with the overcomplete set of six tensors listed in Table \ref{chitens}. 
This will have the consequence that the coefficient functions $f_i$ are determined only up to 
the freedom to choose arbitrary coefficients of the linear combinations (\ref{lindep}), {\it i.e.}
to shift each of the coefficients $f_i$ by an arbitrary scalar function $h$ via the rule,
\bes\bea
&&f_1(k^2; p^2,q^2)\rightarrow f_1(k^2; p^2,q^2) + h(k^2; p^2,q^2)\,,\\
&&f_2(k^2; p^2,q^2)\rightarrow f_2(k^2; p^2,q^2) + h(k^2; p^2,q^2)\,,\\
&&f_3(k^2; p^2,q^2)\rightarrow f_3(k^2; p^2,q^2) - h(k^2; p^2,q^2)\,,
\eea\ees
with the shift in $f_4, f_5, f_6$ determined by (\ref{Bosepairs}) by interchange of $p^2$ and
$q^2$. The arbitrary function $h$ drops out of the final amplitude by use of (\ref{lindep}).

The computation of the finite coefficients given in the literature\,\cite{Rosenb,Adler} amounts to a specific 
choice of the arbitrary function $h$ (by the order of the $\gamma$ matrices when the trace is
performed), and yields
\bes\bea
&& f_1=f_4
=\frac{e^2}{\pi^2}\int_0^1 dx\int_0^{1-x} dy\  \frac{x y}{D}\,\\
&& f_2=\frac{e^2}{\pi^2}\int_0^1 dx\int_0^{1-x} dy\ 
\frac{x(1-x)}{D}\,,\\
&& f_5=\frac{e^2}{\pi^2}\int_0^1 dx\int_0^{1-x} dy\ 
\frac{y(1-y)}{D}\,,\\
&& f_3 = f_6 = 0\,,
\eea \label{chiamps} \ees
where the denominator of the Feynman parameter integral is given by
\be
D \equiv p^2 x(1-x) + q^2 y(1-y) + 2\pq \, xy + m^2 = 
(p^2\,x +q^2\,y)(1-x-y)+ xy\,k^2+m^2\,,
\label{denom}
\ee
strictly positive for $m^2 >0$, and spacelike momenta, $k^2, p^2, q^2 > 0$. Thus each of the 
dimension $-2$ scalar coefficient functions $f_i$ in (\ref{chiamps}) are finite, and free of any 
UV regularization ambiguities, and the full amplitude $\Gamma^{\mu\alpha\beta}(p,q)$ satisfying
\vspace{-0.25cm}
\begin{enumerate}
\item [(i)] Lorentz invariance of the vacuum\,,\vspace{-.3cm}
\item [(ii)] Bose symmetry (\ref{chibose})\,,\vspace{-.3cm}
\item [(iii)] vector current conservation (\ref{chicons})\,,\vspace{-.3cm}
\item [(iv)] unsubtracted dispersion relation of real and imaginary parts\,,\vspace{-.3cm}
\end{enumerate}
\vspace{0.05cm}
with the finite imaginary parts determined by the cut triangle diagram of Fig. \ref{Fig:ImTri}, 
is given by (\ref{tencomp}) and (\ref{chiamps}), without any need of regularization of ultraviolet 
divergent loop integrals at any step. 

Contraction of the finite amplitude $\Gamma^{\mu\alpha\beta}(p,q)$ with the momentum 
$k_{\mu} = (p + q)_{\mu}$ entering at the axial vector vertex can now be computed
unambiguously, and we obtain
\be
k_\mu \, \Gamma^{\mu \alpha\beta}(p,q) = {\cal A} \, \upsilon^{\alpha\beta} (p, q)\,,
\ee
with
\bea
&&{\cal A} = \pq\, f_1 + p^2 f_2 + (p^2 + \pq) f_3 + \pq\, f_4 + q^2 f_5 + (\pq + q^2) f_6\nn
&& \qquad = 2\pq\, f_1 + p^2 f_2 + q^2 f_5\,,
\label{axdiv}
\eea
by (\ref{chiamps}). Substituting the explicit Feynman parameter integrals of (\ref{chiamps}) in (\ref{axdiv}), 
and using (\ref{denom}), (\ref{axdiv}) becomes
\bea
&&{\cal A}(k^2; p^2,q^2) = \frac{e^2}{\pi^2}\int_0^1 dx\int_0^{1-x} dy\  \frac{D - m^2}{D}\nn
&&\qquad = \frac{e^2}{2\pi^2} - \frac{e^2}{\pi^2} \ m^2\,
\int_0^1 dx\int_0^{1-x} dy\  \frac{1}{D}\,.
\label{anomcalc}
\eea
The second term proportional to $m^2$ is what would be expected from the naive axial vector 
divergence (\ref{axclass})\,\cite{Horejb}. The first term in (\ref{anomcalc}) in which the denominator 
$D$ is cancelled in the numerator is
\be
\frac{e^2}{\pi^2} \int_0^1 dx\int_0^{1-x} dy = \frac{e^2}{2\pi^2}\,,
\ee
and which remains finite and non-zero in the limit $m \rightarrow 0$ is the axial anomaly. 

Thus the finite anomalous term is unambiguously determined by our four requirements above, 
and may be clearly identified even for finite $m$, when the chiral symmetry is broken. This
construction of the amplitude from only symmetry principles and its finite parts may be
regarded as a proof that the same finite axial anomaly must arise in {\it any} regularization
of the original triangle amplitude which respects these symmetries and leaves the finite
parts unchanged. Explicit calculations in dimensional regularization and Pauli-Villars 
regularization schemes, which respect these symmetries confirm this\,\cite{Bertl}.

The spectral representations for the triangle amplitude functions,
\be
f_i (k^2; p^2,q^2) = \int_0^\infty ds\ \frac{\rho_i(s; p^2, q^2)}{k^2+s}\,,
\label{defspec}
\ee
used to compute the finite parts also aid in the physical
interpretation of the infrared aspect of the anomaly. If one defines the function,
\be
S(x,y;p^2,q^2) \equiv \frac{(p^2x + q^2y)(1-x-y) + m^2}{xy} = \frac{D}{xy}-k^2\,,
\label{BigS}
\ee
and substitutes the identity,
\be
\frac{1}{D} = \int_0^{\infty} \, \frac{ds}{xy (k^2 +s)}\ \delta (s - S)
\label{sspec}
\ee
valid for $p^2, q^2, m^2 \ge 0$, into the expressions (\ref{chiamps}), interchanging the 
order of the $s$ and $x,y$ integrations, the spectral representation (\ref{defspec}) of
the amplitude is obtained, with
\bes\bea
&&\rho_1(s; p^2, q^2) = \frac{e^2}{\pi^2}\int_0^1 dx\int_0^{1-x} dy 
\,\delta(s-S)\, , \label{chispeca}\\
&&\rho_2(s; p^2, q^2) = \frac{e^2}{\pi^2}\int_0^1 dx\int_0^{1-x} dy\, \frac{1-x}{y}\,
\delta(s-S)\, ,\\
&&\rho_5(s; p^2, q^2)= 
\frac{e^2}{\pi^2}\int_0^1 dx\int_0^{1-x} dy \,\frac{1-y}{x} \,\delta(s-S)
 = \rho_2(s; q^2, p^2) \, .
\eea
\label{chispec}
\ees
From the definition of the function $S(x,y;p^2,q^2)$ in (\ref{BigS}) it follows that
\be
2 \pq + p^2\,\frac{1-x}{y} + q^2\, \frac{1-y}{x} = \frac{D-m^2}{xy} = k^2 + S - \frac{m^2}{xy}\,,
\label{numD}
\ee
and therefore, from (\ref{chispec}),
\be
2\pq\,\rho_1 +p^2\,\rho_2 +q^2\,\rho_5
=(k^2+s)\rho_{_{\cal A}} -m^2\rho_0\,,
\label{chirel}
\ee
where
\bes\bea
&&\rho_{_{\cal A}}(s;p^2,q^2) \equiv \rho_1(s;p^2,q^2)\,, \qquad {\rm and}\label{rhoA0a}\\
&&\rho_0(s;p^2, q^2) \equiv \frac{e^2}{\pi^2} \int_0^1 \, dx\,\int_0^{1-x}\,dy\,\frac{1}{xy}
\ \delta(s-S)\,.
\eea\label{rhoA0}\ees
The relation (\ref{chirel}) with (\ref{rhoA0}) for the imaginary part of the cut
triangle amplitude can be compared to (\ref{axdiv})-(\ref{anomcalc}) for the corresponding real part.
Defined for Euclidean spacelike four-momenta, $k^2, p^2, q^2 > 0$, they are continued
to timelike four-momenta $k^2 <0$, by means of an $-i\epsilon$ prescription
in the denominators of (\ref{defspec}). Then the imaginary part 
of the chiral amplitude (\ref{GJJJ}), corresponding to the cut diagram illustrated in
Fig. \ref{Fig:ImTri} is given by (\ref{chirel}) evaluated at $s= -k^2 > 0$, {\it i.e.}
\be
\left(2\pq\,\rho_1 +p^2\,\rho_2 +q^2\,\rho_5\right)\Big\vert_{s = -k^2}
=-m^2\rho_0\Big\vert_{s = -k^2}\,,
\ee
which shows that $\rho_{_{\cal A}} = \rho_1$ drops out of (\ref{chirel}) for $s=-k^2$ on shell, and 
the finite imaginary part of the amplitude is completely non-anomalous for timelike $k^2$. 

The anomaly in Re $\cal A$ comes about because of the cancellation of the $k^2 + s$ in the denominator
of the unsubtracted dispersion integrals (\ref{defspec}) and the same factor in the spectral function sum,
(\ref{chirel}), resulting in the finite integral,
\be
\int_0^{\infty} \, ds\, \rho_{_{\cal A}}(s; p^2, q^2) = \frac{e^2}{\pi^2} \int_0^1 \, dx\,\int_0^{1-x}\, dy  \, 
\int_0^{\infty} \, ds\, \delta(s-S) = \frac{e^2}{2\pi^2}\,,
\label{chisum}
\ee
independent of $p^2, q^2, m^2 \ge 0$. Thus the anomalous divergence of the axial vector 
current is tied to a ultraviolet finite {\it sum rule} (\ref{chisum}), for the associated spectral density
$\rho_{_{\cal A}} (s) = \rho_1 (s)$. 

The finite sum rule (\ref{chisum}) and relationship (\ref{chirel}) between the
spectral functions is critical to the infrared aspect of the axial anomaly. 
Using $2\pq = k^2-p^2-q^2$, and rearranging (\ref{chirel}) we find 
\be
\rho_{_{\cal A}} = \frac{p^2}{s} \left(\rho_2 - \rho_1\right) 
+ \frac{q^2}{s}\left(\rho_5 - \rho_1\right) + \frac{m^2}{s}\rho_0\,.
\label{rhoA}
\ee
It is easy to see from the Feynman parameter representations (\ref{chispec}) that the differences, 
$\rho_2 - \rho_1$ and $\rho_5 - \rho_1$ are positive for spacelike $p^2, q^2$. Hence the
function obeying the sum rule (\ref{chisum}) is expressed in (\ref{rhoA}) as a sum of non-negative
contributions for spacelike or null $p^2$ and $q^2$. If the limits $p^2, q^2, m^2 \rightarrow 0^+$ 
are taken (in any order), some of the spectral functions $\rho_2 -\rho_1$,  $\rho_2 -\rho_1$ and
$\rho_0$ develop logarithmic singularities, but each term on the right side of (\ref{rhoA}) multiplied 
by $p^2$, $q^2$, or $m^2$ approaches zero. Hence the spectral function $\rho_{_{\cal A}}$ {\it vanishes} pointwise 
for all $s>0$ in this combined limit. In order for this to be consistent with the sum rule (\ref{chisum}), 
$\rho_{_{\cal A}}(s)$ must develop a $\delta (s)$ singularity at $s=0$ in this limit. It is straightforward either 
to calculate the function $\rho_{_{\cal A}} (s; q^2, p^2)$ from the relations above for any $p^2, q^2, m^2$ and 
verify the appearance of a more and more sharply peaked spectral density in the limits, 
$p^2, q^2, m^2 \rightarrow 0^+$, or alternatively, to evaluate $\rho_{_{\cal A}}= \rho_1$ directly from 
(\ref{chispeca}) in this limit, where from (\ref{BigS}) the function $S(x,y; 0,0)$ vanishes identically. 
Then by interchanging the limits and integrations over $x,y$, from (\ref{chispeca}) and 
(\ref{rhoA0a}) we obtain
\be
\lim_{p^2,q^2,m^2 \rightarrow 0^+} \rho_{_{\cal A}} (s; p^2, q^2) = 
\frac{e^2}{\pi^2}\int_0^1 dx\int_0^{1-x} dy \ \delta(s)
= \frac{e^2}{2\pi^2}\,\delta(s)\,.
\label{rhoAdel}
\ee
Hence the appearance of the $\delta (s)$ in this limit is explicit in this representation.
This delta function shows that a massless pseudoscalar appears in the intermediate
state of the cut triangle amplitude when $p^2=q^2=m^2=0$.

To examine this infrared behavior in more detail, it is instructive to consider the
case of $p^2=q^2 =0$, while still retaining $m$ as an infrared regulator. In this case from 
(\ref{BigS}), $S = m^2/xy$, and we easily find\footnote{The conjectured form of
the spectral function of ref.\,\cite{Huang}, eq. (11.50), disagrees with the exact result, (\ref{rhoA00}),
although the qualitative conclusions are unchanged.}

\bea
&&\rho_{_{\cal A}} (s; p^2, q^2)\big\vert_{p^2=q^2=0} =
\frac{e^2}{\pi^2}\int_0^1 dx\int_0^{1-x} dy \ \delta\left(s - \frac{m^2}{xy}\right)
= \frac{m^2}{s}\rho_0(s;0,0) \nn
&&\qquad = \frac{e^2}{\pi^2}\, \frac{m^2}{s^2} \,\ln\left\{\frac{1 + \sqrt{1 - \frac{4m^2}{s}}}
{1 - \sqrt{1 - \frac{4m^2}{s}}}\right\}\,\theta (s-4m^2) \,.
\label{rhoA00}\eea
As expected for $m^2 \rightarrow 0$ the spectral function (\ref{rhoA00}) vanishes pointwise for all $s > 0$.
However by making the change of variables $s = 4m^2/(1-u^2)$ it is straightforward to verify that the integral 
over $s$ of the function (\ref{rhoA00}) is independent of $m$ and given by (\ref{chisum}). Because of 
(\ref{axdiv}) and (\ref{chirel}) with $2p\cdot q = k^2$, the spectral integral (\ref{defspec}) for the full amplitude 
$f_1$ is
\be
f_1(k^2)\big\vert_{p^2=q^2=0} = \frac{1}{k^2}\left[\frac{e^2}{2\pi^2} - m^2\int_0^{\infty} \, ds\, 
\frac{\rho_0(s; 0, 0)}{k^2 +s}\right] = \frac{{\cal A}(k^2;0,0;m^2)}{k^2}\,,
\label{axpole}
\ee
due to the sum rule (\ref{chisum}). This shows that the amplitude $f_1$ develops a pole at $k^2 = 0$ 
when $p^2=q^2=m^2=0$, corresponding to the $\delta(s)$, (\ref{rhoAdel}) in the imaginary part
in the same limit\,\cite{DolZak}.

When the fermion mass is non-zero the amplitude (\ref{axpole}) can also be written in the form,
\be
f_1(k^2)\big\vert_{p^2=q^2=0} = m^2 \int_{4m^2}^{\infty} \frac{ds}{s}\,\frac{\rho_0(s; 0, 0)}{k^2 +s}\,,
\label{a1nopole}
\ee
which shows that the amplitude has no pole divergence as $k^2 \rightarrow 0$ with $m^2 > 0$ fixed\,\cite{Acha}.
Because of the sum rule and relations (\ref{rhoA00}), the residue ${\cal A}(k^2;0,0;m^2)$ of the pole vanishes 
in this limit. This may be understood as a consequence of decoupling, for with no other scales remaining,
the limit $k^2/m^2 \rightarrow 0$ is equivalent to the limit $m^2 \rightarrow \infty$ with $k^2$ fixed,
in which case the entire fluctuation represented by the triangle diagram should vanish on
physical grounds. The spectral function representation and determination of the anomaly by
its finite parts builds in this decoupling limit $m^2 \rightarrow \infty$ automatically.
Conversely, if $m=0$ then the amplitude (\ref{axpole}) behaves like $k^{-2}$ for all $k^2$,
in both the infrared and ultraviolet, as would be expected for a conformal theory with no
intrinsic mass or momentum scale. This decoupling behavior is also inherent in the Pauli-Villars 
regularization of the triangle amplitude, since the first anomalous first term in (\ref{anomcalc}) 
is exactly the negative of the second term in the limit of infinite mass.

For comparison we may consider $\rho_2(s)$, given by
\be
\rho_2(s;0,0) = \rho_5(s;0,0)= \frac{e^2}{2\pi^2}\, \frac{1}{s}\, \sqrt{1 - \frac{4m^2}{s}}\, \theta (s-4m^2)\,.
\ee
Because of its slower fall off with $s$, the integral of $\rho_2(s)$ over $s$ does not converge,
and does not obey a finite sum rule. Its corresponding amplitude,
\bea
&& f_2\big\vert_{p^2=q^2=0} = \frac{e^2}{2\pi^2 k^2} \left\{-2 + \sqrt{1 + \frac{4m^2}{k^2}}
\ln \left[\frac{ \sqrt{1 + \frac{4m^2}{k^2}} + 1}{\sqrt{1 + \frac{4m^2}{k^2}} - 1}\right] \right\}\nonumber\\
&& \qquad \rightarrow \frac{e^2}{2\pi^2 k^2} \ln \left( \frac{k^2}{m^2} \right)\,,
\eea
does not possess a pole or even a finite limit as $m^2 \rightarrow 0$ or $k^2/m^2 \rightarrow \infty$.
Thus this limit, equivalent to $m^2$ fixed and $k^2 \rightarrow \infty$ is purely ultraviolet in character, 
and cannot be interpreted in terms of an infrared massless state with a finite spectral weight.

The full amplitude (\ref{tencomp}) for $p^2=q^2 =0$ becomes 
\bea
&&\Gamma^{\mu\alpha\beta}(p,q)\Big\vert_{p^2=q^2 =0} =
f_1(k^2;0,0)[\tau_1^{\mu\alpha\beta}(p,q) +  \tau_4^{\mu\alpha\beta}(p,q)]
+ f_2(k^2;0,0) [\tau_2^{\mu\alpha\beta}(p,q) +  \tau_5^{\mu\alpha\beta}(p,q)]\nn
&&\qquad\qquad = f_1(k^2;0,0)\ k^{\mu} \upsilon^{\alpha\beta}(p,q)
+ \left[ f_2(k^2;0,0) - f_1(k^2;0,0) \right] 
[\tau_2^{\mu\alpha\beta}(p,q) +  \tau_5^{\mu\alpha\beta}(p,q)]\,,
\label{onshell}
\eea
by use of the identities, (\ref{lindep}). For vanishing $p^2=q^2=0$, the tensors,
\bes\bea
&&\tau_2^{\mu\alpha\beta}(p,q)\Big\vert_{p^2=q^2=0} = p^{\alpha}\upsilon^{\mu\beta}(p,q)\,,\\
&&\tau_5^{\mu\alpha\beta}(p,q)\Big\vert_{p^2=q^2=0} = -q^{\beta}\upsilon^{\mu\alpha}(p,q)\,,
\eea\label{tau25}\ees
have zero contraction with photon wave amplitudes obeying the transversality 
condition, $p^{\alpha} \tilde A_{\alpha} (p) = q^{\beta} \tilde A_{\beta} (q) = 0$. 
Hence the term involving the $\tau_2 + \tau_5$ in (\ref{onshell}) drops out entirely
in the full matrix element of $J_5^{\mu}(0)$ between the vacuum and a physical two-photon state 
$\vert p,q\rangle$, giving simply
\bea
&&\lag 0 \vert J_5^{\mu}(0) \vert p,q\rag = i\Gamma^{\mu\alpha\beta}(p,q)
\tilde A_{\alpha}(p) \tilde A_{\beta}(q)\big\vert_{p^2=q^2=0}\nn
&& \qquad \qquad = if_1(k^2;0,0) \ k^{\mu} \upsilon^{\alpha\beta}(p,q)\,
\tilde A_{\alpha}(p) \tilde A_{\beta}(q)\nn
&& \qquad \qquad \mathop{\longrightarrow}_{_{m \rightarrow 0}} 
\ \frac{ie^2}{2\pi^2 k^2}\ k^{\mu} \upsilon^{\alpha\beta}(p,q)\,
\tilde A_{\alpha}(p) \tilde A_{\beta}(q)\,,
\label{matpol}
\eea
where the last line follows from (\ref{axpole}) for $m=0$. This exhibits the pole at $k^2 = (p+q)^2 = 0$. 
Thus the singular infrared behavior required by the anomaly, survives in the full on shell matrix element 
to physical transverse photons. The residue of the pole is determined by the anomalous divergence,
\be
\lag 0 \vert \partial_{\mu}J_5^{\mu}(0) \vert p,q\rag = ik_{\mu} \lag 0 \vert J_5^{\mu}(0) \vert p,q\rag
= -\frac{e^2}{2\pi^2}\ \upsilon^{\alpha\beta}(p,q)\,\tilde A_{\alpha}(p) \tilde A_{\beta}(q)
\ee
to be non-vanishing when $m=0$.

By examining the expressions above one can see that the full amplitude exhibits propagating 
pole-like behavior for $k^2 \gg p^2, q^2, m^2$, while for finite $p^2, q^2, m^2$, the pole 
appears to soften into a resonance and there is no singularity when $k^2 \le$ min $(|p^2|, |q^2|, m^2)$.
Thus a strict infrared pole at $k^2 =0$ exists in the triangle amplitude only for zero mass fermions, 
and it couples to the physical amplitude only when $p^2 =q^2=m^2= 0$, giving the full answer for the 
on-shell matrix element (\ref{matpol}) of the chiral current to physical transverse photons only in this 
case\,\cite{Acha}. However, because of the sum rule (\ref{chisum}), the pseudoscalar
state implied by the anomaly is present at any momentum or mass scale, while from
(\ref{axpole}) its coupling to $\lag J_5^{\mu}\rag$ and photons simply becomes weaker 
for $k^2 \le$ min $(|p^2|, |q^2|, m^2)$, and decouples entirely as $k^2 \rightarrow 0$ 
for any of $(|p^2|, |q^2|, m^2)$ finite.

This appearance of a massless pseudoscalar in the chiral amplitude (\ref{GJJJ}) in the two-fermion 
intermediate state in the limit of massless fermions is reminiscient of the Schwinger model, {\it i.e.} 
massless electrodynamics in $1 +1$ dimensions, where it is also related to the anomaly\,\cite{Schw}. 
In each case one can use the fermion mass as an infrared regulator to examine the appearance of the 
anomaly pole in the amplitude or delta-function in its imaginary part as the limit $m^2 \rightarrow 0$,
for $k^2 <0$ timelike. In each case when one finally arrives at the limit of null four-momenta,
the intermediate state which gives rise to the pole is a massless electron-positron pair moving
exactly collinearly at the speed of light\,\cite{FSBY}. Thus even the $3+1$ dimensional case becomes
effectively $1+1$ dimensional in this limit, which accounts for the infrared enhancement.
The only essential difference between $d=2$ and $d=4$ dimensions appears to be the
necessity of going to a more complicated three-point amplitude in the $d=4$ case
to reveal the anomaly pole. The special role of the kinematics of the triangle diagram for this 
infrared enhancement in gauge theories in $3+1$ dimensions has been emphasized previously 
in\,\cite{ColGro}.

\section{The Auxiliary Field Description of the Axial Anomaly and Anomalous Current Commutators}

The appearance of a massless pseudoscalar pole in the triangle anomaly amplitude
suggests that this can be described as the propagator of a pseudoscalar field 
which couples to the axial current. Indeed it is not difficult to find the
field description of the pole. To do so let us note first that the axial current 
expectation value $\lag J_5^{\mu}\rag_{_A}$ can be obtained from an extended action principle 
in which we introduce an axial vector field, ${\cal B}_{\mu}$ into the Dirac Lagrangian,
\be
i\bar\psi \gamma^{\mu}\left(\stackrel{\leftrightarrow}{\partial}_{\mu} - ie A_{\mu}\right)\psi
-m \bar\psi \rightarrow  i\bar\psi \gamma^{\mu}\left(\stackrel{\leftrightarrow}{\partial}_{\mu} 
- ie A_{\mu} - ig\gamma^5{\cal B}_{\mu}\right)\psi -m \bar\psi\psi
\label{addax}
\ee
so that the variation of the corresponding action with respect to ${\cal B}_{\mu}$ gives
\be
\frac{\delta {\cal S}} {\delta {\cal B}_{\mu}}= g \lag J_5^{\mu}\rag_{_A} \,.
\label{expect}
\ee
Henceforth we shall set the axial vector coupling $g=1$. Next let us decompose the axial vector 
${\cal B}_{\mu}$ into its transverse and longitudinal parts,
\be
{\cal B}_{\mu} = {\cal B}_{\mu}^{\perp} + \partial_{\mu} {\cal B}
\ee
with $\partial^{\mu} {\cal B}_{\mu}^{\perp} =0$ and $\cal B$ a pseudoscalar. Then, by an integration
by parts in the action corresponding to (\ref{addax}), we have
\be
\partial_{\mu} \lag J_5^{\mu}\rag_{_A} = -\frac{\delta {\cal S}} {\delta {\cal B}}\,
\label{varS}.
\ee
Thus the axial anomaly (\ref{axanom}) implies that there is a term in the one-loop effective action 
in a background $A_{\mu}$ and ${\cal B}_{\mu}$ field, linear in $\cal B$ of the form,
\be
{\cal S}_{eff} = -\frac{e^2}{16\pi^2} \,\int d^4x \, \epsilon^{\mu\nu\rho\sigma}F_{\mu\nu}F_{\rho\sigma}\,{\cal B}\,,
\label{axnonl}
\ee
or since $\partial^{\lambda} {\cal B}_{\lambda} = \sq {\cal B}$,
\be
{\cal S}_{eff} = -\frac{e^2}{16\pi^2} \,\int d^4 x\int d^4y \, [\epsilon^{\mu\nu\rho\sigma}F_{\mu\nu}F_{\rho\sigma}]_x
\sq^{-1}_{xy} \,[\partial^{\lambda} {\cal B}_{\lambda}]_y\,,
\label{nonl}
\ee
where $\sq^{-1}_{xy}$ is the Green's function for the massless scalar wave operator $\sq = \partial_{\mu}\partial^{\mu}$. 
Thus from (\ref{expect}), this non-local action gives\,\cite{SmaiSpal}
\be
\lag J_5^{\mu}\rag_{_A} = \frac{e^2}{16\pi^2} \partial^{\mu} \sq^{-1}
\epsilon^{\alpha\beta\rho\sigma}F_{\alpha\beta}F_{\rho\sigma}\,,
\label{nonlpole}
\ee
which explicitly exhibts the massless scalar pole in the massless limit of (\ref{chipole}), and which agrees with
the explicit calculation of the physical $\lag 0\vert J_5^{\mu}\vert p,q\rag$ triangle amplitude to two photons 
(\ref{matpol}) in the previous section for $p^2=q^2 = m^2 = 0$.

\begin{figure}[htp]
\includegraphics[width=40cm, viewport=130 643 900 720,clip]{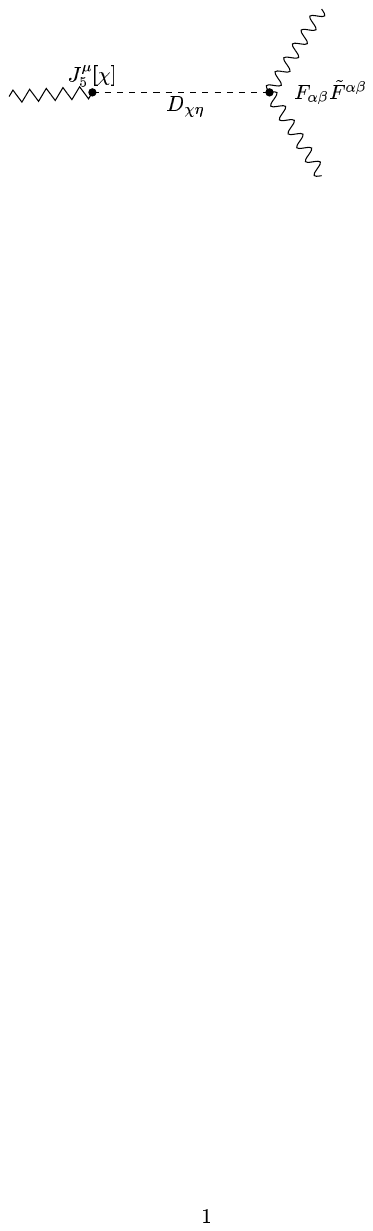}
\caption{Tree Diagram of the Efffective Action (\ref{axeffact}) showing the massless propagator $D_{\chi\eta}$ 
representing the massless $0^-$ state in the triangle amplitude (\ref{matpol}) to physical photons
when the fermion mass $m=0$.}
\label{Fig:xichi}
\end{figure}

The non-local action (\ref{nonl}) can be recast into a local form by the introduction of two pseudoscalar 
auxiliary fields $\eta$ and $\chi$ satisfying the second order linear eqs. of motion,
\bes\bea
&&\sq\,\eta = -\partial^{\lambda}{\cal B}_{\lambda}\,,\\
&&\sq\, \chi = \frac{e^2}{8\pi^2}\, F_{\mu\nu}\tilde F^{\mu\nu} 
= \frac{e^2}{16\pi^2}\,\epsilon^{\mu\nu\rho\sigma} F_{\mu\nu}F_{\rho\sigma}\,.
\label{chiaxeom}
\eea\label{auxaxeom}\ees
Then one can verify that the local quadartic action functional, 
\be
{\cal S}_{eff}[\eta, \chi; A, {\cal B}] =  \int d^4x\, 
\left\{ (\partial^{\mu}\eta)\,(\partial_{\mu}\chi) - \chi\, \partial^{\mu}{\cal B}_{\mu} 
+ \frac{e^2}{8\pi^2}\,\eta\,F_{\mu\nu}\tilde F^{\mu\nu} 
 \right\}
\label{axeffact}
\ee
yields back the eqs. of motion (\ref{auxaxeom}) when freely varied with respect to $\chi$ and $\eta$ respectively, 
while evaluating to (\ref{nonl}) upon using these eqs. of motion to solve for and eliminate the auxiliary fields. 
In the auxiliary field form of the effective action, (\ref{axeffact}) the quantum expectation value of the chiral 
current (\ref{expect}) is given by 
\be
J_5^{\mu}[\chi] = \frac{\delta {\cal S}_{eff}} {\delta {\cal B}_{\mu}} = 
\partial^{\mu} \chi = \frac{e^2}{16\pi^2} \partial^{\mu} \sq^{-1}
\epsilon^{\alpha\beta\rho\sigma}F_{\alpha\beta}F_{\rho\sigma}
\label{chipole}
\ee
at least insofar its anomalous divergence is concerned. The effective action (\ref{axeffact}) reproduces
the anomalous divergence of $\lag J_5^{\mu}\rag$, but not necessarily the non-anomalous parts of the tensor 
amplitude $\Gamma^{\mu\alpha\beta}(p,q)$. The $\eta$ and $\chi$ fields and their propagator are a local field 
representation of a massless $0^-$ state propagating in the physical matrix element (\ref{matpol})
with $p^2=q^2=m^2=0$, which may be represented by the effective tree diagram with source 
$F_{\mu\nu}\tilde F^{\mu\nu}$ on one end, and $\partial^{\lambda}{\cal B}_{\lambda}$ on the other, 
as in Fig. \ref{Fig:xichi}. 

The same diagram also represents the {\it vector} current expectation value $\lag J^{\mu}\rag_{_{\cal B}}$ in the 
presence of a background axial field ${\cal B}_{\lambda}$ and gauge field $A_{\mu}$, also implied by the original 
triangle diagram Fig. \ref{Fig:tri} upon reversing the roles of the axial vertex and one of the vector vertices, 
{\it i.e.}
\be
J^{\mu}[\eta] = \frac{\delta {\cal S}_{eff}} {\delta A_{\mu}} =
- \frac{e^2}{4\pi^2}\,\tilde F^{\nu\mu} \partial_{\nu}\eta
= \frac{e^2}{4\pi^2}\,\tilde F^{\nu\mu}\partial_{\nu} \sq^{-1}\partial^{\lambda} {\cal B}_{\lambda}\,.
\label{Bpole}
\ee
This crossing symmetry or equivalently, the fact that the non-local action (\ref{axnonl}) involves 
a mixed term involving both $F_{\mu\nu}\tilde F^{\mu\nu}$ and $\partial^{\lambda}{\cal B}_{\lambda}$
is the reason why two massless pseudoscalar auxiliary fields rather than just one are required 
to describe the amplitude correctly through a local effective action. A single auxiliary field
would necessarily produce unwanted direct $F_{\mu\nu}\tilde F^{\mu\nu}\sq^{-1}F_{\mu\nu}\tilde F^{\mu\nu}$
and $\partial^{\lambda}{\cal B}_{\lambda}\sq^{-1}\partial^{\lambda}{\cal B}_{\lambda}$ terms in the effective
action, not present in QED. Note that because of the mixed field kinetic term in (\ref{axeffact}), the
only propagator function that can appear is the mixed one, $D_{\chi\eta} = i \lag {\cal T}\chi\eta\rag$.

Several additional remarks concerning the effective action (\ref{axeffact}) are in order. First,
the axial anomaly, and hence the fields $\eta$ and $\chi$ and their propagator $D_{\chi\eta}$ are present 
for any $p^2, q^2, m^2$, although they decouple from the physical amplitude $\Gamma^{\mu\alpha\beta}(p,q)$ 
as $k^2 \rightarrow 0$ if any one of $p^2,q^2,m^2$ are greater than zero \cite{Huang,HorTer}. The massless 
$\eta$ and $\chi$ fields decouple from all physical processes involving electrons in that case, and the 
amplitude has a resonant peak (or peaks) at $s \sim\ (p^2, q^2, m^2)$ as in (\ref{rhoA00}), rather than 
a sharp $\delta(s)$ behavior. Because of the sum rule (\ref{chisum}) the resonance has the same total 
probability when integrated over $s$, but the massless propagator $D_{\chi\eta} = i \lag {\cal T} \chi\eta\rag$ 
saturates the physical on-shell amplitude (\ref{matpol}), and may be substituted in its place only 
when $p^2=q^2 =m^2=0$.

Secondly, since it contains kinetic terms for the auxiliary fields $\eta$ and $\chi$, the effective
action (\ref{axeffact}) describes two massless pseudoscalar degrees of freedom. These degrees of freedom 
are two-particle $0^-$ correlated $e^+e^-$ states, and $\eta$ and $\chi$ are pseudoscalar composite fields 
of bilinears of $\bar \psi$ and $\psi$. In fact, $\eta$ and $\chi$ may be defined by their relations 
to the vector and axial currents $J^{\mu}$ and $J^{\mu}_5$ of the underlying Dirac theory (\ref{currents}) 
by (\ref{Bpole}) and (\ref{chipole}) respectively. Hence varying $\eta$ and $\chi$ and treating them as 
true degrees of freedom is equivalent to varying the bilinear current densities $J^{\mu}$ and $J^{\mu}_5$, 
according to (\ref{Bpole}) and (\ref{chipole}).

Thirdly, these pseudoscalar degrees of freedom are implied also by a canonical operator treatment. Taking 
the effective action (\ref{axeffact}) as defining canonical momenta conjugate to the $\eta$ and $\chi$ 
fields via
\bes\bea
&&\Pi_{\eta} \equiv \frac{\delta {\cal S}_{eff}}{\delta \dot \eta} = -\dot \chi\,,\\
&&\Pi_{\chi} \equiv  \frac{\delta {\cal S}_{eff}}{\delta \dot \chi} = -\dot \eta\,,
\eea\ees
and imposing the equal time canonical commutation relations,
\be
[\eta(t, \vec x), \Pi_{\eta}(t,\vec x')] = i \delta^3(\vec x - \vec x') = [\chi(t, \vec x), \Pi_{\chi}(t,\vec x')] \,,
\label{cancom}
\ee
we find that the currents defined by (\ref{chipole}) and (\ref{Bpole}) satisfy the commutation relations,
\bes\bea
&&[J^0(t, \vec x), J^0_5(t, \vec x')] = -\frac{e^2}{2\pi^2} \tilde F^{0j} \partial_j [\eta(t, \vec x), \dot\chi(t,\vec x')] 
= -\frac{ie^2}{2\pi^2}\, {\bf B \cdot \nabla}\, \delta^3(\vec x - \vec x')\,,\\
&&[J^i(t, \vec x), J^0_5(t, \vec x')] = -\frac{e^2}{2\pi^2} \tilde F^{ij} \partial_j [\eta(t, \vec x), \dot\chi(t,\vec x')] 
= -\frac{ie^2}{2\pi^2}\, ({\bf E \times \nabla})^i \delta^3(\vec x - \vec x')\,,
\eea\label{anomcc}\ees
which are the anomalous commutation relations deducible from the covariant ${\cal T}^*$ time ordering
of the currents required by the axial anomaly\,\cite{Jackcurr,GrosJack}. In other words, the canonical 
commutation relations of the auxiliary fields with the kinetic terms in the effective action (\ref{axeffact}) 
are {\it required} by the anomalous equal time commutators of the currents $J^{\mu}$ and $J^{\nu}_5$. 
This suggests that $\eta$ and $\chi$ should be treated as {\it bona fide} quantum degrees of freedom in 
their own right. Because of the unique kinematic status of the triangle diagram \cite{ColGro}, and the 
non-renormalization of the axial anomaly \cite{AdlBar}, the structure of the effective action (\ref{axeffact})
and commutation relations (\ref{anomcc}) are not modified by any higher order processes.

Fourthly, we observe that the energy corresponding to (\ref{axeffact}) is not positive definite. This 
in itself should not be surprising, since the action (\ref{axeffact}) is a finite effective action in 
which the formally infinite energy of the Dirac sea has been effectively subtracted (by the counterterms
needed to impose gauge invariance, not encountered explicitly in our approach). Under some boundary
conditions, this finite subtracted energy can be negative, as in the Casimir effect. The conditions
under which this is true requires a careful analysis of the surface terms which we have 
neglected so far in our discussion. In fact, because both $\partial_{\mu}{\cal B}^{\mu}$
and $F_{\mu\nu}\tilde F^{\mu\nu} = \partial_{\mu}(\epsilon^{\mu\nu\alpha\beta}A_{\nu}F_{\alpha\beta})$, 
are total derivatives, the action (\ref{axeffact}) changes only by a surface term under constant shifts 
of either $\eta$ or $\chi$, and there are two Noether currents,
\bes\bea
&& K^{\mu}_A \equiv \frac{e^2}{8\pi^2} \epsilon^{\mu\nu\alpha\beta}A_{\nu}F_{\alpha\beta}
- \partial^{\mu}\chi \,,\\
&& K^{\mu}_{\cal B} \equiv {\cal B}^{\mu} + \partial^{\mu}\eta\,,
\eea\ees
\noindent with corresponding Noether charges which are conserved by the eqs. of motion
(\ref{auxaxeom}). The dynamics of the $\eta$ and $\chi$ fields are partly constrained by these
conservation laws, and should be considered together with the dynamics of vector and
axial vector sources $A_{\mu}$ and ${\cal B}_{\mu}$.

Finally, since the effective action (\ref{axeffact}) explicitly exhibiting these two pseudoscalar fields is 
nothing but a rewriting of the non-local form of the effective action for massless QED in the presence 
of an axial vector source, the massless degrees of freedom they represent have not been added in to the 
theory in an {\it ad hoc} manner. They are contained in QED as soon as it is extended by an arbitrary axial 
vector coupling as in (\ref{addax}), and are a necessary consequence of the axial anomaly, which in turn is 
required by imposition of all the other symmetries. 

In condensed matter physics, or electrodynamics at finite temperature or in polarizable media, where 
Lorentz invariance is broken, it is a familiar circumstance that there are low energy collective modes of the 
many-body theory, which are not part of the single particle constituent spectrum. This occurs also {\it in vacuo}
in the two dimensional massless Schwinger model, whose anomaly and longitudinal ``photon" can be described 
by the introduction of an effective scalar field composed of an $e^+ e^-$ pair\,\cite{LSB}. In $3+1$ dimensions,
relativistic kinematics and symmetries severely limit the possibilities for the appearance of such composite massless
scalars, with the triangle anomaly the only known example\,\cite{ColGro}. The fact that the $e^+e^-$ pair becomes 
collinear in the massless limit shows that this effectively reduces the dimensionality back to $1+1$. In the 
well studied $1+1$ dimensional case, the commutation relations of fermion bilinear currents $J^{\mu}$ and 
$J^{\nu}_5$, which create the composite $e^+e^-$ massless state are due to the anomaly\,\cite{AdBertHof}. 
Evidently from (\ref{cancom})-(\ref{anomcc}), a similar phenomenon occurs in the triangle amplitude in $3+1$ 
dimensions.

To conclude this section, one may ask: if there are massless pseudoscalar states in a weakly coupled 
theory like QED, which has been subjected to such exquisitely accurate tests, how could they have escaped 
detection? The answer to this is twofold. First, as we have seen these massless pseudoscalars do not couple 
to real QED with a finite electron mass, except at $k^2 \gg m^2$, so in massive QED they have no effects 
on low energy or long range electromagnetic interactions. Second, and more importantly, they require an axial 
vector source ${\cal B}_{\mu}$ (as well as a non-zero $F\tilde F$). In pure QED there is no axial vector 
coupling, {\it i.e.} $g=0$ in (\ref{addax}). Indeed it is impossible to introduce such a coupling into a 
$U(1)$ gauge theory with a dynamical axial vector ${\cal B}_{\mu}$ field without the breakdown of Ward identities 
necessary to the ultraviolet renormalizability of the theory. While it is theoretically possible to introduce 
a non-dynamical axial vector source, except for $\pi^0$ decay where indeed the axial anomaly with a quark 
triangle amplitude dominates\,\cite{BarFriGM}, it seems to be difficult to realize such a source in nature,
at least on macroscopic scales. In this situation the appearance of a massless pseudoscalar pole in the QED 
triangle amplitude, and its description by massless auxiliary fields is an interesting curiousity, illustrating 
the logical and kinematical possibility that anomalies may lead to unexpected consequences for the long distance 
physics in higher dimensions as well as lower ones, but which does not affect any predictions of QED in four 
dimensions with $g{\cal B}_{\mu}\equiv 0$.

\section{The $\lag TJJ\rag$ Triangle Amplitude in QED}

In this section we consider the amplitude for the trace anomaly in flat space that most closely corresponds 
to the triangle amplitude for the axial current anomaly reviewed in the previous section, and give a complete
calculation of the full $\lag T^{\mu\nu} J^{\alpha} J^{\beta}\rag$ amplitude for all values of the mass and the
off-shell kinematic invariants. Although the tensor structure of this amplitude is more involved than the
axial vector case, the kinematics is essentially the same, and the appearance of the massless pole very 
much analogous to the axial case.

The fundamental quantity of interest is the expectation value of the energy-momentum tensor 
bilinear in the fermion fields in an external electromagnetic potential $A_{\mu}$, 
\be
\lag T^{\mu\nu}\rag_{_A} = \lag T^{\mu\nu}_{free}\rag_{_A} + 
\lag T^{\mu\nu}_{int}\rag_{_A} \,
\label{Texpect}
\ee
where
\bes\bea
&&T^{\mu\nu}_{free} = -i \bar\psi \gamma^{(\mu}\!\!  
\stackrel{\leftrightarrow}{\partial}\!^{\nu)}\psi + g^{\mu\nu}
(i \bar\psi \gamma^{\lambda}\!\!\stackrel{\leftrightarrow}{\partial}\!\!_{\lambda}\psi
- m\bar\psi\psi)\\
&&T^{\mu\nu}_{int} = -\, e J^{(\mu}A^{\nu)} + e g^{\mu\nu}J^{\lambda}A_{\lambda}\,,
\eea\label{EMT}\ees 
are the contributions to the stress tensor of the free and interaction terms of the Dirac 
Lagrangian (\ref{addax}). The notations, $t^{(\mu\nu)} \equiv (t^{\mu\nu} + t^{\nu\mu})/2$ and
$\stackrel{\leftrightarrow}{\partial}\!\!_{\mu} \equiv 
(\stackrel{\rightarrow}{\partial}\!\!_{\mu} - \stackrel{\leftarrow}{\partial}\!\!_{\mu})/2$,
for symmetrization and ant-symmetrization have been used. 
The expectation value $\lag T^{\mu\nu}\rag_{_A}$ satisfies the partial 
conservation equation,
\be
\partial_{\nu} \lag T^{\mu\nu}\rag_{_A} = eF^{\mu\nu} \lag J_{\nu}\rag_{_A}\,,
\label{conspsi}
\ee
upon formal use of the Dirac eq. of motion (\ref{Dirac}). Just as in the
chiral case, the relation is formal because of the {\it a priori} ill-defined nature 
of the bilinear product of Dirac field operators at the same spacetime point in (\ref{EMT}).
Energy-momentum conservation in full QED ({\it i.e.} when the electromagnetic
field $A^{\mu}$ is also quantized) requires adding to the fermionic $T^{\mu\nu}$ of
(\ref{EMT}) the electromagnetic Maxwell stress tensor,
\be
T^{\mu\nu}_{\ Max} = F^{\mu\lambda}F^{\nu}_{\ \ \lambda} - \frac{1}{4} g^{\mu\nu}
F^{\lambda\rho}F_{\lambda\rho}
\label{MaxEMT}
\ee
which satisfies $\partial_{\nu}T^{\mu\nu}_{\ Max} = - F^{\mu\nu} J_{\nu}$. This
cancels (\ref{conspsi}) at the operator level, so that the full stress tensor of QED 
is conserved upon using Maxwell's eqs., $\partial_{\nu} F^{\mu\nu} = J^{\mu}$. Since 
in our present treatment $A_{\mu}$ is an arbitrary external potential, rather than a dynamical
field, we consider only the fermionic parts of the stress tensor (\ref{EMT}) whose
expectation value satisfies (\ref{conspsi}) instead.

At the classical level, {\it i.e.} again formally, upon use of (\ref{Dirac}), the trace of the fermionic 
stress tensor obeys 
\be
T^{\mu\ (cl)}_{\ \mu} \equiv g_{\mu\nu}T^{\mu\nu\, (cl)} = - m \bar\psi\psi \qquad
{\rm (classically)}\,,
\label{trclass}
\ee
analogous to the classical relation for the axial current (\ref{axclass}).
From this it would appear that $\lag T^{\mu\nu}\rag_{_A}$ will become traceless 
in the massless limit $m \rightarrow 0$, corresponding to the global
dilation symmetry of the classical theory with zero mass. However, as in the 
case of the classical chiral symmetry, this symmetry under global scale 
transformations cannot be maintained at the quantum level, without violating 
the conservation law satisfied by a related current, in this case the
partial conservation law (\ref{conspsi}), implied by general 
coordinate invariance. Requiring that (\ref{conspsi}) {\it is} preserved at the
quantum level necessarily leads to a well-defined anomaly in the 
trace,\,\cite{ChanEll,AdlColDun,DrumHath}, namely,
\be
\lag T^{\mu}_{\ \mu}\rag_{_A} \big\vert_{m=0} = -\frac{e^2}{24\pi^2}\, F_{\mu\nu}F^{\mu\nu}\,,
\label{tranom}
\ee
analogous to (\ref{axanom}). It is the infrared consequences of this modified, anomalous
trace identity and the appearance of massless scalar degrees of freedom for vanishing electron mass
$m= 0$ that we wish to study.

\begin{figure}[htp]
\includegraphics*[width=25cm, viewport=150 590 650 670,clip]{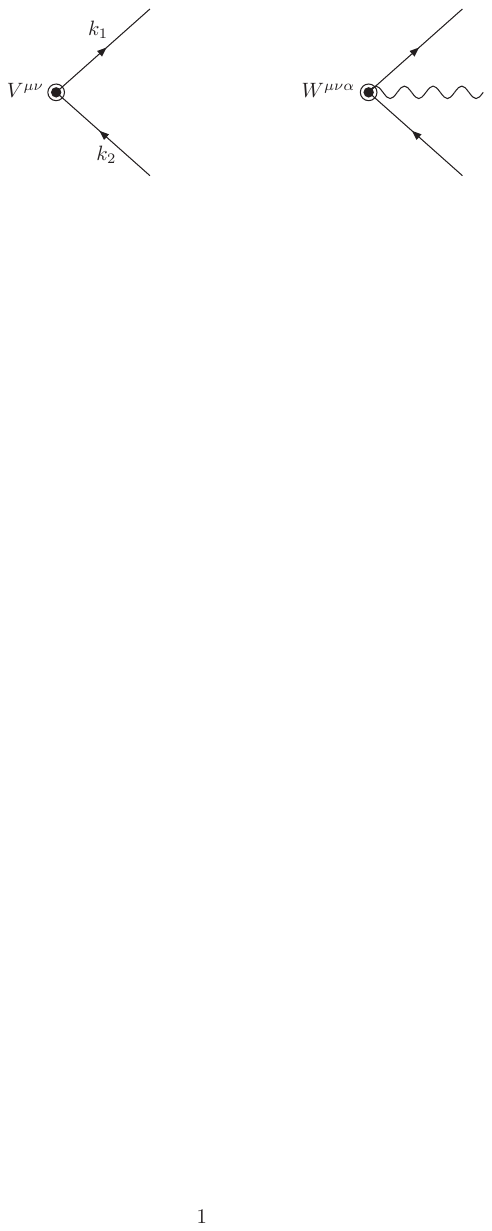}
\caption{The two kinds of vertices contributing to the stress tensor amplitude (\ref{Tampl}).}
\label{Fig:VWvertices}
\end{figure}

Our first task is to evaluate the full amplitude at one-loop order obtained by taking two functional 
variations of the expectation value (\ref{Texpect}) with respect to the external potential, and then 
evaluating at vanishing external field, $A=0$. In position space this is
\bea
&&\Gamma^{\mu\nu\alpha\beta} (z; x, y) \equiv \frac{ \delta^2 \lag T^{\mu\nu} (z) \rag_A}
{\delta A_{\alpha}(x)\delta A_{\beta}(y)} \bigg\vert_{A=0}
= (ie)^2 \lag T_{free}^{\mu\nu} (z) J^{\alpha} (x) J^{\beta} (y) \rag_{A=0} \nn
&& +\delta^4(x-z)g^{\alpha (\mu} \Pi^{\nu )\beta}(z, y)
+ \delta^4 (y-z)g^{\beta(\mu} \Pi^{\nu )\alpha}(z, x)
- g^{\mu\nu}[\delta^4(x-z) -  \delta^4(y-z) ]\Pi^{\alpha\beta}(x, y) \,,
\label{varpos}
\eea
where 
\be
\Pi^{\alpha \beta} (x,y) \equiv -e \frac{\delta \lag J^{\alpha}(x)\rag_{_A}} 
{\delta A_{\beta} (y)}\Big\vert_{_{A=0}}
= -i e^2 \lag J^{\alpha}(x)J^{\beta}(y)\rag_{_{A=0}}
\label{polariz}
\ee
is the electromagnetic polarization tensor in zero external field. Going over to momentum space 
and factoring out the resulting factor of momentum conservation, $(2\pi)^4 \delta^4(k-p-q)$, we obtain
\be
\Gamma^{\mu\nu\alpha\beta} (p,q) = \int\,d^4x\,\int\,d^4y\ e^{ip\cdot x + i q\cdot y}\,
\Gamma^{\mu\nu\alpha\beta} (z= 0; x, y)\,,
\label{Tampl}
\ee
which receives contributions from the two kinds of vertices derived from
the two terms in (\ref{EMT}), namely,
\bes\bea
&& V^{\mu\nu}(k_1, k_2)=\frac{1}{4} \left[\gamma^\mu (k_1 + k_2)^\nu
+\gamma^\nu (k_1 + k_2)^\mu \right] - \frac{1}{2} g^{\mu \nu} 
[\gamma^{\lambda}(k_1 + k_2)_{\lambda} + 2 m]   \,,\\
&& W^{\mu\nu\alpha} = -\frac{1}{2} (\gamma^\mu g^{\nu\alpha} 
+\gamma^\nu g^{\mu\alpha}) + g^{\mu \nu}\gamma^{\alpha}  \,,
\label{verts}
\eea\ees
respectively, represented in Figs. \ref{Fig:VWvertices}.

\begin{figure}[hbp]
\includegraphics*[width=22cm, height=6cm, viewport=190 570 650 700,clip]{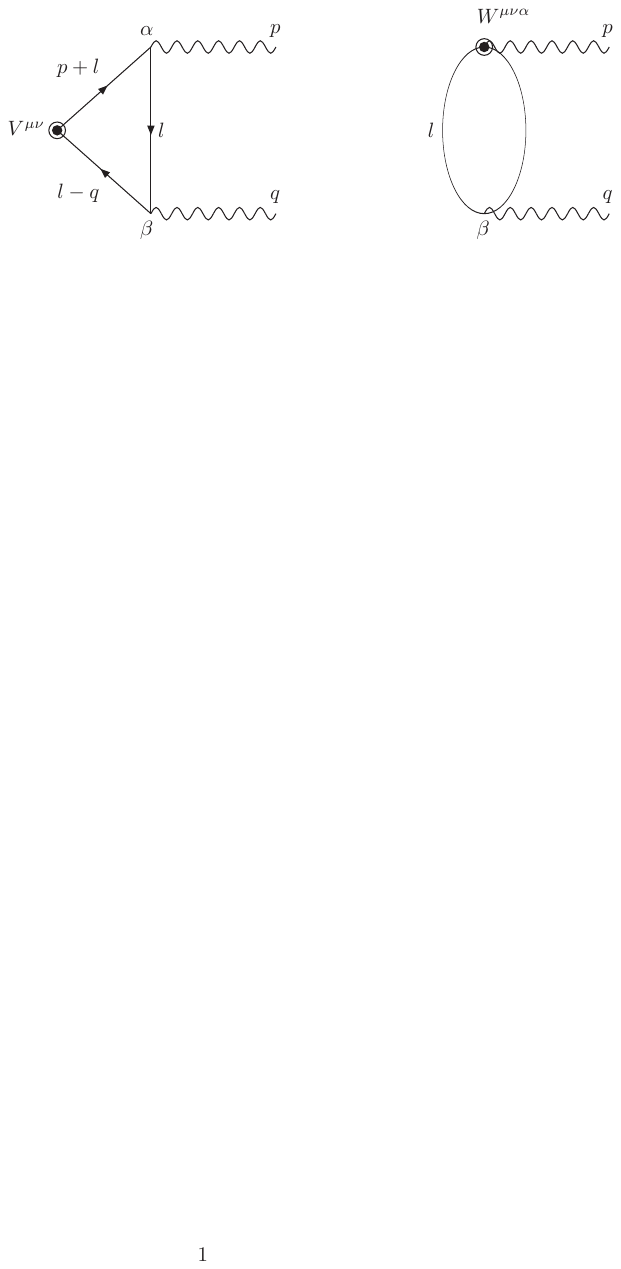}
\caption{The two kinds of amplitudes contributing to the stress tensor amplitude (\ref{Tampl}).}
\label{Fig:VWAmplitudes}
\end{figure}

At the one loop level the amplitude (\ref{Tampl}) is represented by the diagrams 
in Figs. \ref{Fig:VWAmplitudes}, together with those in which the photon legs are interchanged.
The first of these diagrams with the vertex $V^{\mu\nu}$ gives the contribution,
\bea
&&^V\Gamma^{\mu\nu\alpha\beta}(p,q)=(-1)(ie)^2
\int \frac{i\,d^4 l}{(2\pi)^4} \, {\rm tr}\left\{
V^{\mu\nu}(l+p,l-q)\ \frac{-i}{\lsl + \psl + m} \,\gamma^{\alpha}\, \frac{-i}{\lsl + m}
\,\gamma^{\beta} \,\frac{-i}{\lsl - \qsl + m}\right\}  \nn
&& \qquad = -e^2 \int \frac{d^4 l}{(2\pi)^4} \frac{{\rm tr}
\left\{V^{\mu\nu}(l+p,l-q)(-\lsl -\psl +m)\gamma^{\alpha}\,
(-\lsl + m)\, \gamma^{\beta}(-\lsl +\qsl + m)  \right\}}
{[(l+p)^2 + m^2] [(l-q)^2 + m^2] [l^2 + m^2]}\,,
\label{Gamone}
\eea
while the first of the diagrams with the vertex $W^{\mu\nu\alpha}$ gives
\bea
&&^W\Gamma^{\mu\nu\alpha\beta}(p,q)=(-1) i (ie)^2\int \frac{i\,d^4 l}{(2\pi)^4} \,
{\rm tr}\left\{W^{\mu\nu\alpha} \,\frac{-i}{\lsl + m}\gamma^{\beta}
\frac{-i}{\lsl -\qsl + m} \right\} \nn
&&\qquad = e^2 \int \frac{d^4 l}{(2\pi)^4} \frac{{\rm tr}\left\{W^{\mu\nu\alpha} 
\,(\lsl + m)\gamma^{\beta} (\lsl -\qsl +m) \right\}} {[l^2 + m^2][(l-q)^2 + m^2]}\nn
&&\qquad = -\frac{1}{2} g^{\nu\alpha} \Pi^{\mu\alpha} (q)
-\frac{1}{2} g^{\mu\alpha} \Pi^{\nu\alpha} (q)
+ g^{\mu \nu}\Pi^{\alpha\beta} (q)\,,
\label{Gamtwo}
\eea
where
\bea
&&\Pi^{\alpha\beta} (p) = \int d^4 x\, e^{i p\cdot (x-y)}\, \Pi^{\alpha\beta} (x,y)\nn
&&\qquad =e^2 \int \frac{d^4 l}{(2\pi)^4} \frac{{\rm tr}\left\{\gamma^{\alpha} 
\,(\lsl + m)\gamma^{\beta} (\lsl -\psl +m) \right\}} {[l^2 + m^2][(l-p)^2 + m^2]}
\label{polmom}
\eea
is the Fourier transform of (\ref{polariz}). The additional factor of $i$ 
in the loop integration measure of (\ref{Gamone}) and (\ref{Gamtwo}) comes from the 
continuation to Euclidean momenta: $l_0 \rightarrow i l_4$. 

As usual these loop integrals are formal and divergent, since the one-loop polarization requires 
regularization and renormalization, which we postpone for the moment. The second set of diagrams 
with the $W$ interaction vertex give rise to the second set of terms in (\ref{varpos}) 
explicitly proportional to the polarization (\ref{polmom}). As we shall see, we actually
require only the finite parts of the first diagram in Fig. \ref{Fig:VWAmplitudes},
together with the equivalent diagram obtained by interchanging $p$ and $q$ and $\alpha$ and $\beta$.
The contribution of the contact terms of the second kind of diagram in Fig. \ref{Fig:VWAmplitudes}
with the $W^{\mu\nu\alpha}$ vertex will be determined from the Ward identities.

The full one-loop contribution to the amplitude (\ref{Tampl}) is the Bose symmetric sum,
\be
\Gamma^{\mu\nu\alpha\beta} (p,q) =\, ^{V\hspace{-.1cm}}\Gamma^{\mu\nu\alpha\beta}(p,q) +\,
^{V\hspace{-.1cm}}\Gamma^{\mu\nu\beta\alpha}(q,p) +\, ^{W\hspace{-.1cm}}\Gamma^{\mu\nu\alpha\beta}(p,q)
+\, ^{W\hspace{-.1cm}}\Gamma^{\mu\nu\beta\alpha}(q,p) = \Gamma^{\mu\nu\beta\alpha}(q,p)\,.
\label{Tamplsum}
\ee
Vector current conservation $\partial_{\mu} J^{\mu} = 0$ implies that both the polarization, and 
this amplitude should satisfy the Ward identities,
\bes\bea
p_{\alpha} \Pi^{\alpha\beta} (p) = q_{\beta}\Pi^{\alpha\beta} (q) =0\,;\\
p_\alpha \Gamma^{\mu\nu\alpha\beta}(p,q) = q_\beta\Gamma^{\mu\nu\alpha\beta}(p,q)=0  \,.
\eea\label{gauge}\ees
The first of these relations implies that $\Pi^{\alpha\beta} (p)$ is transverse, {\it i.e.}
\be
\Pi^{\alpha\beta} (p) = (p^2 g^{\alpha\beta} - p^{\alpha}p^{\beta}) \Pi (p^2)\,.
\label{trans}
\ee
In addition, the partial conservation law (\ref{conspsi}) implies that the
amplitude in (\ref{varpos}) should also satisfy the Ward identity,
\vspace{-.05cm}
\bea
&&\frac{\partial}{\partial z^{\nu}}\Gamma^{\mu\nu\alpha\beta} (z; x, y) =
\Pi^{\beta\nu}(y,z) (g^{\alpha\mu} \delta^{\lambda}_{\nu}
-\delta^{\alpha}_{\nu}g^{\mu\lambda} )
\frac{\partial}{\partial x^{\lambda}}\delta^4(x-z)\nn
&& \qquad + \Pi^{\alpha\nu}(x,z) (g^{\beta\mu} \delta^{\lambda}_{\nu}
-\delta^{\beta}_{\nu}g^{\mu\lambda})
\frac{\partial}{\partial y^{\lambda}}\delta^4(y-z)\,.
\eea
In momentum space this becomes
\be
k_{\nu}\Gamma^{\mu\nu\alpha\beta} (p,q) =
(g^{\mu\alpha} p_{\nu} -\delta^{\alpha}_{\nu} p^{\mu})\Pi^{\beta\nu}(q) +
(g^{\mu\beta} q_{\nu} -\delta^{\beta}_{\nu}q^{\mu})\Pi^{\alpha\nu}(p)\,,
\ee
or combining with (\ref{trans}), we obtain
\bea
k_\nu\,\Gamma^{\mu\nu\alpha\beta}(p,q)&=& \left(q^\mu p^\alpha p^\beta
-q^\mu g^{\alpha\beta} p^2 +g^{\mu\beta}q^\alpha p^2
-g^{\mu\beta}p^\alpha p\cdot q \right)   \Pi(p^2) \nonumber\\
&&+ \left(p^\mu q^\alpha q^\beta - p^\mu g^{\alpha\beta} q^2
+g^{\mu\alpha}p^\beta q^2 - g^{\mu\alpha}q^\beta p\cdot q\right) \Pi(q^2) \,.
\label{TWard}
\eea
As already remarked, all of these relations are formal since both $\Gamma^{\mu\nu\alpha\beta}(p,q)$ 
and $\Pi^{\alpha\beta}(p)$ are ill-defined {\it a priori} and require some procedure to extract the
finite terms.

Formally one could use (\ref{trclass}) to obtain an identity for the trace on the amplitude 
$g_{\mu\nu}\Gamma^{\mu\nu\alpha\beta}$. Like (\ref{gauge}) and (\ref{TWard}) 
this trace relation can be proven if and only if shifting the loop integration 
variable $l$ in the integrals (\ref{Gamone}) and (\ref{Gamtwo}) is allowed, and 
terms which are odd in the loop integration variable are dropped. Since the loop integrals 
in (\ref{Gamone}), (\ref{Gamtwo}) and (\ref{polmom}) are formally quadratically divergent 
they are not well defined as they stand, so that formal manipulations of this kind can 
yield ambiguous or incorrect conclusions. The standard method of dealing with such ill-defined 
expressions is regularization. Any regularization method that preserves the Ward identities of 
gauge invariance (\ref{gauge}) and coordinate covariance (\ref{TWard}) may be used, such as 
dimensional regularization, Pauli-Villars regularization or the Schwinger-DeWitt heat kernel method. 
It is important to recognize that regularization amounts to supplying additional information which 
serves to {\it define} an ill-defined expression, by requiring that certain symmetries of the classical 
theory be strictly maintained at the quantum level. 

Here we shall follow the alternative approach, exactly parallel to the previous treatment 
of the axial triangle anomaly in Sec. 2, which does not require any explicit choice of covariant 
regularization scheme. Instead we {\it define} the Lorentz covariant tensor amplitude 
$\Gamma^{\mu\nu\alpha\beta}(p,q)$ by its {\it finite} terms, together with the requirement that the 
full amplitude satisfy the Ward identities (\ref{gauge}) and (\ref{TWard}). Then the joint requirements of:
\vspace{-.3cm}
\begin{itemize} 
\item [(i)] Lorentz invariance of the vacuum, \vspace{-.3cm} 
\item [(ii)] Bose symmetry,\vspace{-.3cm}
\item [(iii)] vector current conservation (\ref{gauge}), \vspace{-.3cm}
\item [(iv)] unsubtracted dispersion relation of real and imaginary parts, and\vspace{-.3cm}
\item [(v)]  energy-momentum tensor conservation (\ref{TWard}),  \vspace{-.3cm}
\end{itemize}
are sufficient to determine the full amplitude $\Gamma^{\mu\nu\alpha\beta}(p,q)$
in terms of its explicitly finite pieces, and yield a well-defined finite trace anomaly.
As in the axial anomaly case conisdered previously, this method of constructing the full 
$\Gamma^{\mu\nu\alpha\beta}(p,q)$ may be regarded as a proof that the same finite trace 
anomaly must be obtained in any regularization scheme that respects (i)-(v) above.

Lorentz invariance of the vacuum is assumed first by expanding the amplitude in terms of all the 
possible tensors with four indices depending on $p^{\alpha}$, $q^{\beta}$ and the flat spacetime 
metric $g^{\mu\nu}$. There are $2^4 = 16$ tensors with all four Lorentz indices 
$(\mu ,\nu ,\alpha ,\beta)$ are carried by either $p$ or $q$; $2^2 \times 6 =24$ 
tensors in which two of the four indices are carried by the symmetric metric tensor 
$g^{\mu\nu}$ and the other two by either $p^{\mu}$ or $q^{\mu}$; and just $3$ tensors 
in which the four indices are distributed over a product of two metric tensors 
with no factors of $p$ or $q$. The complete set of these $43$ tensor monomials 
is given in Table \ref{amplbasis}. Lorentz covariance requires that the amplitude
$\Gamma^{\mu\nu\alpha\beta} (p,q)$ must be expandable in this complete set
of $43$ tensors with scalar coefficient functions of the three invariants 
$p^2, q^2$, and $p\cdot q$, or equivalently $p^2, q^2$, and $k^2 = (p+q)^2$.

\begin{table}
$$
\begin{array}{|c @{\hspace{.5cm}}| @{\hspace{.5cm}}c @{\hspace{.5cm}}|
@{\hspace{.5cm}}c @{\hspace{.5cm}}| @{\hspace{.5cm}} c @{\hspace{.5cm}}|
@{\hspace{.5cm}} c @{\hspace{.5cm}}| @{\hspace{.5cm}} c @{\hspace{.5cm}}|}
\hline
& & & & & \\[-.5cm]
\begin{array}[t]{c}
p^{\mu} p^{\nu} p^{\alpha} p^{\beta}\\
q^{\mu} q^{\nu} q^{\alpha} q^{\beta}
\end{array}
&
\begin{array}[t]{c}
p^{\mu} p^{\nu} p^{\alpha} q^{\beta}\\
p^{\mu} p^{\nu} q^{\alpha} p^{\beta}\\
p^{\mu} q^{\nu} p^{\alpha} p^{\beta}\\
q^{\mu} p^{\nu} p^{\alpha} p^{\beta}
\end{array}
&
\begin{array}[t]{c}
p^{\mu} p^{\nu} q^{\alpha} q^{\beta}\\
p^{\mu} q^{\nu} p^{\alpha} q^{\beta}\\
q^{\mu} p^{\nu} p^{\alpha} q^{\beta}
\end{array}
&
\begin{array}[t]{c}
p^{\mu} q^{\nu} q^{\alpha} p^{\beta}\\
q^{\mu} p^{\nu} q^{\alpha} p^{\beta}\\
q^{\mu} q^{\nu} p^{\alpha} p^{\beta}
\end{array}
&
\begin{array}[t]{c}
p^{\mu} q^{\nu} q^{\alpha} q^{\beta}\\
q^{\mu} p^{\nu} q^{\alpha} q^{\beta}\\
q^{\mu} q^{\nu} p^{\alpha} q^{\beta}\\
q^{\mu} q^{\nu} q^{\alpha} p^{\beta}
\end{array}
&
\begin{array}[t]{c}
g^{\mu\nu}g^{\alpha\beta}\\
g^{\alpha\mu}g^{\beta\nu}\\
g^{\alpha\nu}g^{\beta\mu}
\end{array}
\\[2.1cm]
\hline
& & & & & \\[-.5cm]
\begin{array}{c}
p^{\mu} p^{\nu} g^{\alpha\beta}\\
p^{\mu} q^{\nu} g^{\alpha\beta}\\
q^{\mu} p^{\nu} g^{\alpha\beta}\\
q^{\mu} q^{\nu} g^{\alpha\beta}
\end{array}
&
\begin{array}{c}
p^{\beta} p^{\nu} g^{\alpha\mu}\\
p^{\beta} q^{\nu} g^{\alpha\mu}\\
q^{\beta} p^{\nu} g^{\alpha\mu}\\
q^{\beta} q^{\nu} g^{\alpha\mu}
\end{array}
&
\begin{array}{c}
p^{\beta} p^{\mu} g^{\alpha\nu}\\
p^{\beta} q^{\mu} g^{\alpha\nu}\\
q^{\beta} p^{\mu} g^{\alpha\nu}\\
q^{\beta} q^{\mu} g^{\alpha\nu}
\end{array}
&
\begin{array}{c}
p^{\alpha} p^{\nu} g^{\beta\mu}\\
p^{\alpha} q^{\nu} g^{\beta\mu}\\
q^{\alpha} p^{\nu} g^{\beta\mu}\\
q^{\alpha} q^{\nu} g^{\beta\mu}
\end{array}
&
\begin{array}{c}
p^{\mu} p^{\alpha} g^{\beta\nu}\\
p^{\mu} q^{\alpha} g^{\beta\nu}\\
q^{\mu} p^{\alpha} g^{\beta\nu}\\
q^{\mu} q^{\alpha} g^{\beta\nu}
\end{array}
&
\begin{array}{c}
p^{\alpha} p^{\beta} g^{\mu\nu}\\
p^{\alpha} q^{\beta} g^{\mu\nu}\\
q^{\alpha} p^{\beta} g^{\mu\nu}\\
q^{\alpha} q^{\beta} g^{\mu\nu}
\end{array}
\\ [1.2cm]\hline
\end{array}
$$
\caption{The 43 fourth rank tensor monomials into which $\Gamma^{\mu\nu\alpha\beta}(p,q)$ 
can be expanded \label{amplbasis}}
\end{table}

Since the amplitude (\ref{Tampl}) has total mass dimension $2$, the scalar coefficient functions
multiplying the tensors in our list which are homogeneous of degree $4$ in $p$ and $q$ have
mass dimension $-2$. These coefficients can be extracted in terms of loop integrals which are
UV quadratically convergent and finite. Then the coefficients of the remaining tensors
are determined by the Ward identities of vector current and stress-tensor conservation.
Let us define the two-index tensors,
\bes\bea
&&u^{\alpha\beta}(p,q) \equiv (p\cdot q) g^{\alpha\beta} - q^{\alpha}p^{\beta}\,,\\
&&w^{\alpha\beta}(p,q) \equiv p^2 q^2 g^{\alpha\beta} + (p\cdot q) p^{\alpha}q^{\beta}
- q^2 p^{\alpha}p^{\beta} - p^2 q^{\alpha}q^{\beta}\,,
\eea \label{uwdef}\ees
each of which satisfies the conditions of Bose symmetry,
\bes\bea
&&u^{\alpha\beta}(p,q) = u^{\beta\alpha}(q,p)\,,\\
&&w^{\alpha\beta}(p,q) = w^{\beta\alpha}(q,p)\,,
\eea\ees
and vector current conservation,
\bes\bea
&&p_{\alpha} u^{\alpha\beta}(p,q) = 0 = q_{\beta}u^{\alpha\beta}(p,q)\,,\\
&&p_{\alpha} w^{\alpha\beta}(p,q) = 0 = q_{\beta}w^{\alpha\beta}(p,q)\,.
\eea\ees
These tensors may be obtained from the variation of local gauge invariant quantities
$F_{\mu\nu}F^{\mu\nu}$ and $(\partial_{\mu} F^{\mu}_{\ \,\lambda})(\partial_{\nu}F^{\nu\lambda})$
respectively, via
\bes\bea
&&u^{\alpha\beta}(p,q) = -\frac{1}{4}\int\,d^4x\,\int\,d^4y\ e^{ip\cdot x + i q\cdot y}\ 
\frac{\delta^2 \{F_{\mu\nu}F^{\mu\nu}(0)\}} {\delta A_{\alpha}(x) A_{\beta}(y)} \,,
\label{locvaru}\\
&&w^{\alpha\beta}(p,q) = \frac{1}{2} \int\,d^4x\,\int\,d^4y\ e^{ip\cdot x + i q\cdot y}\
\frac{\delta^2 \{\partial_{\mu} F^{\mu}_{\ \,\lambda}\partial_{\nu}F^{\nu\lambda}(0)\}} 
{\delta A_{\alpha}(x) A_{\beta}(y)}\,.\label{locvarw}
\eea\label{locvar}\ees
Making use of $u^{\alpha\beta}(p,q)$ and $w^{\alpha\beta}(p,q)$, one finds that 
of the $43$ tensors in Table \ref{amplbasis}, there are exactly $13$ linearly
independent four-tensors $t_i^{\mu\nu\alpha\beta}(p,q)$, $i=1, \dots, 13$, which satisfy
\be
p_{\alpha} t_i^{\mu\nu\alpha\beta}(p,q) = 0 = q_{\beta} t_i^{\mu\nu\alpha\beta}(p,q) \,,
\qquad i=1, \dots, 13\,.
\label{vcons}
\ee
These $13$ tensors are catalogued in Table \ref{genbasis}. 
\begin{table}
$$
\begin{array}{|c|c|}\hline
i & t_i^{\mu\nu\alpha\beta}(p,q)\\ \hline\hline
1 &
\left(k^2 g^{\mu\nu} - k^{\mu } k^{\nu}\right) u^{\alpha\beta}(p.q)\\ \hline 
2 &
\left(k^2g^{\mu\nu} - k^{\mu} k^{\nu}\right) w^{\alpha\beta}(p.q)  \\ \hline
3 & \left(p^2 g^{\mu\nu} - 4 p^{\mu}  p^{\nu}\right) 
u^{\alpha\beta}(p.q)\\ \hline
4 & \left(p^2 g^{\mu\nu} - 4 p^{\mu} p^{\nu}\right)
w^{\alpha\beta}(p.q)\\ \hline
5 & \left(q^2 g^{\mu\nu} - 4 q^{\mu} q^{\nu}\right) 
u^{\alpha\beta}(p.q)\\ \hline
6 & \left(q^2 g^{\mu\nu} - 4 q^{\mu} q^{\nu}\right) 
w^{\alpha\beta}(p.q) \\ \hline
7 & \left[p\cdot q\, g^{\mu\nu}   
-2 (q^{\mu} p^{\nu} + p^{\mu} q^{\nu})\right] u^{\alpha\beta}(p.q) \\ \hline
8 & \left[p\cdot q\, g^{\mu\nu} 
-2 (q^{\mu} p^{\nu} + p^{\mu} q^{\nu})\right] w^{\alpha\beta}(p.q)\\ \hline
9 & \left(p\cdot q \,p^{\alpha}  - p^2 q^{\alpha}\right) 
\big[p^{\beta} \left(q^{\mu} p^{\nu} + p^{\mu} q^{\nu} \right) - p\cdot q\,
(g^{\beta\nu} p^{\mu} + g^{\beta\mu} p^{\nu})\big]  \\ \hline
10 & \big(p\cdot q \,q^{\beta} - q^2 p^{\beta}\big)\, 
\big[q^{\alpha} \left(q^{\mu} p^{\nu} + p^{\mu} q^{\nu} \right) - p\cdot q\,
(g^{\alpha\nu} q^{\mu} + g^{\alpha\mu} q^{\nu})\big]  \\ \hline
11 & \left(p\cdot q \,p^{\alpha} - p^2 q^{\alpha}\right)
\big[2\, q^{\beta} q^{\mu} q^{\nu} - q^2 (g^{\beta\nu} q^ {\mu} 
+ g^{\beta\mu} q^{\nu})\big]  \\ \hline
12 & \big(p\cdot q \,q^{\beta} - q^2 p^{\beta}\big)\,
\big[2 \, p^{\alpha} p^{\mu} p^{\nu} - p^2 (g^{\alpha\nu} p^ {\mu} 
+ g^{\alpha\mu} p^{\nu})\big] \\ \hline
13 & \big(p^{\mu} q^{\nu} + p^{\nu} q^{\mu}\big)g^{\alpha\beta}
+ p\cdot q\, \big(g^{\alpha\nu} g^{\beta\mu} 
+ g^{\alpha\mu} g^{\beta\nu}\big) - g^{\mu\nu} u^{\alpha\beta} \\
& -\big(g^{\beta\nu} p^{\mu} 
+ g^{\beta\mu} p^{\nu}\big)q^{\alpha} 
- \big (g^{\alpha\nu} q^{\mu} 
+ g^{\alpha\mu} q^{\nu }\big)p^{\beta}  \\ \hline
\end{array}
$$
\caption{The 13 fourth rank tensors satisfying (\ref{vcons}) \label{genbasis}}
\end{table}

This set of $13$ tensors is linearly independent for generic $k^2, p^2, q^2$
different from zero. Five of $13$ are Bose symmetric, namely,
\be
t_i^{\mu\nu\alpha\beta}(p,q) = t_i^{\mu\nu\beta\alpha}(q,p)\,,\qquad i=1,2,7,8,13\,,
\ee
while the remaining eight tensors form four pairs related by Bose symmetry:
\bes\bea
&&t_3^{\mu\nu\alpha\beta}(p,q) = t_5^{\mu\nu\beta\alpha}(q,p)\,,\\
&&t_4^{\mu\nu\alpha\beta}(p,q) = t_6^{\mu\nu\beta\alpha}(q,p)\,,\\
&&t_9^{\mu\nu\alpha\beta}(p,q) = t_{10}^{\mu\nu\beta\alpha}(q,p)\,,\\
&&t_{11}^{\mu\nu\alpha\beta}(p,q) = t_{12}^{\mu\nu\beta\alpha}(q,p)\,.
\eea \label{tpairs}\ees
Expanding the amplitude (\ref{Gamone}) in this basis,
\be
\Gamma^{\mu\nu\alpha\beta}(p,q) = \sum_{i=1}^{13} F_i (k^2; p^2, q^2)\ t_i^{\mu\nu\alpha\beta}(p,q)\,.
\label{Gamt}
\ee
Bose symmetry implies that the scalar functions $F_1, F_2, F_7, F_8$, and $F_{13}$ are symmetric under 
interchange of $p^2$ and $q^2$, while the remaining eight functions form four pairs related by
Bose symmetry,
\bes\bea
&&F_3 (k^2; p^2,q^2) = F_5 (k^2; q^2, p^2)\,,\\
&&F_4 (k^2; p^2,q^2) = F_6 (k^2; q^2, p^2)\,,\\
&&F_9 (k^2; p^2,q^2) = F_{10} (k^2; q^2, p^2)\,,\\
&&F_{11} (k^2; p^2,q^2) = F_{12} (k^2; q^2, p^2)\,,
\eea\ees
corresponding to (\ref{tpairs}). Thus there are $9$ independent scalar functions in the amplitude 
(\ref{Gamt}), $5$ of them completely symmetric, and $4$ of them possessing both symmetric and 
anti-symmetric terms under $p^2 \leftrightarrow q^2$ interchange, for $5 + (4\times 2) = 13$ scalar 
amplitudes in all. We observe that all but $t_{13}$ contain terms which are homogeneous of degree 
four in the external momenta, whose coefficients we may constrain from the finite parts of the amplitude.

We have chosen this basis so that only the first two of the thirteen tensors 
possess a non-zero trace,
\bes\bea
&& g_{\mu\nu}t_1^{\mu\nu\alpha\beta}(p,q)  = 3k^2\, u^{\alpha\beta}(p,q)\,,\\
&& g_{\mu\nu}t_2^{\mu\nu\alpha\beta}(p,q)  = 3k^2\, w^{\alpha\beta}(p,q)\,,
\eea\ees
while the remaining eleven tensors are traceless,
\be
g_{\mu\nu}t_i^{\mu\nu\alpha\beta}(p,q) = 0\,, \qquad i=3, \dots , 13\,.
\ee
Moreover because of eqs. (\ref{tranom}) and (\ref{locvar}), we have chosen the basis $t_i$ 
in anticipation of the result that in the limit of zero fermion mass, the entire trace anomaly 
will reside only in the first amplitude function, $F_1 (k^2; p^2,q^2)$.

To proceed, we now fix as many of the $13$ scalar functions $F_i$ as possible by examining 
the finite terms in the formal expressions (\ref{Gamone}) and (\ref{Gamtwo}). To this end we perform 
the indicated Dirac algebra and introduce the Feynman parameterization (\ref{Feyn}) of the product 
of propagator denominators. Then we make any necessary shifts in the loop integration variable $l$ in 
(\ref{Gamone}) and (\ref{Gamtwo}) in order to extract only those terms for which the four indices 
$(\mu ,\nu ,\alpha ,\beta)$ are carried by the external momentum vectors $p$ and $q$ in various 
combinations, which therefore can be removed from the loop integration. The remaining loop
integration is then finite for these terms and can be extracted unambiguously. The details of this 
computation are given in Appendix A.

To make this procedure completely rigorous, one can calculate first only the discontinuities
of the amplitude, continued to timelike four-momenta, using the Cutkovsky rule replacement, 
\be
\frac{1}{(l+p)^2 + m^2} \rightarrow 2\pi i\, \theta(l^0 + p^0)\,\delta\left((l+p)^2 + m^2\right)
\ee
for the lines cut as in Fig. \ref{Fig:ImTri} in the chiral case. Owing to these delta functions in the
cut diagram, the discontinuity of (\ref{Gamone}) is completely finite. Then for those terms (and only 
those terms) multiplying tensors of degree $4$ in $p$ and $q$, the real parts may be constructed from 
the discontinuous imaginary parts by {\it unsubtracted} dispersion relations,
and are completely finite as well. Since there are $12$ such tensors of degree $4$ in the external 
momenta, listed in Table \ref{tensorcoeff}, there are $12$ finite scalar coefficient functions 
$C_j(k^2;p^2,q^2)$  multiplying them which are defined in this way. The explicit form of their
corresponding imaginary parts $\rho_j$ is given by (\ref{specrep}) in Sec. 6. It is not difficult to 
show from the linear independence of the tensors in Table \ref{tensorcoeff}, and general analyticity 
properties of the amplitude that the finite coefficient functions $C_j$ of mass dimension $-2$ obtained 
in this way from their imaginary parts are identical to those obtained by the recipe of shifting the loop 
integration variable in the original full amplitude, in Feynman parameterized form, and identifying the 
terms multiplying each tensor listed in Table \ref{tensorcoeff}. This is of course also the same result 
for the finite terms that is obtained if the loop integration were regularized in a covariant way, such as 
in the dimensional regularization or Pauli-Villars schemes, in which the shift of the loop integration 
variable is permitted. It is also noteworthy that this procedure of extracting the finite parts of 
(\ref{Gamone}) relies only upon the terms involving the $V^{\mu\nu}$ vertex in the triangle 
diagram of Fig. \ref{Fig:VWAmplitudes},  and not the contributions of the $W^{\mu\nu\alpha}$ 
vertex which are proportional to the polarization tensor (\ref{polmom}), and divergent.

\begin{table}
$$
\begin{array}{|c|c|c|}\hline
j & C_j = {\rm coefficient\ of} & c_j(x,y) \\ \hline\hline
1 & p^{\mu}p^{\nu} p^{\alpha}p^{\beta} & - 4x^2 (1-x) (1-2 x) \\ \hline 
2 &\ \ (p^{\mu}q^{\nu} + q^{\mu}p^{\nu}) p^{\alpha}p^{\beta}\ \ & - x (1-x)(1-4x+8xy) + xy \\ \hline
3 & q^{\mu}q^{\nu} p^{\alpha}p^{\beta} &  2x(1-2y)(1-x-y + 2 x y )\\ \hline
4 & p^{\mu}p^{\nu} p^{\alpha}q^{\beta} & -2x(1-x)(1-2x)(1-2y)\\ \hline
5 &\ \ (p^{\mu}q^{\nu} + q^{\mu}p^{\nu}) p^{\alpha}q^{\beta}\ \ &\ \ \ x(1-x)(1-2y)^2 + y(1-y)(1-2x)^2\ \ \ \\ \hline
6 & q^{\mu}q^{\nu} p^{\alpha}q^{\beta} & - 2y(1-y)(1-2x)(1-2y)\\ \hline
7 & p^{\mu}p^{\nu} q^{\alpha}p^{\beta} &  2 x y(1-2x)^2 \\ \hline
8 &\ \ (p^{\mu}q^{\nu} + q^{\mu} p^{\nu}) q^{\alpha}p^{\beta}\ \ & - 2 x y(1-2x) (1-2y)\\ \hline
9 & q^{\mu}q^{\nu} q^{\alpha}p^{\beta} & 2 x y (1-2y)^2\\ \hline
10 & p^{\mu}p^{\nu} q^{\alpha}q^{\beta} & 2 y(1-2x)(1-x-y + 2xy)\\ \hline
11 &\ \ (p^{\mu}q^{\nu} + q^{\mu}p^{\nu}) q^{\alpha}q^{\beta}\ \ & - y (1-y)(1-4y+8xy) + xy\\ \hline
12 & q^{\mu}q^{\nu} q^{\alpha}q^{\beta} & - 4  y^2\,(1-2 y) (1-y)\\ \hline
\end{array}
$$
\caption{The twelve tensors with four free indices 
$(\mu\nu\alpha\beta)$ on $p,q$ which appear in the amplitude (\ref{Gamt}),
with finite scalar coefficient functions $C_j(k^2;p^2,q^2)$
and corresponding polynomials in the Feynman parameterized form, (\ref{Cj}).}
\label{tensorcoeff}
\end{table}

The $12$ scalar coefficient functions listed in Table \ref{tensorcoeff} are not all 
independent. Owing to the tensor structure in the table imposed by Bose symmetry and 
vector current conservation, two pair of the coefficients are trivially dependent upon 
one other, namely
\bes\bea
&& - (p\cdot q) C_1 = q^2 C_4\,,\label{Crela}\\
&& - (p\cdot q) C_{12} = p^2 C_6\,, \label{Crelb}
\eea\label{Crel}\ees
so that only $10$ independent coefficient functions can be determined from the
finite parts of (\ref{Gamt}). 

Inspection of Table \ref{tensorcoeff} shows also that the coefficients 
$C_5$ and $C_8$ are automatically Bose symmetric, while the remaining ten 
coefficients occur in five Bose conjugate pairs, {\it viz.} $(C_1, C_{12}), 
(C_2, C_{11}), (C_3, C_{10}), (C_4, C_6)$, and $(C_7, C_9)$, so that for example, 
$C_1 (k^2; p^2,q^2) = C_{12} (k^2; q^2,p^2)$. Explicit formulae for all twelve 
finite coefficient functions may be given in the Feynman parameterized form,
\be
C_j(k^2; p^2,q^2) = \frac{e^2}{4\pi^2}
\int_0^1 dx\int_0^{1-x} dy \ \frac{c_j(x,y)}{p^2\, x(1-x)+q^2\, y(1-y)+
2\,xy\,p\cdot q+m^2}\,,
\label{Cj}
\ee
where the polynomials $c_i(x,y)$ for $i=1, \dots, 12$ are listed in Table \ref{tensorcoeff}. 
From (\ref{Cj}) with the help of this Table it is straightforward to verify relations (\ref{Crel}) 
and identify the Bose conjugate pairs of coefficient functions by interchange of $x$ and $y$.
These relations are verified in Appendix B.

{\allowdisplaybreaks
Identifying the coefficients of the finite amplitudes in terms of the tensors of Table \ref{genbasis}
gives $10$ relations, which we group into the following three sets. First we have the three relations,
\bes\bea
&& F_1 + 4F_3 = C_7\,,\\
&& F_1 + 4 F_5 = C_9\,,\\
&& F_1 + 2F_7 = \frac{p^2 C_2 + q^2 C_{11}}{p\cdot q} 
+ \frac{2 p^2 q^2}{(p\cdot q)^2}\ C_5 + C_8 \,, 
\eea \label{tenrel1} \ees
which multiply only the $u^{\alpha\beta}$ tensor. Next we have the three relations,
\bes\bea
&& F_2 + 4F_4 = \frac{C_{10}}{p^2} \,,\\
&& F_2 + 4F_6 = \frac{C_3}{q^2} \,,\\
&& F_2 + 2 F_8 =  -\frac{C_5}{p\cdot q}\,,
\eea \label{tenrel2} \ees
multiplying only the $w^{\alpha\beta}$ tensor. Finally, we have the four relations,
\bes\bea
&& F_9 = \frac{C_2}{p\cdot q} + \frac{q^2\,C_5}{(p\cdot q)^2} \,,\\
&& F_{10}= \frac{p^2\,C_5}{(p\cdot q)^2} + \frac{C_{11}}{p\cdot q}\,,\\
&& F_{11} = \frac{C_3}{2q^2}-\frac{C_{12}}{2p^2}\,,\\
&& F_{12} = \frac{C_{10}}{2p^2}-\frac{C_1}{2q^2} \,,
\eea \label{tenrel3}\ees
\noindent
multiplying tensors that do not appear in either $u^{\alpha\beta}$ or $w^{\alpha\beta}$.
The first three of the relations (\ref{tenrel1}) determine three linear combinations of the
four functions $F_1, F_3, F_5$ and $F_7$ in terms of the finite coefficient functions 
$C_i$, leaving only one of these four functions to be determined. Likewise the second 
three relations (\ref{tenrel2}) determine three linear combinations of the four functions, 
$F_2, F_4, F_6$ and $F_8$ in terms of the $C_i$, leaving only one of these four 
functions to be determined. Finally the last four relations (\ref{tenrel3}) determine the 
four functions $F_9, F_{10}, F_{11}$ and $F_{12}$ completely in terms of the $C_i$, 
and leave only $F_{13}$ to be determined.}

The information needed to fix the remaining three functions comes from our fifth and final
requirement on (\ref{Gamt}), namely the Ward identity (\ref{TWard}). The contraction, 
$k_{\mu}\Gamma^{\mu\nu\alpha\beta}(p,q)$ gives six independent three-tensors obeying 
vector current conservation, and therefore six conditions on the amplitude, 
$\Gamma^{\mu\nu\alpha\beta}(p,q)$,
\bes\bea
&& - p^2 F_3 + (3 q^2 + 4 p\cdot q) F_5 + (2 p^2 + p\cdot q) F_7 - p^2 q^2 F_{10}
- p^2 (p^2 + p\cdot q) F_9 + p^2 q^2 F_{11} = 0\,, \label{WIcons1a}\\
&& p^2 F_4 - (3 q^2 + 4 p\cdot q) F_6 - (2 p^2 + p\cdot q) F_8 - p\cdot q F_{10}
+ (q^2 + 2 p\cdot q) F_{11} = 0\,,\label{WIcons1b}\\
&& -p\cdot q \,(p^2 + p\cdot q) F_9 - q^2 (q^2 + p\cdot q) F_{11} + F_{13} + \Pi(p^2)  =0 \,,
\label{WIcons1c}\eea\label{WIcons1}\ees
and their Bose symmetry conjugates under interchange of $p$ and $q$, 
\bes\bea
&& (3 p^2 + 4 p\cdot q) F_3 - q^2 F_5 + (2 q^2 + p\cdot q) F_7 - p^2 q^2 F_9 
- q^2 (q^2 + p\cdot q) F_{10} + p^2 q^2 F_{12} = 0\,,\label{WIcons2a}\\
&& - (3 p^2 + 4 p\cdot q) F_4 + q^2 F_6 - (2 q^2 + p\cdot q) F_8 - p\cdot q F_9
+ (p^2 + 2 p\cdot q) F_{12}  = 0\,,\label{WIcons2b}\\
&& -p\cdot q \, (q^2 + p\cdot q) F_{10} - p^2 (p^2 + p\cdot q) F_{12} + F_{13} + \Pi(q^2) = 0\,.
\label{WIcons2c}\eea\label{WIcons2}\ees
It is evident that the symmetrized sum of (\ref{WIcons1a}) and (\ref{WIcons2a}) 
provides one new relation between $F_3, F_5$ and $F_7$, needed together with the first three
relations of (\ref{tenrel1}) to determine $F_1, F_3, F_5$ and $F_7$ completely
in terms of the $C_i$. In this way we find
\bes\bea
&& \hspace{-.5cm} F_1 = \frac{C_7 + C_8 + C_9}{3} + \frac{p^2}{3k^2}\,
(-C_1 + C_3 + C_8 - C_9) + \frac{q^2}{3k^2}\, (-C_7 + C_8 + C_{10} - C_{12})\,, \label{F1}\\
&& \hspace{-.5cm} F_3 = \frac{2C_7 -C_8 -C_9}{12} + \frac{p^2}{12k^2}\, 
(C_1 - C_3 -C_8 + C_9) + \frac{q^2}{12k^2}\, (C_7 - C_8 - C_{10} + C_{12})\,,\\
&& \hspace{-.5cm} F_5 = \frac{-C_7 -C_8 + 2 C_9}{12} +\frac{p^2}{12k^2}\,
(C_1 - C_3 - C_8 + C_9) + \frac{q^2}{12k^2}\, (C_7 - C_8 - C_{10} + C_{12})\,,\\
&& \hspace{-.5cm} F_7 = \frac{-C_7 + 2C_8 -C_9}{6} + \frac{p^2}{6k^2}\,
(C_1 - C_3 - C_8 + C_9) + \frac{q^2}{6k^2}\,(C_7 - C_8 - C_{10} + C_{12})\nonumber\\
&& \qquad\qquad\qquad + \frac{\ p^2\,q^2}{(p\cdot q)^2}\ C_5 + \frac{p^2 C_2 + q^2 C_{11}}
{2\,(p\cdot q)} \,.\eea \label{F1357}\ees
Likewise the symmetrized sum of (\ref{WIcons1b}) and (\ref{WIcons2b}) provides one new
relation between $F_4, F_6$ and $F_8$, needed together with the second three
relations of (\ref{tenrel2}) to determine $F_2, F_4, F_6$ and $F_8$ completely
in terms of the $C_i$. This gives
\bes\bea
&& F_2 = \frac{C_1}{3q^2} + \frac{C_{12}}{3p^2} +
\frac{-C_1 + 2C_2 - 2 C_5 + 2 C_{11} - C_{12}}{3k^2}\,, \label{F2}\\
&& F_4 = -\frac{C_1}{12q^2} + \frac{3 C_{10} - C_ {12}}{12p^2} 
+ \frac{C_1 - 2 C_2 + 2 C_5 - 2 C_{11} + C_{12}}{12k^2}\,,\\
&& F_6 = \frac{-C_1 + 3 C_3}{12q^2} - \frac{C_{12}}{12p^2} 
+ \frac{C_1 - 2 C_2 + 2 C_5 - 2 C_{11} + C_{12}}{12k^2}\,,\\
&& F_8 = -\frac{C_5}{2 p\cdot q}  - \frac{C_1}{6q^2} - \frac{C_{12}}{6p^2} 
+ \frac {C_1 - 2 C_2 + 2 C_5 - 2 C_{11} + C_{12}}{6k^2}\,.
\eea \label{F2468}\ees
With $F_9, F_{10}, F_{11}, F_{12}$ given previously by (\ref{tenrel3}) in terms of the finite
coefficient functions $C_i$, the final function $F_{13}$ is determined from the
symmetrized sum of (\ref{WIcons1c}) and (\ref{WIcons2c}) to be
\bea
&&F_{13} = -\frac{\Pi(p^2) + \Pi(q^2)}{2} + \frac{\ p^2\,q^2}{p\cdot q}\ C_5 
+ \frac{p^4 C_4 + q^4 C_6}{4 p\cdot q}
+\frac{p\cdot q}{4}\, (2C_2 + C_3 + C_{10} + 2 C_{11})\nonumber\\
&&\qquad+ \frac{p^2}{4}\,(2C_2 + C_4 + 2C_5 + C_{10}) 
+ \frac{q^2}{4}\, (C_3 + 2 C_5 + C_6+ 2 C_{11})\,.
\label{F13} 
\eea
In this way all the coefficient functions $F_i$ and hence the entire amplitude 
$\Gamma^{\mu\nu\alpha\beta}(p,q)$ is determined from its finite parts $C_i$, 
and the one-loop polarization tensor $\Pi$, by enforcing the conservation Ward identities 
on the amplitude. In particular the trace terms involving $F_1$ and $F_2$ are
determined unambiguously by this procedure, and we shall show in the next section that
$F_1$ contains the anomaly. Clearly if we had not enforced the conservation
Ward identity relations (\ref{WIcons1}) or (\ref{WIcons2}), the trace is not
determined, and could be required to satisfy the corresponding identities of
conformal invariance in the massless limit removing any trace anomaly,
at the price of violating the conservation identities (\ref{WIcons1}) and (\ref{WIcons2}).

Because (\ref{WIcons1}) and (\ref{WIcons2}) potentially overdetermine the coefficients
$F_i$, we note they give three additional conditions on the finite coefficients $C_i$
from their anti-symmetric parts,
\bes\bea
&&p^2 \left(C_1-2 C_2+C_3-2 C_7+2 C_8\right) - 2 p\cdot q \left(C_7-C_9 \right)
-q^2 \left(2 C_8-2 C_9+C_{10} - 2 C_{11}+C_{12}\right)=0\,,\qquad\qquad\\
&&\qquad\qquad -p^4\,C_1 + q^4\,C_{12} - 2\, p\cdot q \left( p^2\,C_1- q^2 \,C_{12} \right) 
+ p^2 \,q^2 \left(-2 C_2+C_3-C_{10}+2 C_{11}\right) = 0\,,\\
&&  p^2 (-2 C_2 + C_4 + 2 C_5 + C_{10})
+ p \cdot q (-2 C_2-C_3+C_{10}+2 C_ {11}) + q^2 (-C_3 - 2 C_5 -C_6 + 2 C_{11})\nonumber  \\
&& \qquad \qquad -\frac{ p^4}{q^2}\,C_1 + \frac{q^4}{p^2}\,C_{12}  = 2 \left[ \Pi (p^2) - \Pi (q^2) \right]\,,
\eea \label{addWI}\ees
which must be satisfied identically, for a consistent solution to exist.
The method for verifying that these three conditions are indeed satisfied by the $C_i$ 
is given in Appendix B.

We note also that the difference,
\be 
\Pi(p^2) - \Pi(q^2) = (q^2-p^2) \int_0^{\infty} \frac{ds}{q^2 + s} \frac{\rho_{\Pi}(s)}{p^2 + s}
\ee
in the last of the relations (\ref{addWI}) is finite, as required by the fact that all
the $C_i$ are finite. Here
\be
\rho_{\Pi}(s) \equiv \frac{1}{\pi}\, {\rm Im}\, \Pi\Big\vert_{p^2 = -s} = 
\frac{e^2}{12\pi^2} \left(1 + \frac{2m^2}{s}\right) \sqrt{1 - \frac{4m^2}{s}} 
\ \theta\left(s-4m^2\right)\,,
\ee
is the familiar spectral function of the one-loop photon polarization in QED,
which tends to a constant for $s \gg 4m^2$. Using the spectral representation of
$\Pi$, the renormalization of Eqs. (\ref{F13}) may be accomplished 
(for non-zero $m$) by defining
\be
\Pi_{_R}(p^2) \equiv \Pi(p^2) - \Pi(0)
= -p^2 \int_0^{\infty} \frac{ds}{s} \frac{\rho_{\Pi}(s)}{p^2 + s} =
-\frac{e^2}{2\pi^2} \int_0^1 dx \, x(1-x) \, \ln \left[ 1 + x(1-x) \frac{p^2}{m^2}\right]\,,
\label{PiR}
\ee
so that
\be
F_{13} = (F_{13})_{_R} + \Pi(0)\,,
\ee
with the logarithmically divergent $\Pi(0)$ removed by charge renormalization,
\be
\frac{1}{e_R^2} = \frac{1}{e^2} \big[1 + \Pi(0)\big]\,.
\label{erenorm}
\ee
Indeed the tensor multiplying the logarithmically divergent $\Pi(0)$ is
\be
t_{13}^{\mu\nu\alpha\beta}(p,q) = - \int d^4x\int d^4y\ e^{ip\cdot x + iq\cdot y} \ 
\frac{\delta^2 T^{\mu\nu}_{Max}(0)}{\delta A_{\alpha}(x) \delta A_{\beta}(y)} \,,
\ee
which must be added to $\Gamma^{\mu\nu\alpha\beta}(p,q)$ in full QED. Thus
unlike the chiral $\langle J_5^{\mu} J^{\alpha}J^{\beta}\rangle$ amplitude 
considered previously, one renormalization of the amplitude $\langle T^{\mu\nu} 
J^{\alpha}J^{\beta}\rangle$ involving the stress tensor is necessary.
However since charge renormalization enters only through the tensor $t_{13}$ 
which is proportional to the stress tensor of the classical electromagnetic
field, which is traceless, the trace of $\Gamma^{\mu\nu\alpha\beta}(p,q)$ 
which resides in the $t_1$ and $t_2$ tensors is finite, and unaffected by 
renormalization.

\section{The Trace Anomaly and Scaling Violation}

Having determined completely the amplitude $\Gamma^{\mu\nu\alpha\beta}(p,q)$
from its finite parts by the principles of Lorentz covariance, gauge
invariance, and general coordinate invariance, we come now to the relations 
that would be expected from the classical conformal
invariance of the theory in the massless limit, $m\rightarrow 0$. 

The $\lag TJJ\rag$ triangle diagram with the first 
vertex replaced by the naive classical trace (\ref{trclass}) is
\bea
&&\Lambda^{\alpha\beta} (p,q) \equiv - m (ie)^2 \int\,d^4x\,\int\,d^4y\, e^{ip\cdot x + i q\cdot y}\,
\langle \bar \psi \psi J^{\alpha}(x) J^{\beta}(y\rangle )\,,\nonumber\\
&& \quad = e^2 \,m\, (-1)\int \frac{i d^4 l}{(2\pi)^4}\ {\rm tr}\ \left\{
	\frac{-i}{\lsl + \psl  + m}\gamma^\alpha \frac{-i}{\lsl + m}
\gamma^\beta \frac{-i}{\lsl - \qsl  + m} +
	\frac{-i}{\lsl + \qsl + m}\gamma^\beta \frac{-i}{\lsl + m}
\gamma^\alpha \frac{-i}{\lsl - \psl + m} \right\}\,. \label{Lamampl}
\eea
This amplitude formally satisfies the conditions of vector current conservation,
\be
p_{\alpha}\Lambda^{\alpha\beta} (p,q) = 0 = q_{\beta}\Lambda^{\alpha\beta} (p,q)\,,
\label{Lamcons}
\ee
if one is free to shift the loop momentum integration variable $l$ in (\ref{Lamampl}).
Although the integral is superficially linearly divergent, it is in fact at worst
only logarithmically divergent because of the Dirac gamma matrix trace, and
one factor of $l$ is replaced by $m$. We may extract the factors of external momenta 
$p$ and $q$, which multiply quadratically convergent integrals in $d=4$ dimensions,
and determine the finite parts in the same manner as the full amplitude.
Then as before, we can determine (\ref{Lamampl}) by Lorentz covariance, and  
vector current consevation, (\ref{Lamcons}), in complete analogy to the case of the 
full amplitude (\ref{Tampl}) considered in the previous section. The evaluation
is given in Appendix A.

Since there are only $2$ two-index tensors composed of $p$, $q$ and the metric tensor,
which satisfy the conservation conditions (\ref{Lamcons}), namely the tensors
$u^{\alpha\beta}$ and $w^{\alpha\beta}$ defined by (\ref{uwdef}), we must have
\be
\Lambda^{\alpha\beta} (p,q) = G_1 u^{\alpha\beta}(p,q) + G_2 w^{\alpha\beta}(p,q)\,,
\label{G1G2}
\ee
with $G_1$ and $G_2$ scalar functions of the three invariants $k^2, p^2, q^2$, and $m^2$.
By identifying the coefficients of the finite terms proportional to the four tensors homogeneous of degree 
two, {\it viz.} $p^{\alpha}p^{\beta}$, $q^{\alpha}q^{\beta}$, $p^{\alpha}q^{\beta}$, and $q^{\alpha}p^{\beta}$, 
we obtain in Appendix A,
\bes\bea
&& G_1 = -\frac{e^2}{2\pi^2} \, m^2\, \int_0^1\,dx\int_0^{1-x}\,dy \, 
\frac{(1-4xy)}{D}\,, \label{G1}\\
&& G_2 = -\frac{e^2} {2\pi^2}\,\frac{m^2}{p \cdot q}\ \int_0^1\,dx\int_0^{1-x}\,dy \, 
\, \frac{(1-2x)(1-2y)}{D}\,, \label{G2}
\eea\label{G12}\ees
with $D$ given by (\ref{denom}). Both of these vanish in the massless limit, and would 
be the expected values of the trace of the full amplitude (\ref{Tampl}), absent any anomalies.

We now compare $\Lambda^{\alpha\beta}(p,q)$ of (\ref{G1G2}) and (\ref{G12}) to the exact trace
of the full amplitude (\ref{Gamt}) 
\be
g_{\mu\nu}\Gamma^{\mu\nu\alpha\beta}(p,q) = 3k^2 F_1 \, u^{\alpha\beta}(p,q)  
+ 3k^2 F_2 \, w^{\alpha\beta}(p,q)\,,
\label{exactrace}
\ee
computed in the previous section by requiring the Ward identities (\ref{WIcons1}) and
(\ref{WIcons2}) of stress tensor conservation. The functions $F_1$ and $F_2$, given
by (\ref{F1}) and (\ref{F2}) respectively are completely determined by that procedure.
Let us consider the second term in (\ref{exactrace}) first. Since the tensor $w^{\alpha\beta}$ 
corresponds to a local dimension six term in the effective action, {\it c.f.} (\ref{locvarw}),
we do not expect to contain any anomaly, {\it i.e.} we expect $3k^2F_2 =G_2$, which 
vanishes in the limit $m\rightarrow 0$. To see this explicitly requires the following
simple algebraic identity satisfied by the $C_i$ coefficients, {\it viz.},
\be
C_2 - C_4 - C_5 - C_6 + C_8 + C_{11} = 0\,,
\label{identF2}
\ee
as is easily verified by direct substitution of Table \ref{tensorcoeff} into (\ref{Cj}). 
Subtracting twice this identity from $3k^2F_2$ given by (\ref{F2}), and using
also eqs. (\ref{Crel}), we find
\be
3k^2 F_2 = -\frac{1}{p\cdot q}\, (p^2 C_4 + q^2 C_6 + 2 p\cdot q C_8)\,.
\ee
In this form one may subsitute the Feynman parameterization of the
coefficients (\ref{Cj}), using the Table \ref{tensorcoeff} to find
\bea
&&3k^2F_2 = \frac{e^2}{2\pi^2\,p \cdot q} \int_0^1\,dx\int_0^{1-x}\,dy \, 
(1-2x)(1-2y)\, \frac{(D-m^2)}{D} \, \nonumber\\
&& \qquad = -\frac{e^2} {2\pi^2}\,\frac{m^2}{p \cdot q}\ \int_0^1\,dx\int_0^{1-x}\,dy \, 
\, \frac{(1-2x)(1-2y)}{D}\,,
\label{trace2}
\eea
since the first integral independent of $m^2$ in which the denominator $D$ is cancelled, 
\be
\int_0^1\,dx\int_0^{1-x}\,dy \, (1-2x)(1-2y) = 0
\label{f2int}
\ee
in fact vanishes identically. Comparing (\ref{trace2}) with (\ref{G2}) we verify that indeed
\be
3k^2 F_2 = G_2 
\ee
is non-anomalous, and vanishes in the $m\rightarrow 0$ limit.

Turning next to the $F_1$ term in the full trace, (\ref{exactrace}),
we need the following identity satisfied by the $C_i$ coefficients,
\bea
&& p^2 (C_1 - 2 C_2 + C_3 + 2 C_4 - 4 C_{8})
+ q^2 (2 C_6 - 4 C_8 + C_{10} - 2 C_{11} + C_{12}) \nn
&& \qquad\qquad + 2 p\cdot q  (C_3 - 2 C_5 - C_7 - C_9 + C_{10} ) =0  \,.
\label{identF1}
\eea
which is verfied in Appendix B. Adding this quantity 
to $3k^2 F_1$ given by (\ref{F1}) yields
\bea
&& 3k^2 F_1 = p^2 (- 2 C_2 + 2 C_3 + 2 C_4 + C_7 - 2 C_8 )
+ q^2 ( 2 C_6 - 2 C_8 + C_9 + 2 C_{10} - 2 C_{11}) \nonumber\\
&& \qquad\qquad + 2 p\cdot q  (C_3 - 2 C_5 + C_8 + C_{10} ) \,,
\label{F1comb}
\eea
By substituting the Feynman parameterization integrals (\ref{Cj}) and
using again Table \ref{tensorcoeff}, we obtain in this case,
\bea
&&3k^2F_1 = \frac{e^2}{2\pi^2} \int_0^1\,dx\int_0^{1-x}\,dy \, 
(1-4xy)\, \frac{(D-m^2)}{D} \nonumber\\
&& \qquad = \frac{e^2}{6\pi^2} - \frac{e^2\ m^2}{2\pi^2}\int_0^1\,dx\int_0^{1-x}\,dy \, 
\frac{(1-4xy)}{D}\,,
\label{trace1}
\eea
since
\be
\int_0^1\,dx\int_0^{1-x}\,dy \, (1-4xy) = \frac{1}{3}\,,
\label{onethird}
\ee
which unlike (\ref{f2int}) does not vanish.

Since the first term on the right side of (\ref{trace1}) in which the
factors of $D$ cancel between numerator and denominator is both
non-vanishing and independent of $m$, the trace is anomalous, and 
we have
\be
3k^2 F_1 = \frac{e^2}{6\pi^2} + G_1\,.
\label{extraterm}
\ee
Hence the coefficient of the $t_1^{\mu\nu\alpha\beta}(p.q)$ tensor in the full amplitude 
may be written
\be 
F_1 (k^2; p^2, q^2) = \frac{e^2}{18\pi^2k^2}\left\{1 - 3 m^2 \int_0^1\,dx\int_0^{1-x}\,dy \, 
\frac{(1-4xy)}{D}\right\}\,,
\label{F1coeff}
\ee
giving rise to the non-zero trace, 
\be
g_{\mu\nu}\Gamma^{\mu\nu\alpha\beta}(p,q)\big\vert_{m=0} = \frac{e^2}{6\pi^2}
\,u^{\alpha\beta}(p,q)\,,
\label{Gamanomtr}
\ee
in the massless limit, which is exactly (\ref{tranom}) in momentum space.

It is clear that this non-vanishing trace is completely determined by
the finite terms in the amplitude together with the imposition of the Ward identities 
of stress tensor conservation, for if we had not made use of (\ref{WIcons1}), $F_1$ 
would be still undetermined, and could be chosen to vanish in the massless limit. 
Of course, with this choice of $F_1$ to satisfy the requirement of naive conformal 
invariance, the conservation identities of (\ref{WIcons1}) or (\ref{WIcons2})
would be violated, there would be an Einstein anomaly, and general coordinate 
invariance of the theory would be lost. This conflict between symmetries is quite 
analogous to the chiral case considered previously, where the naive Ward identity 
of $U_{ch}(1)$ invariance in the massless limit could be maintained by adding an extra term
to the amplitide, at the expense of violating the $U(1)$ conservation
identities (\ref{chicons}).

In the case of the anomalous non-zero trace of the energy-momenum tensor, it is
the conformal invariance of the classical theory with massless fermions that 
cannot be maintained at the quantum level. In $d=4$ flat spacetime the
conformal group is $O(4,2)$ and its $15$ generators consist of global
dilations and the $4$ special conformal transformations together with the
$10$ generators of the Poincar{\'e} group. The Noether dilation current,
\be
D^{\mu} = x^{\nu} T^{\mu}_{\ \nu}
\ee
is divergenceless and the corresponding charge $\int d^3\vec x\, D^{0}$ is conserved if and only if 
$T^{\mu\nu}$ is traceless. A non-zero trace implies instead non-trivial scale dependence\,\cite{ChanEll,AdlColDun}.

To see the effect of global scale transformations implied by the trace anomaly, we consider the 
trace (\ref{exactrace}) at $k = p +q= 0$. Then, since
\be
w^{\alpha\beta}(p,-p) = -p^2u^{\alpha\beta}(p,-p) = p^2(p^2 g^{\alpha\beta} - p^{\alpha}p^{\beta})\,,
\ee
and upon using (\ref{trace2}) and (\ref{trace1}), Eq. (\ref{exactrace}) for the full trace gives
\bea
&&g_{\mu\nu}\Gamma^{\mu\nu\alpha\beta}(p,-p) =  \left[- \frac{e^2}{6\pi^2} + \frac{e^2\ m^2}{2\pi^2}\int_0^1\,dx\int_0^{1-x}\,dy \, 
\frac{(1-4xy) + (1-2x)(1-2y)}{p^2(x+y)(1-x-y) + m^2}\right] (p^2 g^{\alpha\beta} - p^{\alpha}p^{\beta})   \nn
&& \qquad =  \left[- \frac{e^2}{6\pi^2} + \frac{e^2\ m^2}{\pi^2}\int_0^1\,du\ \frac{u(1-u)}{p^2 u(1-u) + m^2}\right] 
(p^2 g^{\alpha\beta} - p^{\alpha}p^{\beta})\,,
\label{tracek0}
\eea
after changing variables to $u= x+y$ and $v=(x-y)/2$, and integrating over $v$. At the same time we note
from (\ref{PiR}) that
\be
2p^2 \frac {d\Pi_{_R}}{dp^2} = - \frac{e^2}{6\pi^2} + \frac{e^2\ m^2}{\pi^2}\int_0^1\,du\ \frac{x(1-x)}{p^2 x(1-x) + m^2} 
= - \frac{\beta(e^2;p^2)}{e^2}
\label{betaPi}
\ee
is related to the $\beta$ function of the electromagnetic coupling $e^2$. Comparing with (\ref{tracek0}), we secure\,\cite{ChanEll}
\be
g_{\mu\nu}\Gamma^{\mu\nu\alpha\beta}(p,-p) = 2p^2 \frac {d\Pi_{_R}}{dp^2}(p^2 g^{\alpha\beta} - p^{\alpha}p^{\beta})
=  -  \frac{\beta(e^2;p^2)}{e^2} (p^2 g^{\alpha\beta} - p^{\alpha}p^{\beta})\,,
\label{tracek01}
\ee
a result that remains valid to all orders in perturbation theory\,\cite{AdlColDun}.  Eq. (\ref{tracek01}) may also be derived by
differentiating the Ward identity (\ref{TWard}) with respect to $q^{\mu}$ (or $p^{\mu}$), and then setting $q=-p$.
Hence the breaking of scale invariance by the trace of the energy-momentum tensor, together with its
conservation, may be regarded as responsible for the $\beta$ function running of the coupling, without any
direct reference to ultraviolet renormalization.

We note also that by combining the two terms in (\ref{betaPi}), the one-loop $\beta$ function here,
\be
\beta(e^2; p^2) = \frac{e^4}{\pi^2}\,p^2 \int_0^1\,dx\ \frac{x^2(1-x)^2}{p^2 x(1-x) + m^2} \rightarrow \frac{e^4}{30\pi^2}\,\frac{p^2}{m^2}\,,
\label{IRbeta}
\ee
vanishes as $p^2 \rightarrow 0$. This is the correct behavior for the {\it infrared} running with $p^2$ of the physical renormalized
coupling at momenta small compared to the electron mass, where vacuum polarization is negligible and decoupling of
the electron loop must hold\,\cite{Mano}. For $p^2 \gg m^2$, $\beta(e^2;p^2) \rightarrow e^4/6\pi^2$, which is then identical to the usual 
ultraviolet $\beta$ function, calculated {\it e.g.} in dimensional regularization where the infrared mass plays no role. We see
then the physical necessity of the trace anomaly in a different way, for if the constant first term in (\ref{betaPi}) determined 
by the trace anomaly at $m=0$ were not present, there would be nothing to cancel the second integral as $p^2 \rightarrow 0$, 
and decoupling of heavy degrees of freedom at large distances ($p^2 \ll m^2$) in (\ref{IRbeta}) would not occur.

\section{Spectral Representation, Sum Rule and the Massless Scalar Pole}

The physical meaning of the anomaly is further exposed by considering the spectral representation of
the amplitude, cut across two of its legs as in Fig. \ref{Fig:ImTri}. Following the pattern of the chiral case considered previously 
in (\ref{chispec}), the spectral representations for the amplitudes $C_i$ may be introduced by using the definition 
(\ref{BigS}) and the identity (\ref{sspec}), to obtain
\be 
C_j(k^2;p^2,q^2) = \int_0^{\infty} \, ds\, \frac{\rho_j(s;p^2,q^2)}{k^2 +s} 
\ee
with
\be
\rho_j(s;p^2,q^2) = \frac{e^2}{4\pi^2}\int_0^1\,dx\int_0^{1-x}\,dy \,\frac{c_j(x,y)}{xy}\ 
\delta\left(s - S(x,y;p^2,q^2)\right) \,.
\label{specrep}
\ee
The $\rho_i$ defined in this way are not necessarily positive, nor are they independent, owing to the relations 
(\ref{addWI}), (\ref{identF2}), and (\ref{identF1}). Indeed the $\rho_j$ satisfy exactly the same identities
as the corresponding $C_j$, of which they are just the discontinuity or imaginary part as $k^2$ is analytically 
continued to $-s$ with $s > 0$. Of interest to us however is only the linear combination which appears in the trace
(\ref{F1comb}). By using (\ref{identF1}), and repeating the steps that led from (\ref{F1}) to (\ref{trace1}), we obtain 
with the help of (\ref{BigS}) and table \ref{tensorcoeff},
\bea
&&k^2\,(\rho_7 + \rho_8 + \rho_9) + p^2\,(-\rho_1 + \rho_3 + \rho_8 - \rho_9) 
+ q^2\, (-\rho_7 + \rho_8 + \rho_{10} - \rho_{12})\nn
&& \quad = p^2 (- 2 \rho_2 + 2 \rho_3 + 2 \rho_4 + \rho_7 - 2 \rho_8 )
+ q^2 ( 2 \rho_6 - 2 \rho_8 + \rho_9 + 2 \rho_{10} - 2 \rho_{11})
+ 2 p\cdot q  (\rho_3 - 2 \rho_5 + \rho_8 + \rho_{10} )\nn
&&\qquad = \frac{e^2}{4\pi^2}\int_0^1\,dx\int_0^{1-x}\,dy \,\frac{\delta (s - S)}{xy}
\left\{p^2 (- 2 c_2 + 2 c_3 + 2 c_4 + c_7 - 2 c_8 )\right.\nn
&& \qquad\qquad\qquad + \left. q^2 ( 2 c_6 - 2 c_8 + c_9 + 2 c_{10} - 2 c_{11})
+ 2 p\cdot q  (c_3 - 2 c_5 + c_8 + c_{10} )\right\}\nn
&& \qquad = \frac{e^2}{2\pi^2}\int_0^1\,dx\int_0^{1-x}\,dy \,\frac{\delta (s - S)}{xy}
\,(1-4xy)\, (D- m^2)\nn
&& \qquad = \frac{e^2}{2\pi^2}\int_0^1\,dx\int_0^{1-x}\,dy \,\delta (s - S )
(1-4xy) \left[ k^2 + S - \frac{m^2}{xy}\right]\nn
&& \qquad = (k^2 + s) \rho_{_T} - m^2 \rho_m\,,
\label{rhorel}
\eea
where
\bes\bea
&&\rho_{_T} (s;p^2,q^2)\equiv \rho_3 - 2\rho_5 + \rho_8 + \rho_{10}\nn
&& \qquad = \frac{e^2}{2\pi^2} \int_0^1\,dx\int_0^{1-x}\,dy \ (1-4xy)
\ \delta \left(s - S(x,y;p^2,q^2)\right)\,, \qquad {\rm and}\label{rhoTdef}\\
&&\rho_m (s;p^2,q^2) \equiv \frac{e^2}{2\pi^2} \int_0^1\,dx\int_0^{1-x}\,dy \, \frac{(1-4xy)}{xy}
\ \delta\left(s - S(x,y;p^2,q^2)\right) \,,
\eea\label{rhoAm}\ees
since $c_3 - 2c_5 + c_8 + c_{10} = 2xy (1-4xy)$. 

These relations may be compared to their somewhat simpler analogs, (\ref{chirel}) and (\ref{rhoA0}),
in the chiral case. Since $1 - 4xy \ge 0$ over the indicated range of $x,y$, both
$\rho_{_T}$ and $\rho_m$ are non-negative functions of $s$ for spacelike $p^2$
and $q^2$. Notice that at $k^2 = -s$ the quantity $\rho_{_T}$ drops out of
(\ref{rhorel}), so that the discontinuity or imaginary part of (\ref{trace1}) 
vanishes in the conformal limit $m\rightarrow 0$, and is non-anomalous.

As in the chiral case (\ref{chisum}), we find that spectral function which determines the anomaly
satisfies a sum rule\,\cite{HorSch},
\be
\int_0^{\infty} \, ds \, \rho_{_T}(s; p^2,q^2) 
= \frac{e^2}{2\pi^2} \int_0^1\,dx\int_0^{1-x}\,dy \, (1-4xy)
\ \int_0^{\infty} \, ds \ \delta (s - S)
= \frac{e^2}{6\pi^2}\,,
\label{sumrule}
\ee
by (\ref{onethird}), which is independent of $p^2 \ge 0, q^2 \ge 0$ and $m^2\ge 0$,
since then $S(x,y; p^2, q^2) \ge 0$ and the $\delta$ function can be satisfied
over the range of $s\ge 0$. On the other hand, using $2p\cdot q= k^2-p^2-q^2$,
and rearranging the second and last lines of (\ref{rhorel}) gives
\bea
&&\rho_{_T} (s; p^2, q^2) = \frac{p^2}{s} (-2 \rho_2 + \rho_3 + 
2\rho_4 + 2\rho_5 + \rho_7 - 3\rho_8 - \rho_{10})\nn
&&\qquad\qquad + \frac{q^2}{s} (-\rho_3 + 2\rho_5 + 2 \rho_6 - 3\rho_8 + \rho_9 + \rho_{10} - 2 \rho_{11})
+ \frac{m^2}{s} \rho_m\,.
\label{pointwise}
\eea
Since the $\rho_i$ develop at worst logarithmic singularities in the combined
limit $p^2, q^2, m^2 \rightarrow 0^+$ (taken in any order), (\ref{pointwise})
shows that $\rho_{_T}$ vanishes pointwise for all $s >0$ in this limit. The only
way that this can be consistent with the sum rule (\ref{sumrule}) is if $\rho_{_T}$
develops a $\delta$ function singularity at $s=0$ in this limit. Indeed
since the function $S(x,y;p^2,q^2)$ defined by (\ref{BigS}) vanishes identically in 
this limit, we see directly from (\ref{rhoTdef}) that
\be
\lim_{p^2,q^2,m^2 \rightarrow 0^+}\rho_{_T} (s;p^2,q^2) = 
\frac{e^2}{2\pi^2} \int_0^1\,dx\int_0^{1-x}\,dy \, (1-4xy)\ \delta (s) =
\frac{e^2}{6\pi^2}\ \delta(s)\,,
\label{specdel}
\ee 
by taking the limits inside the integral. Thus $\rho_{_T}$
exhibits a massless scalar intermediate state in the two-particle cut amplitude.
\footnote{This observation was made in ref.\,\cite{Dolgov} in the context of
photon pair creation by a cosmological gravitational field.}

\begin{figure}[htp]
\includegraphics*[width=12cm, height=7cm, viewport=70 60 410 300,clip]{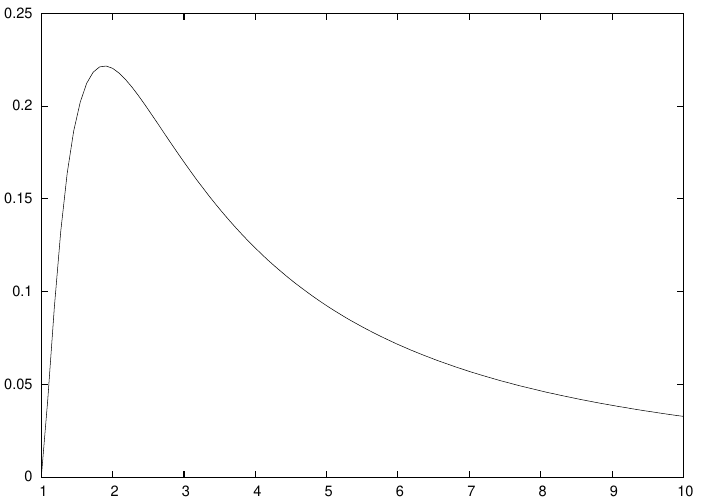}
\caption{The spectral function $\rho_{_T}$ of (\ref{rhoT00}) in units of $\frac{e^2}{32\pi^2}\, m^{-2}$ 
as a function of $\frac{s}{4m^2}$.}
\label{Fig:rhoT}
\end{figure}

It is instructive to retain the non-zero fermion mass $m>0$ as an infrared regulator,
in order to study this intermediate state in more detail. Comparing with (\ref{rhoA00}) from 
the axial anomaly, we find when $p^2=q^2 =0$ that
\bea
&&\rho_{_T}(s; 0,0) =  \frac{m^2}{s} \rho_m(s;0,0) = \frac{1}{2} \left(1 - \frac{4m^2}{s}\right)\rho_{_{\cal A}}(s;0,0)\nn
&&\qquad = \frac{e^2}{2\pi^2}\, \frac{m^2}{s^2} \left(1 - \frac{4m^2}{s}\right)
\,\ln\left\{\frac{1 + \sqrt{1 - \frac{4m^2}{s}}}{1 - \sqrt{1 - \frac{4m^2}{s}}}\right\}\,\theta (s-4m^2) \,.
\label{rhoT00}\eea
This function is plotted in Fig. \ref{Fig:rhoT}.

\noindent
The corresponding amplitude is
\bea
&&F_1(k^2;0,0) = \frac{1}{3k^2}\int_0^{\infty} \, \frac{ds}{k^2 +s} 
\left[(k^2 + s) \rho_{_T} - m^2 \rho_m\right]\nn
&&\qquad = \frac{1}{3k^2}\left[\frac{e^2}{6\pi^2} - m^2\int_0^{\infty} \, ds\, 
\frac{\rho_m(s; 0, 0)}{k^2 +s}\right]
\label{F1pole}
\eea
which exhibits a pole at $k^2 = 0$ when the fermion mass $m=0$, {\it i.e}.
\be
\lim_{p^2,q^2,m^2 \rightarrow 0^+} F_1(k^2;p^2,q^2) = \frac{e^2}{18\pi^2} \frac{1}{k^2} \,.
\label{F1lim}
\ee
When the fermion mass is non-zero the amplitude (\ref{F1pole}) can also be written in the form,
\be
F_1(k^2;0,0)= \frac{m^2}{3} \int_{4m^2}^{\infty} \frac{ds}{s}\ \frac{\rho_m(s; 0, 0)}{k^2 +s}
\label{F1nopole}
\ee
which shows that there is no pole divergence as $k^2 \rightarrow 0$ with $m^2 > 0$ fixed.
This is again the phenomenon of decoupling, as in the cancellation noted previously in (\ref{betaPi}) and
({\ref{IRbeta}), with the place of $p^2 \ll m^2$ being taken here by $k^2 \ll m^2$. Conversely, if $m=0$ the
amplitude (\ref{F1pole}) behaves like $k^{-2}$ for all $k^2$, in both the infrared and ultraviolet, as expected 
in the classical conformal limit of a theory with no intrinsic mass or momentum scale. 

It is also instructive to carry out the same steps for the imaginary part of the trace which is non-anomalous,
{\it i.e.} for the spectral function corresponding to the non-anomalous amplitude $F_2$. Repeating the steps
which led from (\ref{F2}) to (\ref{trace2}), using the identities corresponding to (\ref{Crel}) and (\ref{identF2}) for
their imaginary parts, we obtain
\bea
&&\frac{k^2}{q^2}\,\rho_1 + \frac{k^2}{p^2}\,\rho_{12} 
-\rho_1 + 2\rho_2 - 2 \rho_5 + 2 \rho_{11} - \rho_{12}
 = -\frac{1}{p\cdot q}\, (p^2 \rho_4 + q^2 \rho_6 + 2 p\cdot q \rho_8)\nn
&& \qquad = \frac{e^2}{2\pi^2}\frac{1}{p \cdot q} \int_0^1\,dx\int_0^{1-x}\,dy \,\frac{\delta (s - S)}{xy}
\,(1-2x)(1-2y)\, (D- m^2)\nn
&& \qquad = \frac{e^2}{2\pi^2}\frac{1}{p\cdot q}\int_0^1\,dx\int_0^{1-x}\,dy \,\delta (s - S )
(1-2x)(1-2y) \left[ k^2 + s - \frac{m^2}{xy}\right]\,.
\label{rho2rel}
\eea
In this case the spectral function corresponding to $\rho_{_T}$  in (\ref{rhoTdef}) is
\be
-\frac {\rho_8 (s; p^2,q^2)}{p \cdot q} = \frac{e^2}{2\pi^2}\frac{1}{p\cdot q}\int_0^1\,dx\int_0^{1-x}\,dy \,(1-2x)(1-2y) \,
\delta \big(s - S(x,y;p^2,q^2) \big)\,,
\ee
but unlike (\ref{sumrule}) $\rho_8$ obeys the {\it vanishing} sum rule,
\be
\int_0^{\infty}ds\,\rho_8 (s; p^2,q^2) = -\frac{e^2}{2\pi^2}\int_0^1\,dx\int_0^{1-x}\,dy \,(1-2x)(1-2y) = 0\,.
\ee
Hence, although $\rho_8$ can be expressed in the form,
\be
\rho_8 (s; p^2,q^2) = \frac{p^2}{s} (\rho_4 - \rho_8) + \frac{q^2}{s} (\rho_6 - \rho_8) - \frac{m^2}{s}
\frac{e^2}{2\pi^2}  \int_0^1\,dx\int_0^{1-x}\,dy \,\frac{(1-2x)(1-2y)}{xy}\,\delta (s - S)
\ee
analogous to (\ref{pointwise}), which vanishes pointwise in the combined limit, $p^2, q^2, m^2 \rightarrow 0^+$,
it has no positivity property, and no reason to develop a $\delta$ function singularity at $s=0$ in that limit. Indeed it is not
difficult to see that $\rho_8$ and indeed the corresponding full amplitude $C_8$ vanishes identically in this limit,
{\it c.f.} eq. (\ref{C8m0}) below, consistent with the vanishing of $F_2$, which unlike $F_1$ has no pole as 
$p^2, q^2, m^2 \rightarrow 0^+$\,\cite{Iwas}.

We may use the general amplitude $\Gamma^{\mu\nu\alpha\beta}(p,q)$ to evaluate the matrix element of $T^{\mu\nu}$ 
to physical photons on shell, $p^2=q^2=0$ which are also transverse. In this case all terms with $p^{\alpha}$ and $q^{\beta}$ 
vanish when contracted with the transverse photon polarization states, and the matrix element simplifies considerably. 
The tensor $w^{\alpha\beta}(p,q)$ and hence the tensors $t_2, t_4, t_6$ and $t_8$ vanish upon contraction with
transverse photons, as do the tensors $t_9, t_{10}, t_{11}$ and $t_{12}$. The remaining relevant form factors also simplify 
considerably when $p^2 = q^2 =0$, becoming
\bes\bea
&&F_1(k^2; 0,0) = \frac{2C_7 + C_8}{3}\Big\vert_{p^2=q^2=0}\, \\
&&F_3(k^2; 0,0) = F_5(k^2; 0,0) = \frac{C_7 - C_8}{12}\Big\vert_{p^2=q^2=0}\, \\
&&F_7(k^2; 0,0) = -4 F_3(k^2; 0,0) = \frac{-C_7 + C_8}{3}\Big\vert_{p^2=q^2=0}\, \\
&&F_{13, R}(k^2; 0,0) = \frac{k^2}{4}\, (2C_2 + C_3)\big\vert_{p^2=q^2=0}\,,
\eea\label{Fphysphoton}\ees
for any $m$. The scalar coefficients $C_j$ here are evaluated on the photon mass shell, given by
\be
C_j (k^2; 0, 0) = \frac{e^2}{4\pi^2}
\int_0^1 dx\int_0^{1-x} dy \ \frac{c_j(x,y)}{k^2\,xy + m^2}\,,
\label{Cj00}
\ee
in which case $C_2 = C_{11}$, $C_3 = C_{10}$, and $C_7 = C_9$. Thus, for on-shell photons there are 
only three independent form factors, and we can write the matrix element to physical photons with transverse
field amplitudes $\tilde A_{\alpha} (p)$, $\tilde A_{\beta} (q)$ in the form,
\bea
&&\hspace{-1cm}\lag 0 \vert T^{\mu\nu}(0) \vert p,q\rag = F_1 \left(k^2 g^{\mu\nu} - k^{\mu } k^{\nu}\right)
u^{\alpha\beta}(p,q) \tilde A_{\alpha}(p) \tilde A_{\beta}(q)\nn
&&\quad - 2 F_3 \left[ k^2\, g^{\mu\nu} - 4 (p^{\mu} q^{\nu} + q^{\mu} p^{\nu})
+ 2(p^{\mu}  p^{\nu} + q^{\mu} q^{\nu})\right] u^{\alpha\beta}(p,q) \tilde A_{\alpha}(p) \tilde A_{\beta}(q)\nn
&&\qquad\qquad\qquad + \ F_{13\,R}\ t_{13}^{\mu\nu\alpha\beta}(p,q) \tilde A_{\alpha}(p) \tilde A_{\beta}(q)\,, 
\label{photonmatr}
\eea
with $F_1$, $F_3$ and $F_{13\,R}$ evaluated at $p^2=q^2=0$ given by (\ref{Fphysphoton}), (\ref{Cj00}) 
and Table \ref{tensorcoeff}. The survival of only $3$ independent tensors when both photons are on their mass shell
and have physical transverse polarizations agrees with the literature\,\cite{DrumHath,Iwas}. Each of the three 
tensors remaining in (\ref{photonmatr}) is conserved and their contractions with $k_{\nu}$ vanish for photons 
on shell. Only the first has non-zero trace.  

Taking the $m=0$ limit gives the further simplification that 
\bes\bea
&& C_7 (k^2) \vert_{p^2=q^2=m^2 = 0} = \frac{e^2}{2\pi^2\,k^2}
\int_0^1 dx\int_0^{1-x} dy \ (1-2x)^2 = \frac{e^2}{12\pi^2}\,\frac{1}{k^2}\,\\
&& C_8 (k^2) \vert_{p^2=q^2=m^2 = 0} = 0\,,
\label{C8m0}\eea\ees
while $F_{13\,R}$ contains a $\ln (k^2/m^2)$ behavior in this limit, but no pole (reflecting the need to
renormalize the charge at a mass scale $\mu^2 > 0$ different from $m^2$ in the massless limit).
Thus from (\ref{Fphysphoton}) both the $F_1$ and $F_3$ form factors of the scattering amplitude to physical
on shell photons exhibits a massless scalar pole in the limit of vanishing electron mass,
with $F_1$ given by (\ref{F1lim}) and
\be
\lim_{m \rightarrow 0} 2F_3(k^2; 0, 0) = \frac{e^2}{72\pi^2} \frac{1}{k^2} \,.
\label{F3lim}
\ee
The leading order behavior as $k^2 \rightarrow 0$
of the sum of terms in the amplitude (\ref{photonmatr}) to physical on shell photons is
\be
\lim_{k^2 \rightarrow 0} \lim_{m\rightarrow 0}\lag 0 \vert T^{\mu\nu}(0) \vert p,q\rag \rightarrow
-\frac{e^2}{12\pi^2 k^2}\,(p^{\mu}  p^{\nu} + q^{\mu} q^{\nu})\, u^{\alpha\beta}(p,q) \tilde A_{\alpha}(p) \tilde A_{\beta}(q)
+ \log (k^2) \ {\rm \& \ finite\  terms}\,,
\label{kto0}
\ee
when the limit of vanishing electron mass is taken first. This shows that the singular massless pole behavior survives in
the matrix element of the stress tensor to physical transverse photons (in its tracefree terms), while the 
trace remains finite in the conformal limit of vanishing electron mass and all $3$ 
four-momenta $(k,p,q)$ becoming lightlike.

The kinematics of the state appearing in the imaginary part and spectral function (\ref{specdel})
in this limit is essentially $1+1$ dimensional, and can be represented as the two-particle collinear $e^+e^-$ pair
in Fig. \ref{Fig:epair}. This is the only configuration possible for one particle with four-momentum $k^{\mu}$
converting to two particles of zero mass, $p^2= q^2 = 0$ as $k^2 \rightarrow 0$ as well. A detailed examination 
of the imaginary part of the amplitude, illustrated by the analog of Fig. \ref{Fig:ImTri} shows that there is a cancellation 
between the numerator and Feynman propagator in the denominator of the amplitude from the uncut fermion line 
in the triangle. Thus all particles in the real propagating intermediate state depicted in Fig. \ref{Fig:epair} are 
massless, on shell, and collinear. Although this special collinear kinematics is a set of vanishing measure 
in the two particle phase space, the $\delta (s)$ in the spectral function (\ref{specdel}) and finiteness of the anomaly 
itself shows that this pair state couples to on shell photons on the one hand, and gravitational 
metric perturbations on the other hand, with finite amplitude. When gravitational scattering is considered in
Sec. 8 the four-momentum transfer $k^{\mu}$ may be timelike or spacelike, the pole terms (\ref{F1pole}) and
(\ref{F3lim}) in the real part of the amplitude become relevant, and neither fermion pair nor final state
photons are collinear.

\begin{figure}[htp]
\includegraphics[width=30cm, viewport=20 640 1000 690,clip]{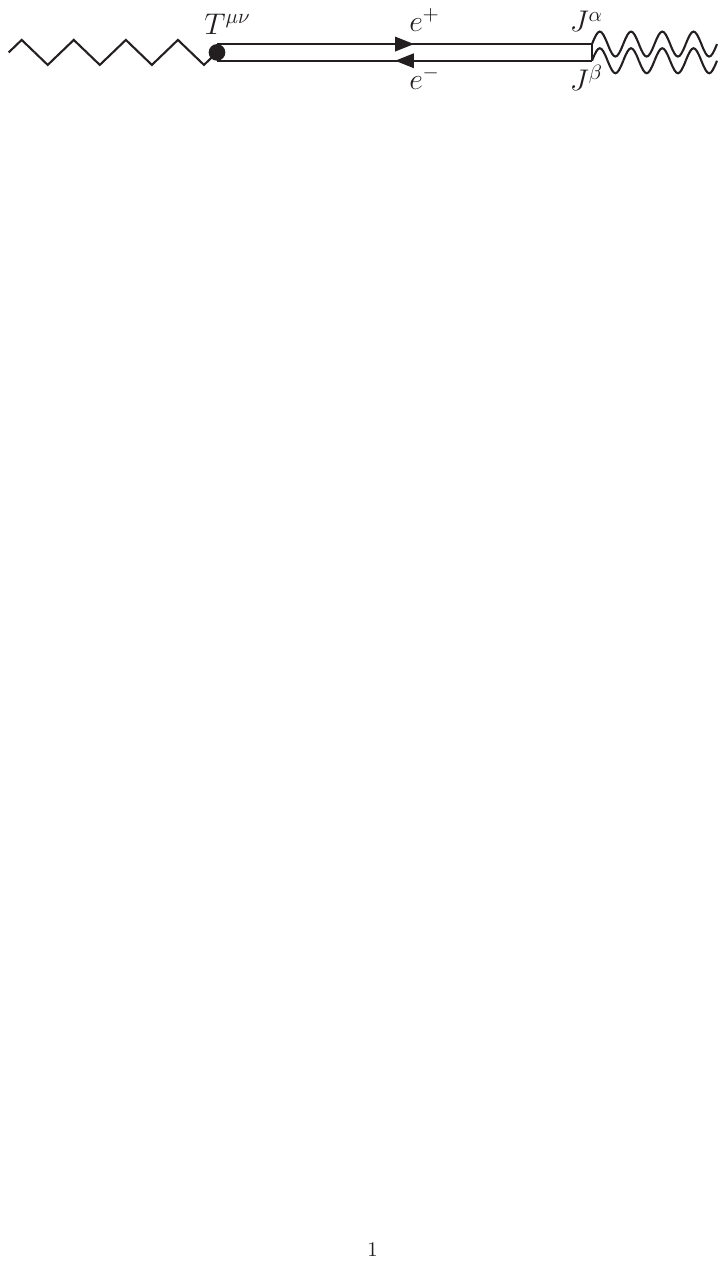}
\caption{The two particle intermediate state of a collinear $e^+e^-$ pair responsible for the $\delta$-fn. in 
(\ref{specdel}).}
\label{Fig:epair}
\end{figure}

\section{Anomaly Effective Action and Massless Scalar Fields}

Having demonstrated the existence of a real massless spin-$0$ intermediate state in the imaginary part of
the triangle amplitude, and a corresponding massless pole in the full amplitude, we turn in this section to
the effective action and scalar fields which describe these massless scalar degrees of freedom. In fact,
a covariant action for the trace anomaly in a general curved space has been given already in several earlier works
\cite{MotVau,Rie,MazMot,AntMazMot}. This general effective action can be presented in the non-local form,
\bea
&& \hspace{-.6cm}S_{anom}[g,A] = \label{Tnonl}\\
&&\frac {1}{8}\int d^4x\sqrt{-g}\int d^4x'\sqrt{-g'} \left(E - \frac{2}{3} \sq R\right)_x
 \Delta_4^{-1} (x,x')\left[ 2b\,C^2 
 + b' \left(E - \frac{2}{3} \sq R\right) + 2c\, F_{\mu\nu}F^{\mu\nu}\right]_{x'} \nonumber
\eea
where the $b$ and $b'$ parameters are the coefficients of the Weyl tensor squared,
$C^2 = C_{\lambda\mu\nu\rho}C^{\lambda\mu\nu\rho} = R_{\lambda\mu\nu\rho}R^{\lambda\mu\nu\rho} 
-2 R_{\mu\nu}R^{\mu\nu}  + \frac{R^2}{3}$ and the Euler density 
$E = ^*\hskip-.2cmR_{\lambda\mu\nu\rho}\,^*\hskip-.1cm R^{\lambda\mu\nu\rho} = 
R_{\lambda\mu\nu\rho}R^{\lambda\mu\nu\rho} - 4R_{\mu\nu}R^{\mu\nu}+ R^2$ respectively
of the trace anomaly in a general background curved spacetime, and the last term in (\ref{Tnonl}) 
takes into account the anomaly in a background gauge field with coefficient $c$.
For the present case of Dirac fermions in a classical gravitational ($g_{\mu\nu}$) and
classical electromagnetic ($A_{\alpha}$) background, $b = 1/320\pi^2$, and $b' = - 11/5760\pi^2$,
and $c= -e^2/24\pi^2$. The notation $\Delta_4^{-1}(x,x')$ denotes the Green's function inverse of the 
conformally covariant fourth order differential operator defined by 
\be
\Delta_4 \equiv  \nabla_\mu\left(\nabla^\mu\nabla^\nu + 2 R^{\mu\nu} - \frac{2}{3} R g^{\mu\nu}\right)
\nabla_\nu = \sq^2 + 2 R^{\mu\nu}\nabla_\mu\nabla_\nu +\frac{1}{3} (\nabla^\mu R)
\nabla_\mu - \frac{2}{3} R \sq\,.
\label{Deldef}
\ee
By varying (\ref{Tnonl}) multiple times with respect to the background metric $g_{\mu\nu}$ and/or
the background gauge fields $A_{\alpha}$ one can derive formulae for the trace anomaly related parts 
of amplitudes involving multiple insertions of the stress  tensor $\lag TTT...JJ\rag$ and $\lag TTT...\rag$ 
in curved or flat space. We emphasize that the effective 
action (\ref{Tnonl}) was obtained by integrating the anomaly, and is determined up to terms which are 
conformally invariant. Therefore one can expect it to yield correct results for the trace related parts of 
amplitudes such as (\ref{Gamone}), while the tracefree parts are not given uniquely by (\ref{Tnonl}).

As detailed in ref.\,\cite{MotVau} we may render the non-local anomaly action (\ref{Tnonl}) into a local form, 
by the introduction of two scalar auxiliary fields $\varphi$ and $\psi$ which satisfy fourth order differential eqs.,
\bes\bea
&& \Delta_4\, \varphi = \frac{1}{2} \left(E - \frac{2}{3} \sq R\right)\,,
\label{auxvarphi}\\
&& \Delta_4\, \psi = \frac{1}{2}C_{\lambda\mu\nu\rho}C^{\lambda\mu\nu\rho} 
+ \frac{c}{2b} F_{\mu\nu}F^{\mu\nu} \,,
\label{auxvarpsi}
\eea
\label{auxeom}
\ees
\hspace{-.35cm}
where we have added the last term in (\ref{auxvarpsi}) to take account of the background gauge field. This
local effective action corresponding to (\ref{Tnonl}) in a general curved space is given by
\be
S_{anom} = b' S^{(E)}_{anom} + b S^{(F)}_{anom} + \frac{c}{2} \int\,d^4x\,\sqrt{-g}\ 
F_{\mu\nu}F^{\mu\nu} \varphi\,,
\label{allanom}
\ee
where
\bea
&& S^{(E)}_{anom} \equiv \frac{1}{2} \int\,d^4x\,\sqrt{-g}\ \left\{
-\left(\sq \varphi\right)^2 + 2\left(R^{\mu\nu} - \frac{R}{3} g^{\mu\nu}\right)(\nabla_{\mu} \varphi)
(\nabla_{\nu} \varphi) + \left(E - \frac{2}{3} \sq R\right) \varphi\right\}\,;\nonumber\\
&& S^{(F)}_{anom} \equiv \,\int\,d^4x\,\sqrt{-g}\ \left\{ -\left(\sq \varphi\right)
\left(\sq \psi\right) + 2\left(R^{\mu\nu} - \frac{R}{3}g^{\mu\nu}\right)(\nabla_{\mu} \varphi)
(\nabla_{\nu} \psi)\right.\nonumber\\
&& \qquad\qquad\qquad + \left.\frac{1}{2} C_{\lambda\mu\nu\rho}C^{\lambda\mu\nu\rho}\varphi +
\frac{1}{2} \left(E - \frac{2}{3} \sq R\right) \psi \right\}\,.
\label{SEF}
\eea
\vspace{-.4cm}

\noindent
The free variation of the local action (\ref{allanom})-(\ref{SEF}) with respect to $\psi$ and $\varphi$ yields
the eqs. of motion (\ref{auxeom}).
Each of these terms when varied with respect to the background metric gives a stress-energy tensor
in terms of the auxiliary fields satisfying eqs. (\ref{auxeom}). Since we are interested here in only the
first variation of the action with respect to $g_{\mu\nu}$, we may drop all terms in (\ref{allanom}) which are
second order or higher in the metric deviations from flat space. Also, if we solve (\ref{auxvarpsi}) formally
for $\psi$ in flat space, we find a $\sq^{-2} \rightarrow k^{-4}$ pole in this stress tensor. The
simplest way to eliminate this higher order pole is to assume that $\varphi$ is also first order in
metric deviations from flat space, so that the entire $b'S^{(E)}$ contribution to (\ref{allanom}) can be
neglected as well. These reductions are equivalent to replacing the general non-local effective action 
of the anomaly (\ref{Tnonl}) by the much simpler form,
\be
S_{anom}[g,A]  \rightarrow  -\frac{c}{6}\int d^4x\sqrt{-g}\int d^4x'\sqrt{-g'}\, R_x
\, \sq^{-1}_{x,x'}\, [F_{\alpha\beta}F^{\alpha\beta}]_{x'}\,,
\label{SSimple}
 \ee
valid to first order in metric variations around flat space, or its local equivalent (\ref{allanom}) by
\be
S_{anom} [g,A;\varphi,\psi'] =  \int\,d^4x\,\sqrt{-g} 
\left[ -\psi'\sq\,\varphi - \frac{R}{3}\, \psi'  + \frac{c}{2} F_{\alpha\beta}F^{\alpha\beta} \varphi\right]\,,
\label{effact}
\ee
where
\bes\bea
&&\psi' \equiv  b \sq\, \psi\,, \label{tpsidef}\\
&&\sq\,\psi' =  \frac{c}{2}\, F_{\alpha\beta}F^{\alpha\beta} \label{tpsieom}\,,\\
&&\sq\, \varphi = -\frac{R}{3}\,.
\label{phieom}\eea  
\ees
Then after variation we may set $\varphi = 0$ in flat space, and the only terms which remain in the
stress tensor derived from (\ref{allanom}) are those linear in $\psi'$, {\it viz.}
\be
T^{\mu\nu}[\psi'(z)]=  \frac{2}{\sqrt{-g}}\frac{\delta S_{anom}}{\delta g_{\mu\nu}(z)}\Bigg\vert_{flat, \varphi=0} = 
\frac{2}{3}\, (g^{\mu\nu}\, \sq - \partial^{\mu}\partial^{\nu})\psi' (z)\,,
\label{anomten}
\ee
which is independent of $b$ and $b'$, and contain only second order differential
operators, after the definition (\ref{tpsidef}). Solving (\ref{tpsieom}) formally for $\psi'$ 
and substituting in (\ref{anomten}), we find
\be
T^{\mu\nu}_{anom}(z) =
\frac{c}{3}  \left(g^{\mu\nu}\sq - \partial^{\mu}\partial^{\nu}\right)_z \int\,d^4x'\, \sq_{z,x'}^{-1}
\left[F_{\alpha\beta}F^{\alpha\beta}\right]_{x'}\,,
\label{Tanom}
\ee
a result that may be derived directly from (\ref{SSimple}) as well.

By varying (\ref{Tanom}) again with respect to the background gauge potentials,
making use of (\ref{locvaru}) and Fourier transforming, we obtain
\be
\Gamma_{anom}^{\mu\nu\alpha\beta}(p,q) = \int\,d^4x\,\int\,d^4y\, e^{ip\cdot x + i q\cdot y}\,\frac{\delta^2 T^{\mu\nu}_{anom}(0)}
{\delta A_{\alpha}(x) A_{\beta}(y)} = \frac{e^2}{18\pi^2} \frac{1}{k^2} \left(g^{\mu\nu}k^2 - k^{\mu}k^{\nu}\right)u^{\alpha\beta}(p,q)\,,
\label{Gamanom}
\ee
which coincides with the first term of (\ref{Gamone}), with (\ref{F1lim}), and gives the full trace for massless fermions,
\be
g_{\mu\nu}T^{\mu\nu}_{anom} = c F_{\alpha\beta}F^{\alpha\beta} = -\frac{e^2}{24\pi^2} F_{\alpha\beta}F^{\alpha\beta}\,,
\ee
in agreement with (\ref{tranom}). We observe that as in the chiral case, the strict $1/k^2$ pole in the anomalous 
amplitude $F_1$ obtained from the $D_{\psi'\varphi} = i\lag {\cal T}\psi'\varphi \rag$ propagator auxiliary field 
applies only in the limit of (\ref{F1lim}), or equivalently for $k^2 \gg (|p^2|, |q^2|, m^2)$. The spectral representations 
and sum rule of the previous section show that when this condition is not satisfied, the two-particle intermediate 
state in the anomalous amplitude becomes a broad resonance instead of an isolated pole, as in Fig. \ref{Fig:rhoT},
and the residue of the pole at $k^2=0$ vanishes when any of $p^2, q^2$ or $m^2$ are non-zero. The tree amplitude of the
effective action (\ref{effact}) which reproduces the pole in the trace part of the $\lag TJJ\rag$ triangle
amplitude is illustrated in Fig. \ref{Fig:psiphi}.

\begin{figure}[htp]
\includegraphics[width=40cm, viewport=100 630 1000 720,clip]{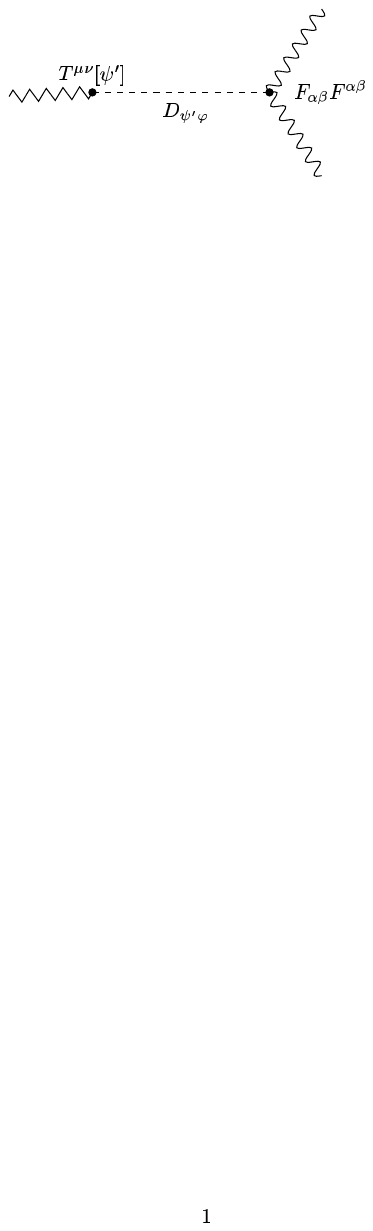}
\caption{Tree Diagram of the Effective Action (\ref{effact}), which reproduces the trace of the triangle anomaly.
The dashed line denotes the propagator $D_{\psi'\varphi} = \sq^{-1}$ of the scalar intermediate state, while 
as in Fig. \ref{Fig:epair} the jagged line denotes the gravitational metric field variation 
$h_{\mu\nu} = \delta g_{\mu\nu}$.}
\label{Fig:psiphi}
\end{figure}

Most of the remarks about the auxiliary field description of the axial anomaly at the end of Sec. 3 apply also
to the trace anomaly case. As in the case of the axial anomaly, the effective action (\ref{SEF}) or (\ref{effact}) 
explicitly exhibiting these two scalar fields is a rewriting of a part of the non-local form of the effective 
action (\ref{Tnonl}) or (\ref{SSimple}) for massless QED in curved spacetime, with the reduction to (\ref{SSimple}) 
correct to leading order in the metric deviation from flat space, $h_{\mu\nu} = \delta g_{\mu\nu}$.
The massless degrees of freedom $\varphi$ and $\psi'$ are a necessary consequence of the trace anomaly, 
required by imposition of all the other symmetries. In this case these are scalar rather than pseudoscalar 
degrees of freedom. As in the chiral case, two independent fields are required, and the propagator appearing 
in the intermediate state of the triangle amplitude is a certain off-diagonal $D_{\psi'\varphi} = 
i\lag {\cal T}\psi' \varphi\rag$ term. Unlike the chiral case the general effective action (\ref{SEF}) 
or (\ref{effact}) requires the fourth order differential operator $\Delta_4$ of (\ref{Deldef}), implying 
that higher order amplitudes such as $\lag TT JJ\rag$ should have double poles.

An important physical difference with the axial case is that the introduction of a chiral current $J_5^{\mu}$
and axial vector source ${\cal B}_{\mu}$ corresponding to it appear rather artificial, and difficult to realize in 
nature, whereas the trace of the stress tensor obtained by a conformal variation of the effective action
is simply a particular metric variation already present in the QED Lagrangian in curved space, required
by general coordinate invariance and the Equivalence Principle, without any additional couplings or 
extraneous fields. Since the stress-energy tensor couples to the universal force of gravity, we should expect 
that physical processes can excite the scalar $\varphi$ and $\psi'$ scalar degrees of freedom required by the 
trace anomaly with a gravitational coupling strength. If $m=0$ these produce effects of arbitrarily long
range. An example of this coupling to a gravitational scattering amplitude is given in the next section. 

Finally we remark that strictly speaking, the anomaly action (\ref{Tnonl}) or (\ref{effact}) and stress 
tensor derived from it contain no information about the non-anomalous or tracefree amplitudes 
$F_i, i = 3, \dots, 13$, although in certain circumstances the addition of homogeneous solutions to 
the wave eqs. (\ref{auxeom}) can give tracefree terms in the stress tensor which have physical 
consequences\,\cite{MotVau}. Our detailed computation of the full amplitude (\ref{Gamone}) shows 
that there is also a massless pole appearing in the traceless part of the physical amplitude to two 
photons, (\ref{photonmatr}) and (\ref{F3lim}). This traceless pole 
term corresponds to a term in the effective action of the form,
\be
\frac{c}{6} \int d^4x\sqrt{-g}\int d^4x'\sqrt{-g'}\, h_{\mu\nu}(x)
\, \sq^{-1}_{x,x'}\, \left[3\,(\partial^{\mu}\partial^{\nu}F_{\alpha\beta})F^{\alpha\beta}
+ \frac{1}{4} \left( g^{\mu\nu} \sq - 4\, \partial^{\mu}\partial^{\nu}\right)F_{\alpha\beta}F^{\alpha\beta}\right]_{x'}\,.
\ee
The tensor structure of this term precludes writing it as a scalar particle exchange. The pole in this
amplitude with non-trivial tensor structure is clearly connected with the possible non-zero values of 
$\partial^{\mu}\partial^{\nu}F_{\alpha\beta}$ in the background electromagnetic field, which breaks 
Lorentz invariance. Thus it appears that in this case of a non-vanishing background field which is 
non-gravitational in origin, the Ward identities obeyed by the full amplitude (\ref{Gamone}) implies 
the existence of additional massless intermediate states which do not transform as spacetime scalars, 
and therefore cannot in general be described by the anomaly induced effective action (\ref{SEF}) or
(\ref{allanom}). Instead these massless modes are associated with longitudinal components 
of the metric perturbation in a Lorentz non-invariant background, analogous to longitudinal plasmon
excitations in a finite temperature electromagnetic plasma. 

\section{Scalar Anomaly Pole Contribution to Gravitational Scattering} 

In order to verify the existence of the massless scalar pole in a physical process, we consider the simple tree
diagram of gravitational exchange between an arbitrary conserved stress-energy source $T^{\prime\,\mu\nu}$
and photons illustrated in Fig. \ref{Fig:GravScat}. 

\begin{figure}[htp]
\includegraphics[width=30cm, viewport=0 560 800 710,clip]{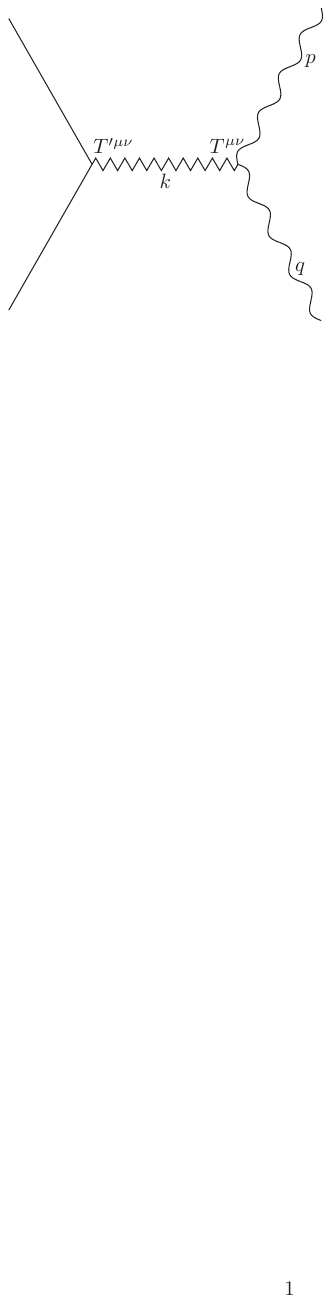}
\caption{Tree Level Gravitational Scattering Amplitude}
\label{Fig:GravScat}
\end{figure}

This process is described by the scattering amplitude\,\cite{Feyn},
\be
{\cal M} = 8\pi G \int d^4x'\int d^4x\, \left[ T^{\prime\,\mu\nu}(x')\, 
\left(\frac{\hspace{.05cm}1}{\sq}\right)_{\hspace{-.1cm}x',x}\hspace{-.2cm}T_{\mu\nu}(x) -
\frac{1}{2}\,T^{\prime\,\mu}_{\ \ \mu}(x')\,
\left(\frac{\hspace{.05cm}1}{\sq}\right)_{\hspace{-.1cm}x',x}\hspace{-.2cm}T^{\nu}_{\ \nu}(x)\right]
\label{ScatM}
\ee
The relative factor of $-\frac{1}{2}$ between the two terms is dictated by the requirement that there be no 
scalar or ghost state exchanged between the two sources, and is exactly the prediction of General Relativity, 
linearized about flat space. That only a spin-$2$ propagating degree of freedom is exchanged between the two 
sources in Fig. \ref{Fig:GravScat} can be verified by introducing the following $3+1$ decomposition for each 
of the conserved stress tensors,
\bes\bea
&& T^{00} = T_{00}\,,\\
&& T^{0i} = -V^{\perp\,i} - \partial^i \frac {1\hspace{.1cm}}{\nabla^2}\, \dot T_{00}\,,\\
&& T^{ij} = T^{\perp\,ij} + \partial^i \frac {1\hspace{.1cm}}{\nabla^2}\, \dot V^{\perp\,j}  
+ \partial^j \frac {1\hspace{.1cm}}{\nabla^2}\, \dot V^{\perp\,i} 
+ \frac{1}{2} \left(g^{ij} - \partial^i  \frac {1\hspace{.1cm}}{\nabla^2}\,\partial^j\right) (T^{\mu}_{\ \mu} + T_{00}) \nn
&& \qquad\qquad - \frac{1}{2} \left(g^{ij} - 3\,\partial^i  \frac {1\hspace{.1cm}}{\nabla^2}\,\partial^j\right)
\frac {1\hspace{.1cm}}{\nabla^2}\,\ddot T_{00}\,,
\eea \label{decomp}\ees
where $\partial_iV^{\perp\,i} = 0$, $\partial_iT^{\perp\,ij} = T^{\perp\,i}_{\ i} = 0$, and $\nabla^{-2}$ 
denotes the static Green's function of the Laplacian operator, $\nabla^2 = \partial^i\partial_i$ in flat space.
This parameterization assumes only the conservation of the stress-tensor source(s), {\it i.e.}
$\partial_{\mu} T^{\mu\nu} = 0$, so that there remain six independent components of $T^{\mu\nu}$
which must be specified, and we have chosen these six to be $T_{00}, V^{\perp\,i}, T^{\perp\,ij}$
and the total trace $T^{\mu}_{\ \mu}$, which is a spacetime scalar. Substituting the decomposition 
(\ref{decomp}) into (\ref{ScatM}) gives
\bea
&&{\cal M} = 8\pi G \int d^4x'\int d^4x\, \left[ T^{\prime\perp}_{ij} \left(\frac{1}{\sq}\right)_{\hspace{-.1cm}x',x}\hspace{-.2cm}T^{\perp}_{ij}
-2\, V^{\prime\perp}_i  \left(\frac {1\hspace{.1cm}}{\nabla^2}\right)_{\hspace{-.1cm}x',x}\hspace{-.2cm}V^{\perp}_i\right. \nn
&&\qquad\qquad \left. + \frac{3}{2}\, T'_{00}\left( \frac {1\hspace{.2cm}}{(\nabla^2)^2}\right)_{\hspace{-.1cm}x',x}\hspace{-.2cm}\sq \,T_{00}
+ \frac{1}{2}\,T'_{00} \left(\frac {1\hspace{.1cm}}{\nabla^2}\right)_{\hspace{-.1cm}x',x}\hspace{-.2cm}\,T^{\mu}_{\ \mu}
+ \frac{1}{2}\,T^{\prime\mu}_{\ \ \mu}\,\left(\frac {1\hspace{.1cm}}{\nabla^2}\right)_{\hspace{-.1cm}x',x}\hspace{-.2cm}T_{00}\right]\,,
\label{Scatpos}
\eea
which becomes
\be
{\cal M} \rightarrow -8\pi G \left[ T^{\prime\perp}_{ij} \frac{1}{k^2}\,T^{\perp}_{ij}\,
-2\, V^{\prime\perp}_i\, \frac{1}{\vec k^2}\,V^{\perp}_i +  \frac{3}{2}\, T'_{00}\,\frac{k^2}{(\vec k^2)^2}\, \,T_{00}
+ \frac{1}{2}T'_{00}\, \frac{1}{\vec k^2}\,T^{\mu}_{\ \mu}  + \frac{1}{2}\,T^{\prime\mu}_{\ \ \mu}\, \frac{1}{\vec k^2}\,T_{00}\right]\,,
\label{Scatdecom}
\ee
in momentum space. These expressions show that only the spatially transverse and tracefree components
of the stress tensor, $T^{\perp}_{ij}$ exchange a physical propagating helicity $\pm 2$ graviton in the intermediate state,
characterized by a Feynman (or for classical interactions, a retarded) massless propagator $-\sq^{-1} \rightarrow k^{-2}$ 
pole in the first term of (\ref{Scatpos}) or (\ref{Scatdecom}). All the other terms in either expression contain only an 
instantaneous Coulomb-like interaction $-\nabla^{-2} \rightarrow \vec k^{-2}$ or $\nabla^{-4} \rightarrow \vec k^{-4}$ 
between the sources, in which no propagating physical particle appears in the intermediate state of the cut diagram. 
This is the gravitational analog of the decomposition,
\bes\bea
&& J^0 = \rho\,,\\
&& J^i = J^{\perp\,i} - \partial^i \frac {1\hspace{.1cm}}{\nabla^2}\, \dot \rho\,,
\eea\ees
of the conserved electromagnetic current and corresponding tree level scattering amplitude, 
\be
\int d^4x'\int d^4x\, J^{\prime\mu}(x') \left(\frac{1}{\sq}\right)_{\hspace{-.1cm}x',x} \hspace{-.2cm}J_{\mu}(x) 
\rightarrow -J^{\prime\mu} \frac{1}{k^2} J_{\mu} = -J^{\prime\perp}_i\, \frac{1}{k^2}\,J^{\perp}_i 
+ \rho'\, \frac{1}{\vec k^2}\,\rho \,,
\label{Coul}
\ee
which shows that only a helicity $\pm 1$ photon is exchanged between the transverse components of the current,
the last term in (\ref{Coul}) being the instantaneous Coulomb interaction between the charge densities.

\begin{figure}[htp]
\includegraphics[width=10cm, height=6cm, viewport=120 560 380 720,clip]{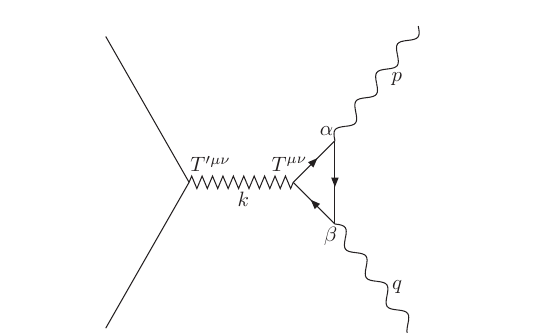}
\caption{Gravitational Scattering of Photons from the source $T^{\prime\mu\nu}$ via the triangle amplitude}
\label{Fig:GravPhoTri}
\end{figure}

We now replace one of the stress tensor sources by the matrix element (\ref{photonmatr}) of the
one-loop anomalous amplitude, considering first the trace term with the anomaly pole in $F_1$.
This corresponds to the diagram in Fig. \ref{Fig:GravPhoTri}. We find for this term,
\bes\bea
&& \lag 0\vert T_{00} \vert p,q\rag_{_1}= - \vec k^2 F_1(k^2)\, u^{\alpha\beta}(p,q)\tilde A_{\alpha}(p) \tilde A_{\beta}(q)\,\\
&& \lag 0\vert T^{\mu}_{\ \mu}  \vert p,q\rag_{_1} = 3k^2 F_1(k^2)\, u^{\alpha\beta}(p,q)\tilde A_{\alpha}(p) \tilde A_{\beta}(q)\,\\
&& \lag 0\vert V^{\perp}_i \vert p,q\rag_{_1} = \lag 0\vert T^{\perp}_{ij} \vert p,q\rag_{_1} = 0\,.
\eea\ees
Hence the scattering amplitude (\ref{Scatdecom}) becomes simply,
\be
{\cal M}_1 = 4\pi G \, T^{\prime\mu}_{\ \ \mu}\,F_1(k^2)\,u^{\alpha\beta}(p,q)\tilde A_{\alpha}(p) \tilde A_{\beta}(q)
=  \frac{4\pi G}{3} \, T^{\prime\mu}_{\ \ \mu}\,\frac{1}{k^2}\,\lag 0\vert T^{\nu}_{\ \nu}  \vert p,q\rag_{_1}\,
\label{scat1}
\ee
where (\ref{F1pole}) has been used. Thus for massless fermions the pole in the anomaly amplitude becomes 
a scalar pole in the gravitational scattering amplitude, appearing in the intermediate state as a massless scalar 
exchange between the traces of the energy-momentum tensors on each side. The standard gravitational 
interaction with the source has produced an effective interaction between the scalar auxiliary field $\psi'$ 
and the trace $T^{\prime\mu}_{\ \ \mu}$ with a well defined gravitational coupling. Thus we may equally well 
represent the scattering as Fig. \ref{Fig:GravPhoTri} involving the fermion triangle, or as the tree level diagram 
Fig. \ref{Fig:GravScal} of the effective theory, with a scalar particle exchange. 

\begin{figure}[hbp]
\includegraphics[width=30cm, viewport=30 600 800 755,clip]{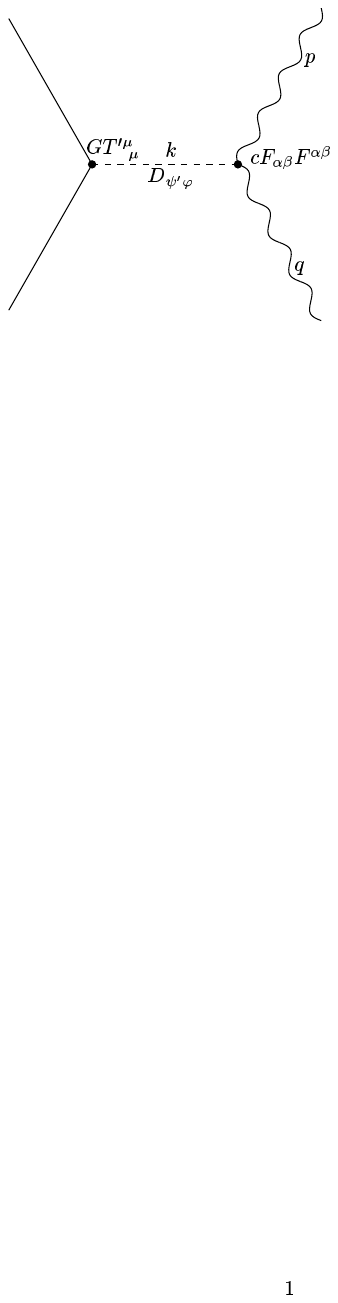}
\caption{Gravitational scattering of photons from the trace of a source $T^{\prime\mu}_{\ \ \mu}$
via massless scalar exchange in the effective theory of (\ref{effactT}).}
\label{Fig:GravScal}
\end{figure}

This diagram is generated by the effective action in flat space modified from (\ref{effact}) to
\be
S_{eff} [g,A;\varphi,\psi'] =  \int\,d^4x\,\sqrt{-g} 
\left[ -\psi'\sq\,\varphi + \frac{8\pi G}{3}\, T^{\prime\mu}_{\ \ \mu}\, \psi'
+ \frac{c}{2}\, F_{\alpha\beta}F^{\alpha\beta} \varphi
\right]\,,
\label{effactT}
\ee
to include the coupling to the trace of the energy-momentum tensor of any matter $T^{\prime\mu}_{\ \ \mu}$
source. Correspondingly the eq. (\ref{phieom}) for $\varphi$ becomes
\be
\sq \varphi = \frac{8\pi G}{3} \,T^{\prime\mu}_{\ \ \mu}\,,
\label{newphieom}
\ee
instead (\ref{phieom}).
The eq. of motion for $\psi'$ remains (\ref{tpsieom}) and is unaffected. We note that if the source
$T^{\prime\mu\nu}$ generates the curvature $R$ by Einstein's eqs., then $R = -8\pi G\,T^{\prime\mu}_{\ \ \mu}$,
so that (\ref{effactT}) and (\ref{newphieom}) are equivalent to (\ref{effact}) and (\ref{phieom}) at leading order
in $G$.

We conclude that in the conformal limit of massless electrons, the pole in the trace sector of the anomaly
amplitude contributes to gravitational scattering amplitudes as would a scalar field coupled to the trace
of the energy-momentum tensor of classical sources. The gravitationally coupled intermediate
scalar can be understood as arising from collinear $e^+e^-$ correlated pairs in a total spin $0^+$
state. Although the result appears similar in some respects to a Jordan-Brans-Dicke scalar\,\cite{JBD}, 
the coupling induced by the anomaly involves two scalar fields each coupling to a {\it different}
source, with an off-diagonal propagator, $D_{\psi'\varphi}$. Hence the phenomenology of this scalar coupling 
will be quite different, and the observational limits on a Jordan-Brans-Dicke scalar do not apply\,\cite{Will}.
In particular there is no direct coupling of classical energy-momentum sources $T^{\prime\mu}_{\ \mu}$
to $T^{\prime\nu}_{\ \nu}$ via scalar exchange as there would be in a classical scalar-tensor theory.

Another important difference is that as we have seen, the anomaly pole is a necessary consequence of quantum
fluctuations and low energy symmetries, whereas in classical scalar-tensor theories a postulated scalar field is 
simply added to Einstein's General Relativity. As a consequence there are one or more free parameters 
introduced in such an approach, whereas the effective action (\ref{effactT}) is completely specified 
without any arbitrariness or free parameters, once the underlying quantum theory's matter content and 
couplings are given. It will be interesting to study the consequences for astrophysics and cosmology
of this effective action derived from quantum first principles and fundamental low energy symmetries.

\section{Summary}

We have presented a complete calculation of the $\lag TJJ\rag$ triangle amplitude in QED, for all values of the
kinematic invariants and electron mass. As a consequence of the trace anomaly, this amplitude exhibits a 
massless pole in the conformal limit, which contributes to long range gravitational interactions, and is associated 
with the exchange of a massless $0^+$ degree of freedom. This scalar exchange is described by a low energy 
local effective action (\ref{effactT}) with two massless dynamical scalar fields $\varphi$ and $\psi'$.

For the benefit of the reader we provide here a summary of the main results to be found in each section of the paper.

We reviewed in Sec. 2 the derivation of the axial anomaly in QED, showing how the finite parts of the triangle
amplitude, together with the symmetry principles of Lorentz invariance, gauge invariance, and Bose
exchange symmetry are sufficient to yield the complete amplitude, (\ref{GJJJ}) for any mass and any value 
of the kinematic invariants, and determine the axial anomaly, without any explicit need of regularization 
of ultraviolet divergent integrals. We showed that the anomaly is closely connected to a finite sum rule of 
the spectral density (\ref{chisum}) obtained by cutting the amplitude as in Fig. \ref{Fig:ImTri}. For physical, 
transverse photons on shell, this spectral density vanishes pointwise for all $s>0$, becoming 
proportional to $\delta (s)$ in the conformal limit of massless fermions, (\ref{rhoAdel}). Corresponding to the 
$\delta (s)$ in the spectral weight is a massless pseudoscalar pole singularity in the full amplitude and matrix 
element to physical photons (\ref{matpol}). This illustrates the infrared aspect of the anomaly, and the 
appearance of massless states in a theory with anomalies in $3 + 1$ dimensions.

We showed next in Sec. 3 that the anomaly pole in the chiral case implies the existence of a non-local effective 
action, (\ref{nonl}) which can be brought into a local form by the introduction of two pseudoscalar auxiliary 
fields (\ref{auxaxeom})-(\ref{axeffact}). These fields and the anomaly pole can be understood as arising from 
a certain correlated two-particle collinear $e^+e^-$ state in the massless limit. The local effective action 
of the auxiliary fields has kinetic terms, and their canonical commutation relations reproduces the anomalous 
current commutation relations of the underlying fermionic theory (\ref{anomcc}). Thus the auxiliary fields appear 
to be {\it bona fide} massless pseudoscalar degrees of freedom, required by the chiral anomaly.

In Sec. 4 we presented a full computation of the $\lag T JJ\rag$ triangle amplitude in QED, where the
chiral current is replaced by the fermionic energy-momentum-stress tensor $T^{\mu\nu}$. Following the
same method as in the chiral case, we showed how the finite parts of the triangle amplitude, together with 
the same symmetry principles of Lorentz invariance, gauge invariance, and Bose exchange symmetry,
and the additional Ward identity following from general coordinate invariance are sufficient to yield the 
complete $\lag T JJ\rag$ amplitude, (\ref{Gamt}), given by eqs. (\ref{F1357})-(\ref{F13}) with
(\ref{tenrel3}), (\ref{Cj}) and Tables \ref{genbasis} and \ref{tensorcoeff}, for any value of mass or 
the kinematic invariants, without any need of regularization of ultraviolet divergent integrals.

In Sec. 5 we computed the trace and found the finite anomaly (\ref{Gamanomtr}), equivalent to (\ref{tranom}).
The coefficient of the second possible tensor in the trace defined in (\ref{uwdef}) is non-anomalous, but
both are needed to determine the scaling violation $\beta$ function at finite momentum and finite
electron mass. This infrared $\beta$ function is given by (\ref{betaPi}) in terms of the photon polarization
which vanishes when $p^2 \ll m^2$, consistent with decoupling, and approaches the more commonly 
considered ultraviolet $\beta$ function only in the opposite limit $p^2 \gg m^2$.

In Sec. 6 we gave the spectral representation (\ref{specrep}) for the imaginary part of the triangle amplitude, 
cut as in Fig. \ref{Fig:ImTri}, for $k^2 = -s < 0$ timelike. We showed that the imaginary part of the amplitude is
non-anomalous, with the anomaly in the real part arising from a cancellation between factors of $k^2 + s$
in both the numerator and denominator of (\ref{specrep}). For the particular linear combination of 
spectral functions appearing in the anomalous trace, $\rho_{_T}(s)$ defined by (\ref{rhoTdef}), we derived
on the one hand the finite sum rule (\ref{sumrule}), and on the other hand the representation (\ref{pointwise}),
which shows that $\rho_{_T}$ must develop a $\delta(s)$ singularity when the fermion mass, and photon
virtualities $p^2$ and $q^2$ vanish. We also exhibited this $\delta(s)$ explicitly in this limit, (\ref{specdel}).
Corresponding to the $\delta (s)$ in the spectral function $\rho_{_T}$, the corresponding full amplitude
(\ref{F1pole}) has a pole at $k^2 = 0$, indicating the presence of a massless scalar propagating state
in the matrix element of the stress tensor to physical photons (\ref{photonmatr}). As in the chiral case
the massless anomaly pole can be understood as arising from a correlated two-particle collinear $e^+e^-$ 
state, which because of the kinematics is essentially $1 +1$ dimensional, {\it c.f.} Fig. \ref{Fig:epair}. 

In Sec. 7 we showed that the trace part of the $\lag TJJ\rag$ triangle amplitude is identical with that
predicted by the covariant effective action (\ref{Tnonl}), obtained in earlier work by integrating the anomaly.
In particular, the variation of the simplified effective action (\ref{effact}) in terms of the two scalar
auxiliary fields $\psi'$ and $\varphi$ yields the amplitude (\ref{Gamanom}) which coincides with
the first term of (\ref{Gamone}) which gives its full trace in the massless limit. In the effective action
the massless scalar two particle state of the triangle amplitude is replaced by scalar fields,
whose propagator gives rise to the anomaly pole at $k^2 = 0$.

Finally in Sec. 8 we considered the tree level gravitational scattering amplitude (\ref{ScatM}), 
Fig. \ref{Fig:GravScat}, with one vertex replaced by the triangle amplitude $\lag TJJ\rag$,
and showed that the massless scalar pole in the latter survives in the physical scattering
amplitude in the conformal limit of massless electrons. In the effective theory (\ref{effactT}) 
it is described as a propagating massless scalar interaction, Fig. \ref{Fig:GravScal} between 
the trace parts of the energy-momentum sources. Abstracting from the axial and trace anomaly
QED examples presented here in detail, we conclude that the trace anomaly and anomaly pole
of conformal fields imply the existence of new long range scalar interactions with a 
gravitational coupling strength to ordinary matter.

\vspace{.5cm}
\centerline{\bf Acknowledgements}
\vspace{.2cm}

We are very much indebted to Albert Roura for extensive discussions during the course of this work, which
helped clarify a number of key points. We also thank L. S. Brown for several entlightening conversations,
and P. O. Mazur for communicating to us his discussions with A. Casher concerning the correlated 
two-particle state appearing in the triangle amplitude.

\vspace{.5cm}

\appendix
\section{Extraction of Finite Parts of $\lag TJJ\rag$}
 
For the amplitude (\ref{Gamone}), in order to extract the finite terms for
which each of the indices $(\mu\nu\alpha\beta)$ is associated with an
external momentum $p$ or $q$, we may drop the $g^{\mu\nu}$ terms in the
vertex $V^{\mu\nu}$, and consider only
\be
-\frac{e^2}{4} \int \frac{d^4 l}{(2\pi)^4} \frac{{\rm tr}
\left\{[\gamma^{\mu}(2l + p-q)^{\nu} + \gamma^{\mu} (2l + p-q)^{\mu}]
\,(-\lsl -\psl +m)\,\gamma^{\alpha}\,(-\lsl + m)\,\gamma^{\beta}
\, (-\lsl +\qsl + m)  \right\}}
{[(l+p)^2 + m^2] [(l-q)^2 + m^2] [l^2 + m^2]}\,,
\label{Gamrel}
\ee
where the continuation to Euclidean $l$ has already been performed.
Introducing the Feynman parameterization,
\be
\frac{1}{(l+p)^2 + m^2}\ \frac{1}{(l-q)^2 + m^2} \ \frac{1}{l^2 + m^2}
= 2 \int_0^1 dx \int_0^{1-x} dy\, \frac{1}{(l^{\prime\,2} + D)^3}\,,
\label{Feyn}
\ee
with $l^{\prime} = l + px - qy$ and $D$ given by (\ref{denom}), we
shift the integration variable in (\ref{Gamrel}) from $l$ to $l^{\prime}$.
Dropping the terms involving either powers of $l^{\prime}$ or $m$ in
the numerator, since these cannot give rise to terms which are homogeneous
of degree $4$ in $p$ and $q$ in $^V\Gamma^{\mu\nu\alpha\beta}$, and 
evaluating the finite Euclidean integral,
\be
\int \frac{d^4 l^{\prime}}{(2\pi)^4}\,\frac{1}{(l^{\prime\,2} + D)^3} 
= \frac{1}{32\pi^2} \frac{1}{D}\,,
\label{momint}
\ee
we obtain from (\ref{Gamrel}),
\be 
\frac{e^2}{32\pi^2} \int_0^1 dx \int_0^{1-x} dy\,\frac{1}{D}\, [p(1-2x) -q(1-2y)]^{(\mu}
[p(1-x) + qy]_{\lambda} [px-qy]_{\rho} [px + q(1-y)]_{\sigma} {\rm tr} \{ \gamma^{\nu)} 
\gamma^{\lambda} \gamma^{\alpha} \gamma^{\rho} \gamma^{\beta}\gamma^{\sigma}\}\,.
\label{GamVnext}
\ee
Of the $15$ terms in the $\gamma$-matrix trace, only the $6$ terms,
\be
\frac{1}{4}{\rm tr} \{ \gamma^{\nu} 
\gamma^{\lambda} \gamma^{\alpha} \gamma^{\rho} \gamma^{\beta}\gamma^{\sigma}\}
= - g^{\nu\lambda}(g^{\alpha\rho} g^{\beta\sigma} + g^{\alpha\sigma} g^{\beta\rho})
 - g^{\nu\rho}(g^{\alpha\lambda} g^{\beta\sigma} - g^{\alpha\sigma} g^{\beta\lambda})
- g^{\nu\sigma}(g^{\alpha\lambda} g^{\beta\rho} + g^{\alpha\rho} g^{\beta\lambda}) + \dots
\ee
need to be retained, since the other $9$ contract at least two of the free indices
$(\mu\nu\alpha\beta)$ and do not give rise to terms in which all four indices are
carried by $p$ or $q$. Thus we retain from (\ref{GamVnext}) only the terms,
\bea
&&\frac{e^2}{8\pi^2} \int_0^1 dx \int_0^{1-x} dy\,\frac{1}{D}\, \left\{ -[p(1-2x) -q(1-2y)]^{(\mu}
[p(1-x) + qy]^{\nu)} [px-qy]^{\alpha} [px + q(1-y)]^{\beta}\right.\nn
&& \qquad\qquad
-[p(1-2x) -q(1-2y)]^{(\mu} [p(1-x) + qy]^{\nu)} [px-qy]^{\beta} [px + q(1-y)]^{\alpha}\nn  
&& \qquad\qquad
-[p(1-2x) -q(1-2y)]^{(\mu} [p(1-x) + qy]^{\alpha} [px-qy]^{\nu)} [px + q(1-y)]^{\beta}\nn
&& \qquad\qquad
+[p(1-2x) -q(1-2y)]^{(\mu} [p(1-x) + qy]^{\beta} [px-qy]^{\nu)} [px + q(1-y)]^{\alpha}\nn
&& \qquad\qquad
-[p(1-2x) -q(1-2y)]^{(\mu} [p(1-x) + qy]^{\alpha} [px-qy]^{\beta} [px + q(1-y)]^{\nu)}\nn
&& \qquad\qquad
\left.-[p(1-2x) -q(1-2y)]^{(\mu} [p(1-x) + qy]^{\beta} [px-qy]^{\alpha} [px + q(1-y)]^{\nu)}\right\}
\,.
\label{TJJfinite}
\eea
In this form it is straightforward to collect the terms which multiply each of the $12$ tensors of degree
$4$ which are listed in Table \ref{tensorcoeff} of the text. For example, the coefficient of
the tensor $p^{\mu}p^{\nu}p^{\alpha}p^{\beta}$ from (\ref{TJJfinite}) is
\be
\frac{e^2}{8\pi^2} \int_0^1 dx \int_0^{1-x} dy\,\frac{1}{D}\, \left\{-4 (1-2x)(1-x)x^2\right\}\,.
\ee
When we add the Bose symmetric contribution to (\ref{TJJfinite}) with $p$ replaced by $q$
and $\alpha$ replaced by $\beta$, the coefficient of $q^{\mu}q^{\nu}q^{\alpha}q^{\beta}$ will 
give an equal contribution to the coefficient $C_1$, after also interchanging the
parameter integration variables $x$ and $y$. Thus,
\be
C_1(k^2;p^2,q^2) = \frac{e^2}{4\pi^2} \int_0^1 dx \int_0^{1-x} dy\,\frac{c_1(x,y)}{D} \,,
\ee
with
\be
c_1(x,y) = - 4 x^2(1-x)(1-2x)\,,
\ee
and we have verified (\ref{Cj}) for the first entry of Table \ref{tensorcoeff} of the text. The
remainder of the Table \ref{tensorcoeff} may be derived from (\ref{TJJfinite}) in the same way.

For the amplitude $\Lambda^{\alpha\beta}$ the calculation is similar. Beginning with (\ref{Lamampl}),
we have
\be
\Lambda^{\alpha\beta}(p,q) = e^2 m \int \frac{d^4 l}{(2\pi)^4} \frac{{\rm tr}
\left\{(-\lsl -\psl +m)\,\gamma^{\alpha}\,(-\lsl + m)\,\gamma^{\beta}
\, (-\lsl +\qsl + m)  \right\}} {[(l+p)^2 + m^2] [(l-q)^2 + m^2] [l^2 + m^2]}
+ (p \leftrightarrow q, \alpha \leftrightarrow \beta)\,.
\label{Lamnext}
\ee
Since the trace of an odd number of $\gamma$ matrices vanishes, only those terms with at
least one additional factor of $m$ in the numerator survive. Since we wish to extract
only those finite terms homogeneous of degree $2$ in the external momenta, namely
$p^{\alpha}p^{\beta}, p^{\alpha}q^{\beta}, q^{\alpha}p^{\beta}$ or $q^{\alpha}q^{\beta}$,
determining the other terms by vector current conservation (\ref{Lamcons}), we focus
only on those terms with exactly one additional factor of $m$. Using the Feynman
parameterization (\ref{Feyn}), shifting integration variables from $l$ to
$l^{\prime} = l + px - qy$, and evaluating the momentum integral (\ref{momint}) as
before, we find from the first term of (\ref{Lamnext}),
\bea
&&\frac{e^2m^2}{16\pi^2}  \int_0^1 dx \int_0^{1-x} dy\,\frac{1}{D}\, \left\{(px -qy)_{\lambda}
[px + q(1-y)]_{\mu} {\rm tr}( \gamma^{\alpha}\gamma^{\lambda}\gamma^{\beta}\gamma^{\mu})\right.\nn
&&\quad \left.- [p(1-x)+qy]_{\lambda} [px + q(1-y)]_{\mu} {\rm tr}(\gamma^{\lambda}\gamma^{\alpha}\gamma^{\beta}\gamma^{\mu})
- [p(1-x) + qy]_{\lambda} (px-qy)_{\mu} {\rm tr}( \gamma^{\lambda}\gamma^{\alpha}\gamma^{\mu}\gamma^{\beta})\right\}\,.
\eea
In the $\gamma$ matrix traces we may further discard all terms involving $g^{\alpha\beta}$, which
leaves the remaining terms,
\bea
&&\frac{e^2m^2}{4\pi^2}  \int_0^1 dx \int_0^{1-x} dy\,\frac{1}{D}\, \left\{(px -qy)^{\alpha}
[px + q(1-y)]^{\beta} + (px -qy)^{\beta} [px + q(1-y)]^{\alpha}\right.\nn
&&\qquad -[p(1-x)+qy]^{\alpha} [px + q(1-y)]^{\beta} + [p(1-x)+qy]^{\beta} [px + q(1-y)]^{\alpha}\nn
&& \qquad \left. -[p(1-x)+qy]^{\alpha} (px -qy)^{\beta} - [p(1-x)+qy]^{\beta}(px -qy)^{\alpha}\right\}\,.
\eea
Adding the Bose symmetrized term with $p \leftrightarrow q$ and $\alpha \leftrightarrow \beta$, we
obtain for these finite terms,
\bea
\hspace{-.8cm} -\frac{e^2m^2}{2\pi^2}\int_0^1 dx \int_0^{1-x} dy\,\frac{1}{D}\,\left\{2 p^{\alpha}p^{\beta} x(1-2x)
+ p^{\alpha}q^{\beta} (1-2x)(1-2y) - q^{\alpha}p^{\beta} (1-4xy)
+ 2 q^{\alpha}q^{\beta} y(1-2y) \right\}\,.
\label{Lameval}
\eea
From the definitions of $G_1$ and $G_2$ in (\ref{G1G2}) and $u^{\alpha\beta}(p,q)$ and
$w^{\alpha\beta}(p,q)$ in (\ref{uwdef}), it follows that the coefficient of $-q^{\alpha}p^{\beta}$ is
\be
G_1 (k^2; p^2, q^2) = - \frac{e^2m^2}{2\pi^2}\int_0^1 dx \int_0^{1-x} dy\,\frac{(1-4xy)}{D}\,,
\ee
and the coefficient of $(p\cdot q) p^{\alpha}q^{\beta}$ is
\be
G_2 (k^2; p^2, q^2) = - \frac{e^2}{2\pi^2} \frac{m^2}{p\cdot q}\int_0^1 dx \int_0^{1-x} dy\,\frac{(1-2x)(1-2y)}{D}\,,
\label{G2app}
\ee
which are Eqs. (\ref{G12}) of the text. 

\section{Proof of Identities}

The coefficients of the $p^{\alpha}p^{\beta}$ and $q^{\alpha}q^{\beta}$ terms in (\ref{Lameval})
apparently do not match those of $w^{\alpha\beta}(p,q)$ with the identification of $G_2$ given.
This mismatch is only apparent, because of the identities,
\bes\bea
&&\int_0^1 dx \int_0^{1-x} dy\,\frac{2x(1-2x)}{D} = - \frac{q^2}{p\cdot q}
\int_0^1 dx \int_0^{1-x} dy\,\frac{(1-2x)(1-2y)}{D}\,, \label{identone}\\
&&\int_0^1 dx \int_0^{1-x} dy\,\frac{2y(1-2y)}{D} = - \frac{p^2}{p\cdot q}
\int_0^1 dx \int_0^{1-x} dy\,\frac{(1-2x)(1-2y)}{D} \label{identtwo}\,.
\eea\ees
These identities are most easily proven by considering integrals of the kind,
\be
\int_0^1 dx \int_0^{1-x} dy\,(1-2x)\,\frac{\partial \ln D}{\partial y}
= \int_0^1 dx \int_0^{1-x} dy\,(1-2x)\,\frac{1}{D}\,\frac{\partial D}{\partial y}\,,
\label{intone}
\ee
which on the one hand vanishes, because
\be
\int_0^1 dx \,(1-2x)\,\ln D\Big\vert_{y=0}^{y=1-x} =
\int_0^1 dx \,(1-2x)\,\ln \left\{ \frac{x(1-x)k^2 + m^2}{x(1-x)p^2 + m^2}\right\} = 0\,,
\label{evenodd}
\ee
due to the fact that $(1-2x)\rightarrow -(1-2x)$ is odd upon reflection about the midpoint of the integral,
$x \rightarrow 1-x$, whereas $x(1-x) \rightarrow x(1-x)$ is even; while on the other hand, (\ref{intone}) 
is equal to
\be
\int_0^1 dx \int_0^{1-x} dy\,(1-2x)\,\left[\frac{(1-2y)q^2 + 2x p\cdot q}{D}\right]\,.
\ee
Setting this expression to zero and rearranging gives (\ref{identone}). The
second identity (\ref{identtwo}) is proven in a similar manner by exchanging $x$ and $y$.
When these two identities are substituted into (\ref{Lameval}), the
$p^{\alpha}p^{\beta}, p^{\alpha}q^{\beta}$ and $q^{\alpha}q^{\beta}$ terms become
proportional to the corresponding terms in $w^{\alpha\beta}$ defined in (\ref{uwdef}),
and the coefficient function (\ref{G2app}) is obtained in every case. We remark also that
despite appearances, $G_2$ has no pole at $p\cdot q =0$, since the integral in (\ref{G2app}) 
multiplying it vanishes if we set $p\cdot q =0$ in the denominator $D$, by an argument 
similar to that leading to (\ref{evenodd}).

The identities (\ref{Crel}) are proven in a similar way. For example, from Table \ref{tensorcoeff} and (\ref{Cj}),
\bea
&& (p\cdot q) C_1 + q^2 C_4 = -\frac{e^2}{2\pi^2}\int_0^1 dx \int_0^{1-x} dy \ \frac{x(1-x)(1-2x)}{D} 
\left[ 2 x p\cdot q + (1-2y) q^2 \right]\nn
&& \qquad \qquad = -\frac{e^2}{2\pi^2}\int_0^1 dx \ x(1-x)(1-2x)\int_0^{1-x} dy\ \frac{\partial \ln D}{\partial y}\nn
&& \qquad\qquad = -\frac{e^2}{2\pi^2}\int_0^1 dx\ x(1-x)(1-2x)\ 
\ln\left\{\frac{x(1-x)k^2 + m^2}{x(1-x)p^2 + m^2}\right\} = 0\,,
\eea
for the same reason (\ref{evenodd}) vanishes. This proves (\ref{Crela}), with (\ref{Crelb}) proven in exactly the
same manner after interchanging $p^2$ and $q^2$, and $x$ and $y$. 

For the first of identities (\ref{addWI}), we employ a similar method. In the Feynman parameterized 
integral representation, using (\ref{Cj}) with Table \ref{tensorcoeff}, we find that the linear combination,
\be
p^2 \left(C_1-2 C_2+C_3-2 C_7+2 C_8\right) - 2 p\cdot q \left(C_7-C_9 \right)
-q^2 \left(2 C_8-2 C_9+C_{10} - 2 C_{11}+C_{12}\right)\nonumber
\ee
is proportional to
\bea
&&\int_0^1dx\int_0^{1-x}dy\, \frac{(1-x-y)}{D} \big\{ x(x+3y-1)(1-2x)p^2 - 4xy (x-y) p\cdot q -y(3x+y -1)(1-2y)q^2\big\}\nn
&&= \int_0^1dx\int_0^{1-x}dy\, (1-x-y) \left\{ x(x+3y-1)\frac{\partial \ln D}{\partial x} - y(3x+y -1)\frac{\partial \ln D}{\partial y}\right\}\nn
&& = \int_0^1dx\int_0^{1-x}dy\, \ln D \left\{ \frac{\partial}{\partial x} \Big[x(1-x-y)(1-x-3y)\Big] 
- \frac{\partial}{\partial y}\Big[y(1-x-y)(1-3x-y)\Big]\right\} = 0\,,
\eea
which vanishes identically.

{\allowdisplaybreaks
Lastly, the linear combination of terms in (\ref{identF1}), 
\be
p^2 (C_1 - 2 C_2 + C_3 + 2 C_4 - 4 C_{8}) + 2 p\cdot q  (C_3 - 2 C_5 - C_7 - C_9 + C_{10})
+ q^2 (2 C_6 - 4 C_8 + C_{10} - 2 C_{11} + C_{12}) \nonumber  \,,
\ee
after substituting for the $C_j$ from (\ref{Cj}) with Table \ref{tensorcoeff} is proportional to
\bea
&&  \int_0^1dx\int_0^{1-x}dy\,\frac{1}{D}\, \bigg\{  x \Big[x(1-x)+y(3 y-2)\Big]  (1-2x)\, p^2 -  2 x y \Big[ x(1-2x)+y(1-2y) \Big] \pq  \nn
&& \hspace{4cm}  +  \ y  \Big[ y(1-y)+x (3 x-2) \Big]  (1-2y) \,q^2  \bigg\}  \\
&&  = \int_0^1 dy \int_0^{1-y} dx\ x  \Big[x(1-x)+y (3 y-2)\Big]  \frac{\partial \ln D}{\partial\,x} 
+ \int_0^1 dx \int_0^{1-x} dy\ y  \Big[y(1-y)+x (3 x-2)\Big]  \frac{\partial \ln D}{\partial\,y} \nn
&&  = -\int_0^1 dy   \,  y (1-y)(1-2y) \ln \left[y(1-y) k^2 + m^2\right] 
 -\int_0^1 dx    \, x (1-x) (1-2x) \ln \left[x(1-x) k^2 + m^2\right]  \nn 
&&+ \int_0^1 dy  \int_0^{1-y} dx\, (x-y)(3y+3x-2) \ln  D
 - \int_0^1 dx  \int_0^{1-x} dy\, (x-y)(3y+3x-2) \ln D  = 0\,,\nonumber
\eea
which also vanishes because the last two terms cancel, while the first and second terms are each 
separately zero by their odd parity under reflection through the midpoint of the remaining integral.}
The other identities can be checked by similar methods, and with the help of algebraic
manipulation software such as Mathematica.

\end{document}